\documentclass{article}

\usepackage[utf8]{inputenc}
\usepackage[LGR,T1]{fontenc}
\usepackage[greek,british]{babel}

\usepackage{amsmath}

\usepackage[margin=1in]{geometry}
\usepackage{natbib}

\bibpunct{(}{)}{;}{a}{}{,} 

\usepackage{wasysym}
\usepackage{amssymb}
\usepackage{siunitx}
\DeclareSIUnit\mearth{M_\oplus}
\DeclareSIUnit\rearth{R_\oplus}
\DeclareSIUnit\mj{M_{\textrm{\jupiter}}}
\DeclareSIUnit\rj{R_{\textrm{\jupiter}}}
\DeclareSIUnit\msun{M_\odot}
\DeclareSIUnit\year{yr}
\DeclareSIUnit\au{au}

\def\beq{\begin{equation}}
\def\eeq{\end{equation}}
\usepackage{graphicx}
\usepackage{url}
\usepackage{hyperref}             
\hypersetup{pdfauthor=Christoph Mordasini}
\hypersetup{backref=true, pagebackref=true, hyperindex=true, breaklinks=true,colorlinks=true,urlcolor=blue, linkcolor=blue,  citecolor=blue,pagecolor=red, bookmarks=true, bookmarksopen=true}

\usepackage{xcolor}

\title{Planet formation - observational constraints, physical processes, and compositional patterns}
\author{Christoph Mordasini$^1$ \& Remo Burn$^2$}
\date{\small{
    $^1$Division of Space Research and Planetary Sciences, University of Bern, Switzerland\\%
    $^2$Max Planck Institute for Astronomy, Heidelberg, Germany\\[2ex]}%
}

\begin{document}

\maketitle

\section{Introduction}
\subsection{Motivation and chapter content}
A general theory of planet formation has been a topic of intense study over many years. The interest in such a theory emerges naturally from asking the question of where our planet came from. However, a general picture is also required which explains not only the planet Earth but the whole Solar System with its diverse planets. Moreover, after the first discovery of an exoplanet around a Sun-like star \citep{1995NatureMayorQueloz}, the exoplanet revolution added numerous additional constraints from thousands of systems of planets orbiting stars different than the Sun.

The goal of planet formation as a field of study is not only to provide the understanding of how planets come into existence. It is also an interdisciplinary bridge which links astronomy to geology and mineralogy. Recent observations of young stars accompanied by their protoplanetary disks \citep{2022PPVIIManara} provide direct insights into the conditions at which planets are forming. These astronomical observations can  be taken as initial conditions for the models of planet formation. In particular, we are currently gaining insights into the composition of the inner, planet-forming region within the disks thanks to observations from the James Webb Space Telescope \citep{Grant2023,Perotti2023}. The task for planet formation modelling is then to link these observational properties and compositional content to the physical properties, and also the  elementary inventory of meteorites, the Moon, Earth, and the other planets. If successful, this global approach can provide useful constraints for geological studies.

Here, we will review some of the recent steps that have been undertaken in this task. In particular, we will provide a short overview of observational constraints from the Solar System and from exoplanets, discusing also recent shifts of paradigms (Sect. \ref{sec:obs_constraints}). This is followed by a simple, educational review of a number of relevant physical processes (Sect. \ref{sec:physical_processes}) before we put the pieces together to show some state-of-the art model results regarding the physical (Sect. \ref{sec:results_phys}) and compositional (Sect. \ref{sec:results_compo}) properties of model planetary systems.

\subsection{Challenges in planet formation theory}
In the classical planet formation theory based on the Solar System alone, it was expected that there should be no massive planets inside of about 3 AU for physical reasons that we will discuss below, and that the overall architecture of the Solar System with lower mass inner rocky and massive outer gaseous and icy planets should be universal.

However, in the last nearly three decades, a very large population of exoplanets has been found exactly where the Solar System-based formation theory did not predict them. This pointed towards a serious gap in the understanding of planet formation derived from our planetary system alone and pointed at the necessity of developing a general theory that is not only based on a handful, or even a single planetary system, but can address both the Solar System and the exoplanets in a general way.

\begin{figure*}[h]
	\centering
	\includegraphics[width=1\linewidth]{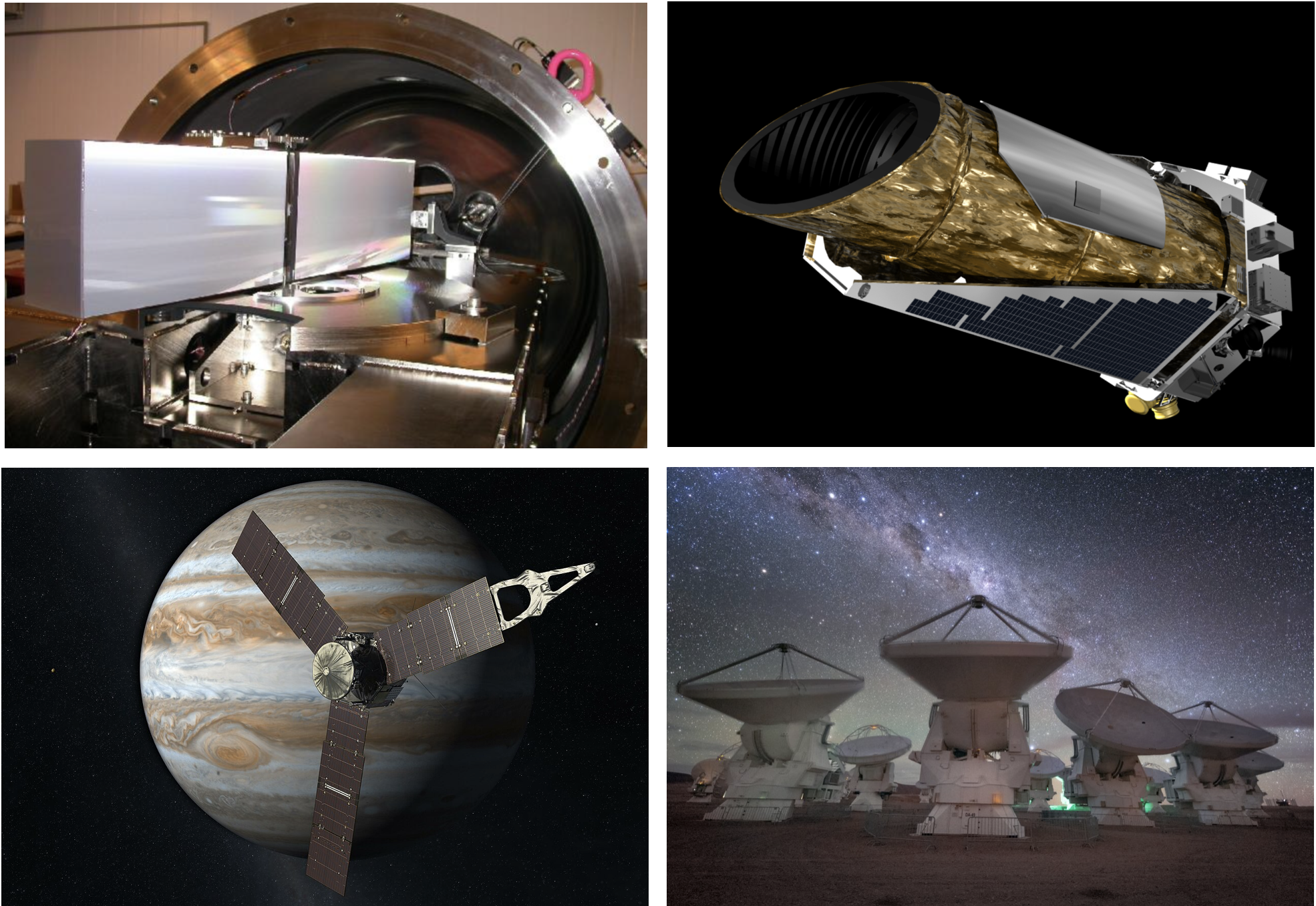}
	\caption{Observational facilities yielding key constraints for planet formation theory. Top left: the HARPS spectrograph \citep{2003MsngrMayor} installed at ESO's 3.6 Meter telescope in Chile. Through its unprecedented radial velocity precision of 1 m/s, it allowed for the first time the discovery of low-mass extrasolar planets and the study of exoplanetary system architectures in the mass-distance diagram. Top right: the NASA Kepler satellite \citep{2010ScienceBorucki}. With ultra-precise transit photometry it has unveiled the demographics of the transiting exoplanet population. Bottom left: The NASA Juno satellite in orbit around Jupiter has deeply changed our picture of the interior of giant planets. Bottom right: the joint American-European-Japanese Atacama Large Millimeter/Submillimeter Array (ALMA) has started a new era in the study of protostellar and protoplanetary disks, the cradles of planets (Image credits: HARPS: ESO; ALMA: CC BY 4.0 ESO/Beletsky; Kepler: NASA; Juno: NASA/JPL-Caltech)}
	\label{fig:instruments}
\end{figure*}

While considerable progress has been made including the shift of several paradigms (see Sect. \ref{sect:predictionsclassiccaltheorypardigmshifts}), it has proven very challenging to come up with such a general theory based on the first principles of physics alone (conservation laws of mass, energy, and momentum), and the field is currently still far from a complete understanding of how planets from and evolve. This implies a current absence of a highly predictive theory of planet formation - the field is rather driven by observational discoveries, often made possible thanks to new observational facilities on the ground or in space. Observational facilities that have played a key role are depicted in Figure \ref{fig:instruments}. 

Developing a theory of planet formation is challenging for a number of reasons. First, planet formation is a process covering a huge range in spatial scales from  dust grains with a size on the order of $10^{-4}$ cm to giant planets with a radius on the order of $10^{10}$ cm. At each size scale, there are different governing physical processes involving multiple input physics like gravity, hydrodynamics, thermodynamics, relative transport, magnetic fields, high pressure physics, and so on.

Second, there are many strong nonlinear mechanisms and feedbacks. For example, we now understand that in the classical picture from the Solar System, where giant planets were predicted to exist only at large orbital distances, the effect of orbital migration (the radial motion of protoplanets through the natal gas disk) was missing. Orbital migration is caused by an exchange of angular momentum between the embedded protoplanet and the disk. This process does, however, not only change the planet's orbit and thus its angular momentum, but (because overall, angular moment needs to be conserved) also changes the structure of the gas disk, leading in particular to the formation of gaps in the disk. This implies that the different processes acting during planet formation (here the co-evolution of the planets' orbit and of the disks' structure) must be described in a self-consistently coupled way.

Third, in contrast to other fields where laboratory experiments can be used as a source of ground truth, for planet formation, laboratory experiments are only possible for special regimes, namely for the early growth regimes from dust to pebbles. But also the approach via numerical simulations has its limits: global three-dimensional high-resolution radiation-magneto-hydrodynamic simulations covering the entire growth of a planetary system would be much too computationally expensive. Taken together, these challenges demonstrate the importance of observational guidance to develop a conclusive theory of planet formation and the necessity of quantitatively confronting theoretical and observational results, which is by itself not trivial. It is also important to take into account many different observational constraints as they concern and constrain different aspects of the theory. 

\section{Overview of observational constraints}
\label{sec:obs_constraints}
Our current understanding of the origin and evolution of planets is mainly based on three different data sets: the Solar System, the extrasolar planets, and especially since the advent of ALMA observations, the protoplanetary disks surrounding young stellar objects. The latter yield crucial observational constraints on the initial and boundary conditions of planet formation. In a few instances over the past few years, ongoing planet formation can even be observed directly (see \citealt{2018AAKeppler} for the discovery of the first bona fide extrasolar gas giant planet undergoing active gas accretion). For an overview of the constraints coming from disk observations, see \citet{2020ARAAAndrews,2022PPVIIMiotello,2022PPVIIManara}. In the following, we first discuss selected constraints from the Solar System and then from the extrasolar planet population. 

\subsection{Selected constraint from the Solar System}
Before the detection of the exoplanets and protoplanetary disks, the classical theory of planet formation  \citep{1969BookSafronov,1981PThPSHayashi,1986IcarusBodenheimerPollack,1993IcarusWetherillStewart,Lissauer1993,1993IcarusIdaMakino,1996IcarusPollack} was based on centuries of Solar System studies going back to \citet{Kant1755} who first proposed that planets form in a flattened  disk revolving about the Sun. These studies include remote and in situ observations of major and minor bodies (planets, minor planets, asteroids, comets), laboratory measurements of meteorites, theoretical studies, as well as sample return missions. Today, the Solar System remains the benchmark system for planet formation theory, because only here we have access to a very high number of unique detailed constraints not accessible for exoplanets: the interior structures of the planets, the properties and dynamics of small bodies not directly observable around other stars, and cosmochemical constraints obtained from the meteoritic record and the interior and atmospheric composition and structure of the various bodies. The latter yields the abundances of elements and isotopes providing information on composition and origin of planetary building blocks, accretion processes, the timing, and early planetary-wide chemical differentiation. 

A non-exhaustive list of central constraints for planet formation theory including in particular some elements not available for exoplanets (or contrasting them) is given by the following points. There are, of course, many more constraints, see reviews of, e.g., \citealt{cameron1988,Lissauer1993ARAA,dunlop2013solar,Spohn2015trge,encrenaz2021solar}):
\begin{itemize}
\item The chronology of important events during the formation of the Solar System like the formation of calcium-aluminium-rich inclusions as first solids in the Solar system representing the temporal zero point \citep{AnandMezger2023}, the formation of chondrites \citep{Kleine2009}, the accretion of the terrestrial planets \citep{Yin2002Nature,Onyett2023Nature}, core formation (\citealt{2002NaturKleine}), the moon forming impact \citep{1989IcarusBenz,2001NatureCanup}, the formation timescale of Jupiter \citep{2017PNASKruijer}, the lifetime of the Solar gas nebula \citep{2017ScienceWang} or the timing of a potential late dynamical instability leading to a re-rise of impacts (the late heavy bombardment, \citealt{BottkeNorman2017}).

\item The orbits of the terrestrial planets, in particular their small eccentricities (varying over Milankovich cycles, see \citealp{Berger1991}, with present-day values of 0.017, 0.007, 0.094, and 0.2 for Earth, Venus, Mars, and Mercury) and the absence of planets inside of 0.39 AU (the orbit of Mercury). The latter is in stark contrast to very frequent extrasolar systems with close-in (inside of 0.1 AU) compact systems of multiple super-Earth and sub-Neptune planets \citep{2018AJPetigura,2022PPVIIWeiss}.

\item The masses of the terrestrial planets, in particular Mars’ small mass and its short formation timescale \citep{DauphasPourmand2011}. Based on classical formation models \citep{Lissauer1993ARAA} adopting a Minimum-Mass-Solar-Nebula (MMSN)-like distribution of solid building block (definition see below), one would rather expect that planet mass increases with semimajor distance. Likely, the small mass is an imprint of the important role of Jupiter in shaping the entire Solar System, as supposed for example in the ``Grand tack model'' \citep{2011NatureWalsh}. In this model, it is assumed, based on the Masset-Snellgrove mechanism of outward migrating pairs of resonant giant planets \citep{2001ApJMasset}, that Jupiter migrated first to 1.5 AU before tacking to migrate again outwards after Saturn had caught up, thereby truncating the planetesimal disk and depriving Mars from further building blocks. Mars does thus probably represent the protoplanet stage rather than a fully grown terrestrial planet emerging from the final giant impact growth phase. 

\item The structure of the asteroid belt including the absence of big bodies and the presence of a radial compositional gradient \citep{DeMeoCarry2014} from relatively dry ordinary chondrites (less than 0.1\% water by mass) to carbonaceous chondrites with 5–20\% water by mass \citep{2017IcarusRaymondIzidoro}, as well as the size distribution of the asteroids which gives clues about the likely typical primordial size of planetesimals (probably $\sim$100 km, \citealt{2009IcarusMorbidelli}).

\item Earth’s total water content, usually thought to have been delivered in the form of water-rich primitive asteroidal material \citep{2000MAPSMorbidelli}. Alternatively, it might have formed to some extent intrinsically from magma ocean - primordial atmosphere interactions \citep{Young2023Nature} or was delivered in parts by Theia, the impactor of the moon forming impact \citep{Budde2019,Mezger2021}.

\item The masses and orbits (semimajor axes, eccentricities, inclinations) of the gas and ice giant planets, characterised and contrasting the extrasolar planets by the absence of mean motion resonances and low eccentricities and inclinations. It has been proposed that the orbital distances of Jupiter and Saturn (5.2 and 9.5 AU) are larger than what is typical for extrasolar giant planet systems (confined to rather 1-3 AU) which would mean that other systems are typically more compact than the Solar system \citep{2019ApJFernandes,2021ApJSFulton}. However, this question is not yet settled because of the detection bias against distant exoplanets, and some other studies suggest that current data on exoplanets does not suggest a decline in frequency outside of 3 to 8 AU \citep{Lagrange2023}.
    
\item The giant planets' internal structure and their bulk and atmospheric composition as found from measurement of the gravitational moments (in particular by the JUNO spacecraft, \citealt{2017GRLWahl}) or by entry probes and remote observations as well as their intrinsic luminosity \citep{GuillotGautier2015,Helled2022Icarus}. This includes the over-luminosity of Saturn and the under-luminosity of Uranus relative to predictions of fully-convective interior models \citep{Nettelmann2013,2019AALinder}. Particularly important is Jupiter’s complex structure as informed by the Juno probe \citep{2017GRLWahl} with a diluted core and non-convecting regions with entropy and compositional gradients \citep{DebrasChabrier2019,Miguel2022,2023HowardGuillot}. This  contrasts the traditional view of a single fully convective adiabatic interior of giant planets (see \citealt{Guillot2023} for a review). 

\item The structure, dynamics, and evolution of the Kuiper belt which constrains the size of the nascent planetesimals disk and which might have played a crucial role triggering a dynamical instability among the giant planets as supposed in the Nice Model for the formation of the Solar System \citep{Levison2008Icarus}.

\item Further compositional and isotopic constrains (e.g., for deuterium and noble gases) for various bodies (Earth's atmosphere, asteroids, comets like 67P), which are of key interest for example for the origin of Earth’s water and its atmospheres \citep{Altwegg2015,Bekaert2020}.
\end{itemize}

\subsection{Statistical constraints: the demographics of extrasolar planets}
On the other hand, there is the continuously growing population of extrasolar planets. The situation for exoplanets is quite different from the Solar System. We have typically little knowledge about an individual exoplanetary system, although there are of course, also notable exceptions. But there is a very large number of exoplanets known (currently about 5000), meaning that the exoplanet population yields statistical constraints on, for example, the frequencies of certain planet types, the distributions of fundamental properties, or on numerous correlations between different planetary properties, between stellar properties, and between planetary system properties. For quantitative statistical constraints, large surveys play a primary role, as for them it is possible to account for the observational bias and to correct for it. Important examples are the
HARPS survey \citep{2011MayorArxiv} and the California Legacy survey \citep{2021ApJSRosenthal} for the radial velocity method; the Kepler satellite for photometric transits \citep{2010ScienceBorucki,2018AJPetigura}; direct imaging surveys like GEMINI \citep{2019AJNielsenA} or SPHERE \citep{2021AAVigan}; and finally  microlensing surveys \citep[][]{2016ApJSuzuki}. These different techniques probe planets with different properties (like close-in versus distant planets). This is important, as different sub-populations are most important in constraining different aspects of planet formation theory. This  also implies that for a general theory, the constraints of all different methods should be considered together self-consistently. In the following, we present a selection of important results obtained from these observations. We start with an overview, and then address the frequency of different types of exoplanet, followed by the distribution of  planetary characteristics, and finally a number of correlations, for example with the properties of the host star. More detailed and comprehensive information can be found in the reviews of \citep{2015ARA&AWinn,2021ARAAZhuDong,2022PPVIIWeiss,Lissauer2023PPVII}.  

\begin{figure*}[htb]
	\centering
	\includegraphics[width=0.8\linewidth]{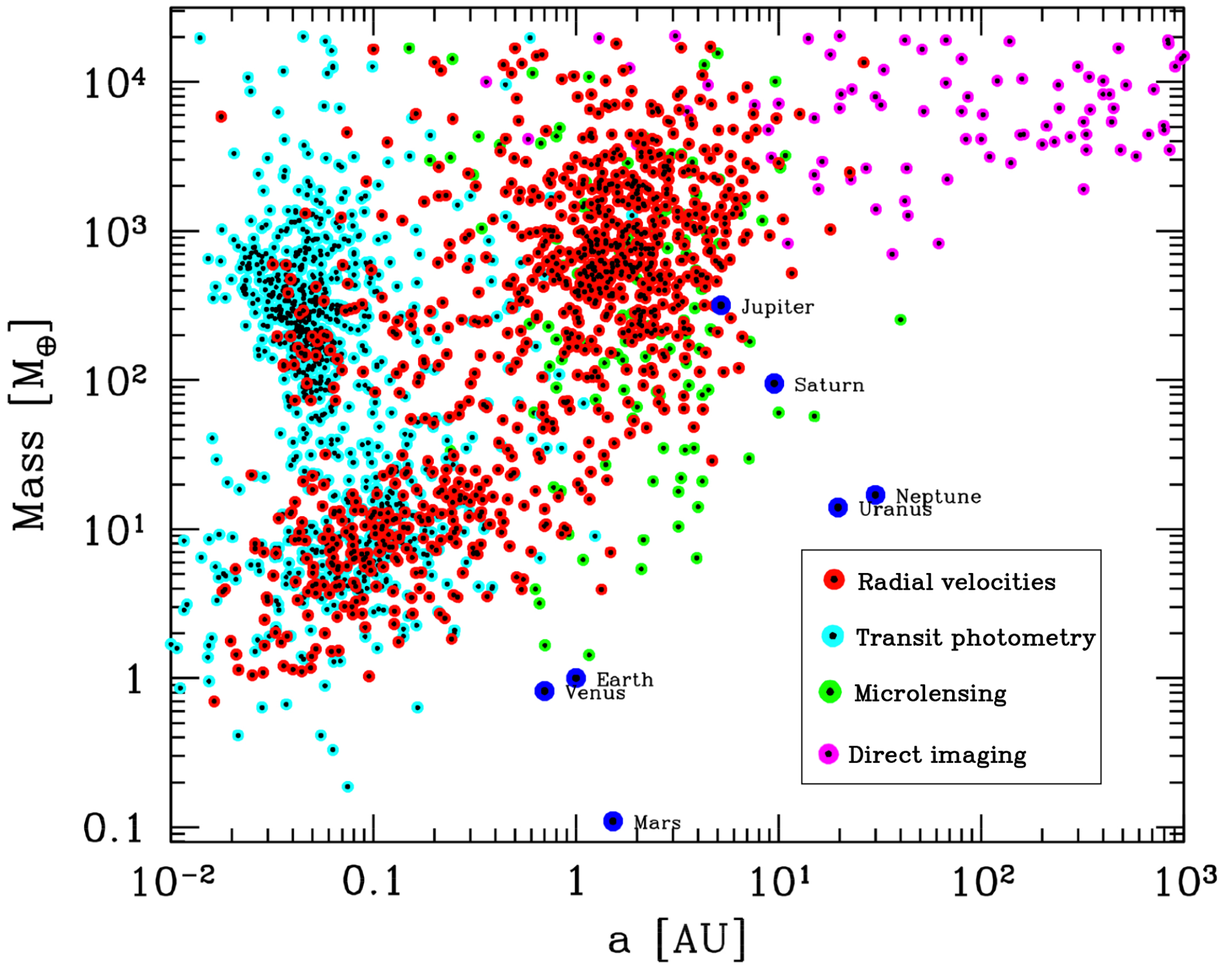}
	\caption{The observed mass-distance diagram of extrasolar planets. The colors indicate the detection method. The planets of the Solar System are also included for comparison. While illustrating the diversity of extrasolar planets and pointing at the existence of some structure (voids, over-densities), the diagram also gives a highly biased view of the underlying exoplanet population. }
	\label{fig:amobserved}
\end{figure*}

\subsubsection{The mass-distance diagram}
As an overview, Figure \ref{fig:amobserved} shows one of the most important diagrams when it comes to extrasolar planets, which is the observed mass-distance diagram. It importance is comparable to the Hertzsprung-Russel diagram for stars. The coloured points show the exoplanets discovered by the radial velocity (red), transit photometry (cyan), direct imaging (magenta), and  microlensing (green) techniques.

While giving an impression of the diversity of the outcome of the planet formation process that theoretical models must reproduce, this observed diagram also gives a highly biased view, making for example Hot Jupiters appear frequent while when correcting for the detection bias, they are a rare planet type. The absence of planets in the bottom right of the diagram is also simply caused by the current observational bias, that is, the current inability to detect distant low-mass planets. Nevertheless, besides the diversity, one can identify in the diagram some structures in the form of over- and under-densities as well as the aforementioned complementarity of the different detection methods.

\subsubsection{Frequency of different types of planets}\label{sect:frequenciesobs}
Despite their prominence in the observed mass-distance diagram, close-in giant planets (Hot Jupiters, also known as pegasids after the prototypical first exoplanet 51 Pegasi b \citealt{1995NatureMayorQueloz} of this kind) have a bias-corrected frequency of occurrence of only about 0.5-1\% for Solar-like stars \citep{2010ScienceHoward,2011MayorArxiv,2018ARAADawson}. This is much less than the frequency of giant planets at larger orbital distances, where an upturn in frequency is seen at about 1 AU \citep{2007ARAAUdrySantos}. Within 5-10 AU, about 10 to 20\% of FGK stars have giant planets  \citep{2008PASPCumming,2011MayorArxiv,2021ApJSFulton}. This means that the clear majority of Solar-like stars does not a giant planet companion, which might have had profound consequences for the Solar System \citep{2015IcarusMorbidelli}. For lower stellar masses, there is a clear trend towards less frequent occurrence of giant planets (e.g. \citealp{Gaidos2013}, see also Section 
\ref{sect:obscorrstellarprops}).

About half of the extrasolar giant planets are members of multi-giant systems  \citep{2016ApJBryan} like our Jupiter-Saturn system. As  shown initially by high precision radial velocity surveys \citep{2006NatureLovis} and then the Kepler transit survey, low-mass respectively small close-in planets are a very frequent type of planet. Depending on the exact criteria such super-Earth and sub-Neptunian planets at orbital distances of fractions of an AU with masses of about 1 to 30 Earth masses, or radii of $R\lesssim4 R_\oplus $ have a high frequency (20-50\%) for Solar-like stars \citep{2011MayorArxiv, 2013ApJFressin,2013PNASPetigura,ZhuPetrovich2018}. These planets are frequently members of multiple systems with compact architectures \citep{2022PPVIIWeiss}. This implies again that many Solar-like stars have a planetary system that differs clearly from the Solar System. Moving to (very) large orbital distances of tens to hundreds of AU, there is a lower frequency on a level of about 1-5\% of stars with detectable, i.e., sufficiently luminous distant giant planets as probed by direct imaging surveys \cite{2021AAVigan}. An implication of this is that the frequency of giant planets must decrease with distance by a factor of several as we move outwards. This puts constraints on the efficiency of planet formation by gravitational instability which would be the prime mechanism to populate this region \citep{2018MNRASForgan}. The occurrence rate of such planets probably scales with the mass of the host star \citep{2016PASPBowler}. As visible in Fig. \ref{fig:amobserved}, microlensing surveys mainly probe an intermediate orbital distance range of cold, roughly Neptunian-mass planets around M dwarfs, which are found to be a abundant population \citep{CassanKubas2012,2016AASuzuki}. Finally, if we consider the overall fraction of stars with planets that can be detected by a high-precision radial velocity survey at about 1 m/s precision, a value of about 75\% is found \citep{2011MayorArxiv}. A similarly high number applies to M-dwarfs \citep{2013AABonfils}.

\subsubsection{Distributions of important planetary properties}\label{sect:distribobsprop}

\begin{figure*}[h!]
	\centering
	\includegraphics[width=0.95\linewidth]{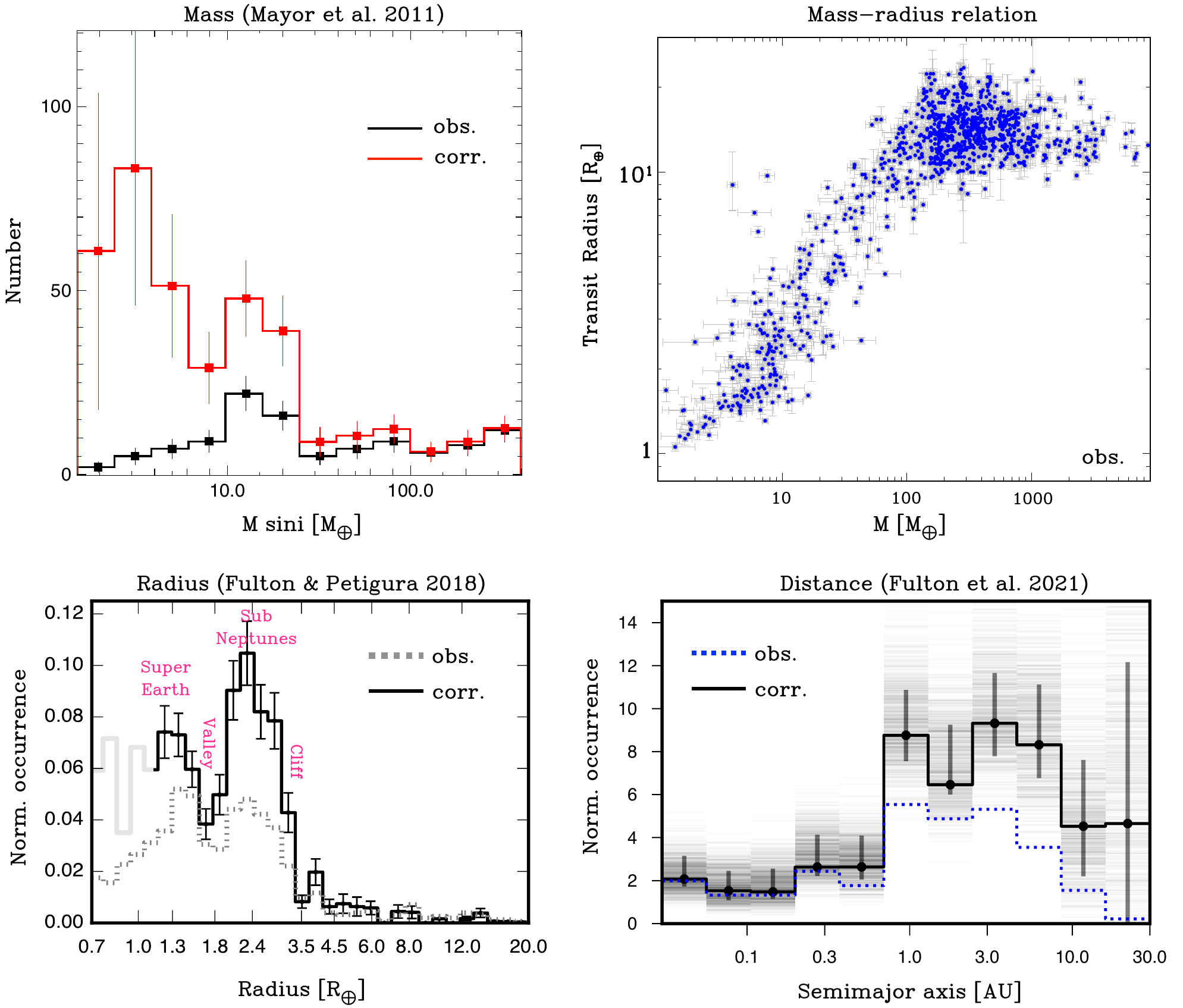}
	\caption{Statistical observational constraints from the exoplanet population. Top left (modified from \citealt{2011MayorArxiv}): mass distribution. Top right: observed (biased) mass-radius relation. Data from the NASA exoplanet archive, including planets with relative mean errors of less than 50\% in mass and radius and with known stellar mass and orbital period. Bottom left (modified from \citealt{2018AJFultonPetigura}): radius distribution. Some important features are labelled. Bottom right (modified from \citealt{2021ApJSFulton}): semimajor axis distribution for planets more massive than 30 $M_\oplus$. Raw observed and bias-corrected results are given.}
	\label{fig:exoplanetconstraints}
\end{figure*}

There are several essential distributions like the mass, radius, eccentricity, and orbital distance distributions that characterise the exoplanet population (Fig. \ref{fig:exoplanetconstraints}). The planetary mass function, or PMF, is known from radial velocity surveys mainly for Solar-like stars and from microlensing for M dwarfs. It is an important distribution that has been addressed and predicted by various theoretical works \citep[e.g.][]{2005ApJIdaLin,2009A&AMordasinib,2021AAEmsenhuberB}. Focusing on Solar-like stars, the observed distribution as found by radial velocity surveys scales with mass $M$ approximately as $1/M$ \citep{2005PThPSMarcy} in the mass range of about 30 $M_\oplus$ to roughly 4 $M_{\rm J}$ (with $M_{\rm J}$ being the mass of Jupiter). The distribution is thus bottom heavy in the sense that there are more low mass giants than there are high mass giant planets. Above this value, the frequency decreases even stronger \citep{2017A&ASantosA}. Whether there is an clear upper end of the PMF is unknown, but there are only very few brown dwarfs around Solar-like stars inside of about 5 AU \citep{GretherLineweaver2006}. The driest part, and thus a potential upper end of the PMF might be at around 30 $M_{\rm J}$ \citep{Sahlmann2011}. Moving to small masses, there is a break in the PMF at around 30 $M_\oplus$ below which the frequency increases strongly \citep{2010ScienceHoward,2011MayorArxiv} as had been predicted by models based on the core accretion theory, where at a similar total mass (core + envelope), gas accretion becomes important \citep{2009A&AMordasinib}. Thus, the break would correspond to the transition from a slope that is governed by solid accretion to one governed by gas accretion. Microlensing observations, however, challenge this picture \citep{2018ApJSuzuki}.

Regarding the semimajor axes of giant planets, there is a local maximum at a period of about 3-4 days that is formed by the Hot Jupiters. It is followed by a less populated region known as the period valley \citep{Udry2003}. At about 1 AU, the frequency increases by about a factor 4 \citep{2021ApJSFulton}. Already outside of about 3-10 AU, the frequency could be decreasing \citep{2016ApJBryan,2019ApJFernandes,2021ApJSFulton} but this might be an effect of the observational bias \citep{Lagrange2023}.

While the eccentricities of the Solar System planets is very low, there is a broad distribution of eccentricities for extrasolar giant planets with about half of them having an eccentricity in excess of 0.3 \citep{2008ApJFordRasio}. Some exoplanets even have eccentricities exceeding 0.9 \citep{2001AANaef}. The distribution resembles for the higher values a Rayleigh distribution which is the distribution expected from planet-planet scattering \citep{2008ApJJuric}. This is an indication that in some extrasolar systems, strong dynamical interactions have happened \citep{2015ARA&AWinn}, an effect that is also seen in numerical simulations \citep{1996ScienceRasioFord,Emsenhuber2023}. When taking into account that radial velocity measurements are biased towards overestimating eccentricities \citep{Lucy1971AJ,2019MNRASHara}, then a significant number of orbits are also consistent with being circular. Planets detected via radial velocity with lower masses (less than about 30 $M_\oplus$) tend to have lower eccentricities of less than about 0.5 \citep{2011MayorArxiv}. Transiting planets detected by the Kepler mission in multiple systems planets have orbits that are almost circular (average of about 0.02-0.05). They also have low mutual inclinations of about 1.4 degrees or less scaling  inversely with the number of planets in the system \citep{2022PPVIIWeiss}. Eccentricity and inclination are correlated with each other \citep{2016PNASXie}.

The distribution of radii of planets detected by the Kepler satellite shows a peak at about the same size as Jupiter, which is consistent with the theoretical relation between mass and radius \citep{2012A&AMordasiniC}. The radius distribution is roughly constant in $\log(R)$ for planets between 4 and 10 $R_\oplus$. For smaller planets, the number of planets increases sharply \citep{2013ApJFressin} which is sometimes referred to as the ``radius cliff'' \citep{kite2019}. There is a local minimum in the distribution at around 1.7 $R_\oplus$ known as the radius gap or radius valley \citep{2017AJFulton,vaneylen2018}. It separates smaller super-Earth planets that are mostly rocky from larger Sub-Neptunes that have thick H/He atmospheres or water layers \citep{2020Mousis}. This could be because of the loss of the primordial H/He atmospheres due to stellar radiation \citep{2013ApJOwenWu,2018ApJJin} or different formation pathways for these two types of planets \citep{Venturini2020,Burn2024}.

\subsubsection{Correlations with host star properties, age, and intra-system architecture}\label{sect:obscorrstellarprops}
The most well-known correlation with stellar properties is the higher frequency of giant planets around stars with higher metallicity which was noted already early on \citep[][]{1997MNRASGonzalez,2004A&ASantos,2005ApJFischer,DongXie2018}. In the domain of metallicities above the Solar value, the frequency of giant planets increases by about an order of magnitude from [Fe/H]=0 to [Fe/H]=0.5. This is often seen as evidence that core accretion is the main mode of giant planet formation \citep{2004ApJIda1,2012A&AMordasiniA}. For lower mass or smaller planets, the dependency on  metallicity becomes weaker and weaker, and for planets with the smallest radii currently known ($\lesssim1.7 R_\oplus$), there seems even to be a preference for slightly sub-Solar values \citep{2011MayorArxiv,2018AJPetigura,2018AJNarang}.

The occurrence of giant planets around host stars with lower masses (M-dwarfs) is about 2-6\%  which is lower than for Solar-like stars \citep{2013AABonfils,2018ApJGhezzi,2021Sabotta,Ribas2023}. The discovery of giant planets around stars of even very low mass ($\sim0.1$ Solar masses, $M_\odot$) is seen as a challenge to both core accretion and gravitational instability models \citep{2019ScienceMorales,2021AABurn,2022AASchlecker}, but recent simulations combining growth by pebble accretion and giant impacts find for special conditions (high disk masses, high turbulence level) giant planet formation also around such stars \citep{Pan2023}. In contrast to giant planets, low-mass stars have 2-3 times more low-mass planets (around 10 Earth masses) than G-type stars as found via the transit method using Kepler \citep{2015ApJMuldersB} and later supported using radial velocity \citep{2021Sabotta}. However, a recent re-analysis of the Kepler data challenges this trend \citep{Bergsten2023}. For stars with masses above 1 $M_\odot$, the occurrence of giant planets detected by radial velocity first rises to a peak at about 1.7 to 2 Solar masses, and then declines for even more massive stars \citep{Reffert2015,Wolthoff2022}. 

The relationship between stellar age and planetary properties and their statistics is now being explored with various surveys \citep[e.g.,][]{Mann2016,Grandjean2021,Capistrant2024}. For these studies, it is important that Gaia has allowed to greatly expanded our knowledge of young stellar groups \citep{bouma2022}. Several close-in planets have been found around young (T-Tauri, pre-main-sequence and young main sequence) stars \citep{David2016Nature,Donati2016Nature,Plavchan2020Nature}. They reveal that some planets close to their stars can form within a few million years, probably due to orbital migration caused by planet-disk interactions but that dynamical effects (high eccentricity migration) also plays a role \citep{Dai2023}. T-Tauri stars may have more hot Jupiters than main sequence stars \citep{YuDonati2017}. At larger separations, direct imaging can also inform about the properties of young planets with ages of a few tens of million years \citep{2010ScienceLagrange,2015ScienceMacintosh,2019ApJWagner,2021AAVigan}. First direct detections of planets that are still in the formation process have occurred revealing emission not only from the planets' photosphere but also from the gas accretion shock and from the surrounding circumplanetary disk \citep{2018AAKeppler,2019NatAsHaffert,Benisty2021,Wang2021}. Such observations have the potential to put much more direct constraints on physical processes like gas accretion \citep{Hashimoto2020,Marleau2022}. Furthermore, there are indirect indications of the presence of forming planets via their impact on their parent protoplanetary disk (gaps, kinks in the rotational velocities) \citep{Teague2018ApJ,HuangAndrews2018,Pinte2020ApJ}. In the coming years, the PLATO transit survey \citep{Rauer2014ExA} will provide much more statistical information on how the population of transiting planets changes over time.

 As illustrated by the mass-distance diagram, the exoplanet population shows a lot of variation and diversity among different systems, but interestingly, there is in contrast intra-system uniformity within individual systems of small close-in planets, called the peas-in-a-pod pattern \citep{2018AJWeiss,2017ApJMillholland,2022PPVIIWeiss}: planets in the same system have comparable sizes, masses, and relative distances from each other. The study of extrasolar planetary systems and their architectural patterns, as opposed to individual planets, is currently a field of active research \citep[e.g.,][]{2011ApJSLissauer,Alibert2019,Adams2020,Bashi2021,2023AAMishraA,2023AAMishraB,Emsenhuber2023}.

\subsection{Predictions from original Solar System theory and the shift of paradigms}\label{sect:predictionsclassiccaltheorypardigmshifts}
The classical theory of planetary formation based on the Solar System alone had predicted that the fundamental architecture of the Solar System (refractory low-mass terrestrial planets inside, then (two) massive gas giant planets outside of the iceline, and finally ice giants of intermediate mass, all on nearly circular orbits) should be universal \citep{Dole1970}. Assuming that planets form in situ (i.e., grow at the position where they are found now), it was thought that giant planets are always found outside of the ice line, i.e., the distance outside of which it is cold enough ($\sim$170 K) for water to condense in the protoplanetary nebula surrounding the young Sun (which is at about 2.7 AU in optically thin irradiated disk models, \citealt{2004ApJIda1}). The amount of planetary building blocks (assumed to be only kilometer-sized planetesimals) thus increases there, providing the material to form massive cores (about 10 Earth masses) necessary to start efficient gas accretion \citep{1996IcarusPollack}. At even larger distances $r$, the collisional growth timescale is long since it is proportional to the local Keplerian frequency 
\beq
\Omega_{\rm K}(r) = \sqrt{\frac{G M_\star}{r^3}},
\eeq
where $M_\star$ and $G$ are the stellar mass and Newtonian gravitational constant, respectively. Thus, growth is reduced such that the outwardly decreasing surface density of planetesimals would only allow the formation of lower-mass core-dominated ice giants during the finite lifetime of the Solar nebula of a few million years \citep{1996IcarusPollack}. Finally, inside of the water iceline, due to limited availability of building blocks, only low-mass refractory terrestrial planets should form.    

An important  concept in this in situ theory is the Minimum Mass Solar Nebula MMSN \citep{1977ApSSWeidenschilling,1981PThPSHayashi}. In this approach, the masses in solids (refractory rocky material and volatile ices) contained in the Solar System planets (quantities which can only be inferred indirectly for the giant planets) are radially spread over touching annulli, and the gas is complemented in a Solar-composition ratio. The result is that this minimum mass Solar nebula should have contained about 80 $M_\oplus$ of solids, and 0.013 $M_\odot$ of gas, and that the surface density of planetesimals and gas should have decreased with orbital distance as a smooth power law proportional to distance $\propto r^{-1.5}$  with a jump (increase) by a factor 4.2 at the water condensation front (ice line) at 2.7 AU for the planetesimals \citep{1981PThPSHayashi}. This indirect inference of the nebula properties contrasts today's approach of obtaining the properties of protoplanetary disks from direct astronomical observations in star forming regions \citep{2009ApJAndrews,2022PPVIIMiotello}. While the MMSN still remains a point of reference for modern models, the importance of structured disk instead of smooth power-law disks \citep{2020ARAAAndrews}, and high diversity in the sense of a large spread in inferred masses and sizes are stressed today \citep{2022PPVIIManara}.

\begin{figure*}[h]
\begin{minipage}{0.5\textwidth}
\centering
\includegraphics[width=1.0\textwidth]{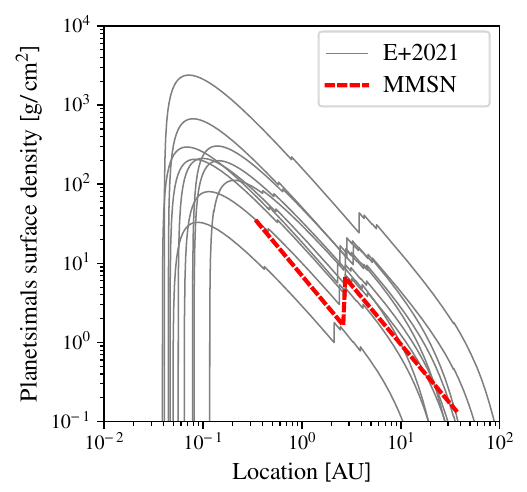}          \end{minipage}
\begin{minipage}{0.5\textwidth}
\centering
\includegraphics[width=0.87\textwidth]{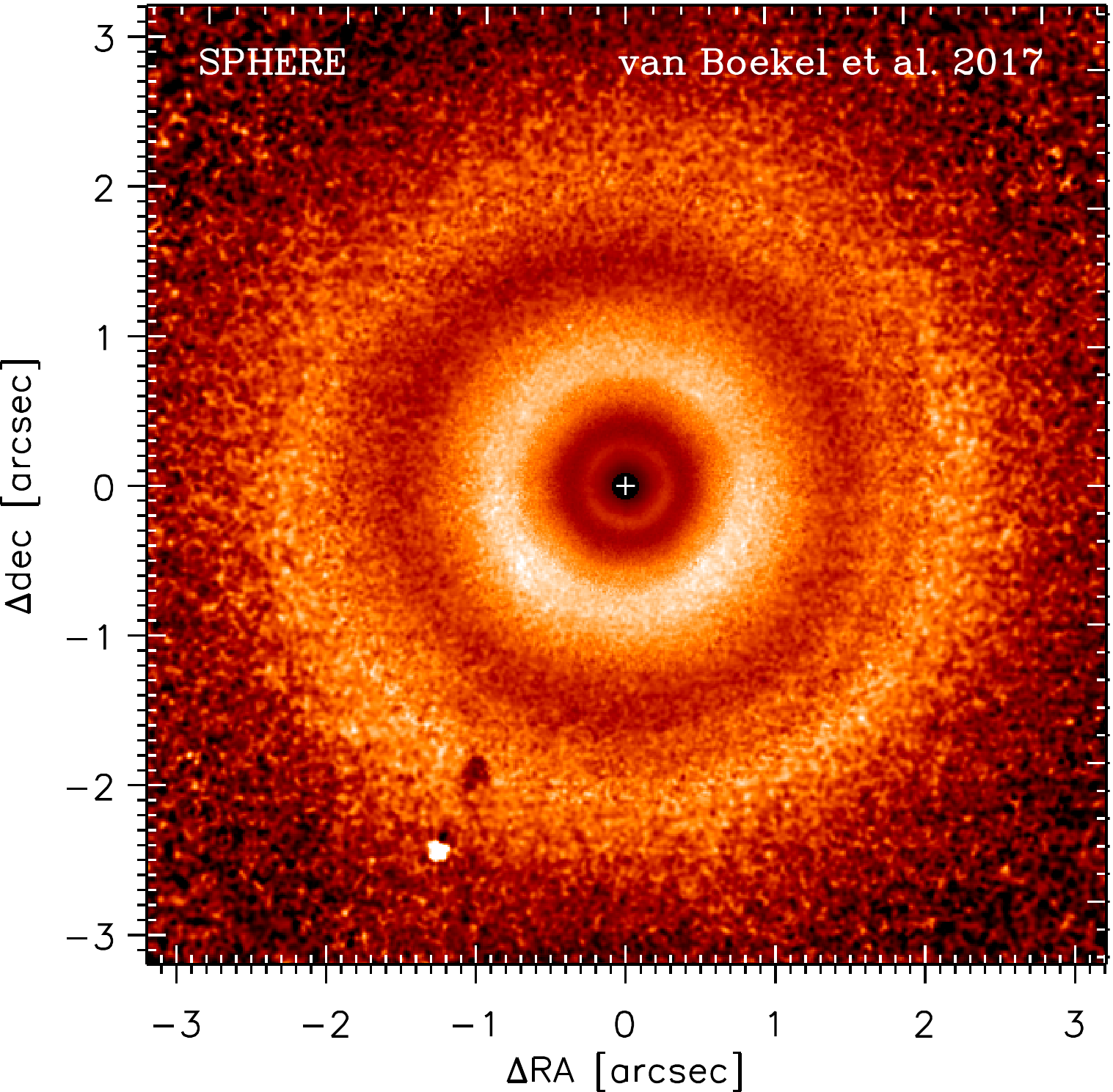}          \end{minipage}

    \caption{Left panel: Theoretical surface densities of planetesimals as a function of orbital distance. The red dashed line is the Minimum Mass Solar Nebula from the classical work of \citet{1981PThPSHayashi}. The jump at 2.7 AU represents the condensation front of water (the water iceline), leading to an increase of the surface density. Apart from the jump, the MMSN follows a smooth unstructured power-law. The gray lines show surface densities informed by observations of protoplanetary disks \citep{2018ApJSTychoniec}. Jumps correspond to icelines of different species. While observationally informed by actual disks, these surface density profiles still assume a unstructured power-law profile (Figure adapted from \citealt{2021AAEmsenhuberA}). This contrasts the observations of numerous disk structures as illustrated in the right panel \citep[figure adapted from][]{vanBoekel2017}. It shows the scattered light image of the TW Hya disk, tracing sub-micron dust grains. The structures indicate  gas surface density variations of 50-80\% across the three radial gaps. }

    \label{fig:mmsn} 
\end{figure*}

However, the discovery of the diversity of  extrasolar planets, but also new theoretical studies and disk observations have shattered the static picture derived from the Solar System, profoundly shifting several paradigms of planet formation. One can distinguish three, partially interlinked paradigm shifts:

\textit{1. Mobility of the building blocks:} Already the first extrasolar planet 51 Peg b \citep{1995NatureMayorQueloz} as a giant planet orbiting at a few days period was diametrically opposed to the predictions of the in situ theory. The subsequently revealed proprieties of exoplanets has in their sum drawn a much more dynamic picture of planet formation, i.e., a process where the radial mobility and redistribution, but also trapping, as well as dynamical interactions of building blocks at different growth scales are key. This paradigm of mobility and dynamics regards both roughly cm-sized drifting pebbles and large migrating protoplanets. 

\textit{2. The nature of protoplanetary disks: structures and accretion mechanism:} More recently, also the view of the planet's formation environment, the protoplanetary disks, has shifted in two ways: first, to a complex structured picture \citep{2020ARAAAndrews,2022PPVIIBae}, away from the smooth power-laws inspired by the Solar System. The strong impact of disk structures on planet formation \citep{2022AALau,JiangOrmel2023} had already been stressed by several early works \citep{Matsumura2007,Hasegawa2013,Coleman2016}. Second, the assumed fundamental mechanism driving accretion in protoplanetary disks is also shifting away from the traditional picture of viscous disks \citep{1952ZNatALust,1974NMRASLyndenBellPringle} where a turbulent viscosity redistributes angular momentum \citep{1973A&AShakuraSunyaev} to MHD-wind driven disk evolution where magnetic fields extract angular momentum \citep{2022PPVIILesur,Weder2023}. As disk structures, this has profound impacts on planet formation \citep{2018AAOgihara,Aoyama2023}.

\textit{3. Solid accretion:} Finally, as an additional shift of paradigm related more directly to the planets, it is now understood that planetesimals are not the only potential agents of planetary solid growth, but also smaller pebbles \citep{2010AAOrmelKlahr}, in particular for the growth of the cores of (distant) giant planets \citep{2012AALambrechtsJohansen}. 

It should be noted that these paradigm shifts are ongoing processes and no convergence of the relative importance of the different concepts has been reached at this time. It could well be that eventually, a more complex picture will arise that synthesises aspects of both the classical and new paradigms \citep{2022AAVoelkel}.

In the context of the paradigm of  mobility of the building blocks and protoplanets, it is interesting to note that both processes, the early radial drift of pebbles \citep{1977ApSSWeidenschilling} as well as gas-driven orbital migration of protoplanets \citep{1980ApJGoldreichTremaine,1986ApJLinPapaloizouA} were theoretical concepts derived well before the discovery of 51 Peg b. Orbital migration, because of its seeming absence or inefficiency as judged from the Solar System alone, was however not seen as of high import or was even dismissed completely until the exoplanets were discovered.

\section{Physical processes governing planet formation}
\label{sec:physical_processes}
\begin{figure*}[h]
	\centering
	\includegraphics[width=1\linewidth]{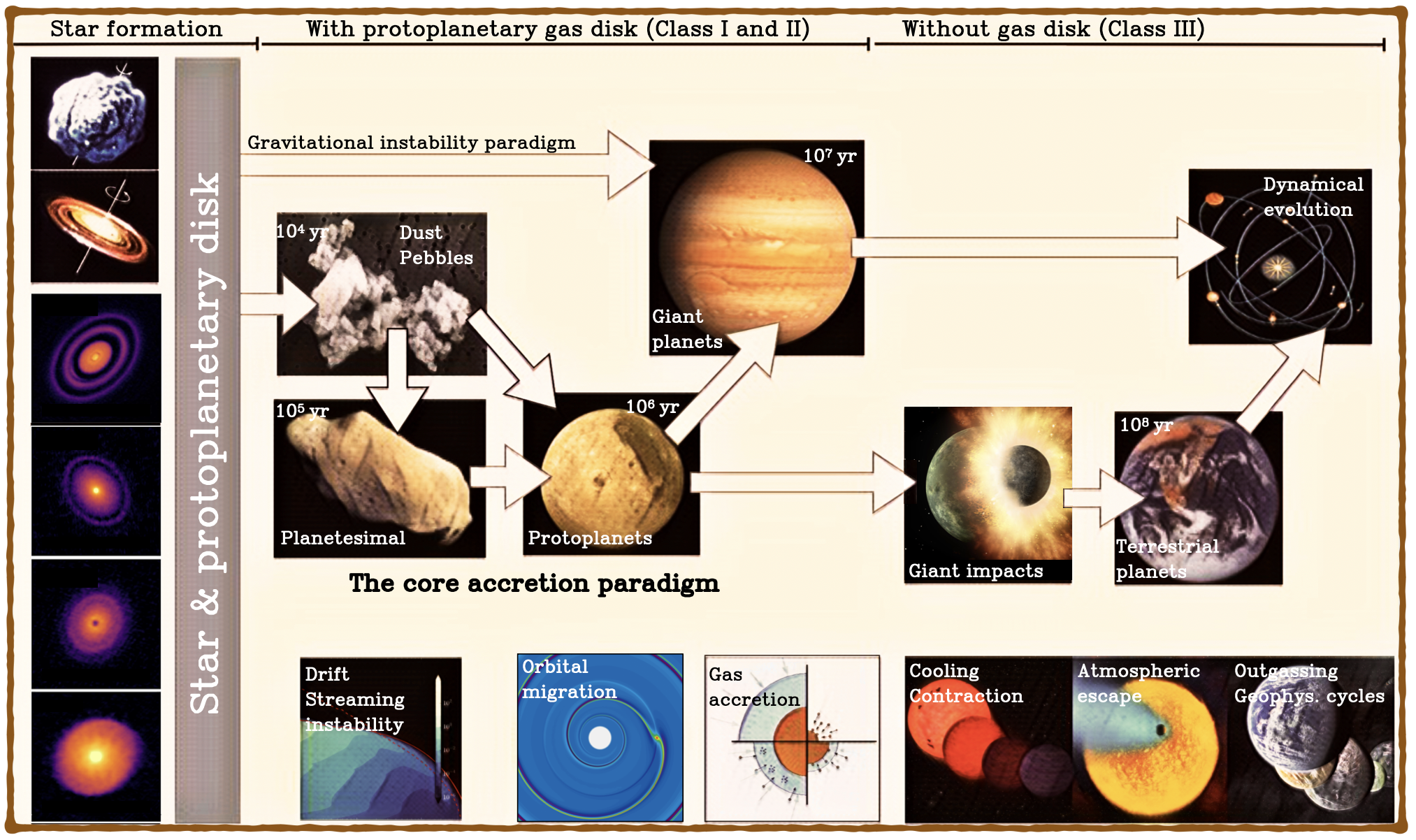}
	\caption{A cartoon view of the different stages of planet formation and evolution and the processes involved. At the top, the main classes of disk evolution are indicated by black horizontal bars. In the middle, the panels connected by white arrows show the main growth stages and involved processes of growth in the core accretion paradigm which is a bottom up process. The gravitational instability paradigm for giant planet formation, a top-down process, is also shown. The bottom panels show various additional physical processes governing at the different stages. A number of typical timescales are also given. (Image Credits: W. Benz/C. Mordasini (composite). Sub-figures from \citet{2018ApJAndrewsB}; \citet{2014PPVIBaruteau}; \citet{2012A&ABirnstiel}; NASA; ESA; NASA/JPL; ESA, A. Vidal-Madjar (IAP, CNRS), NASA.)}
\label{fig:schematic}
\end{figure*}

While the emergence of planets from a protoplanetary disk is undisputed, several competing or complementary paradigms to form the  bodies exist. Similar to stars, giant gaseous planets may form directly in a top-down fashion from the self-gravitational collapse of a large unstable part of a protoplanetary disk \citep{1997ScienceBoss} - the so-called disk instability or gravitational instability mechanism \citep{Kratter2016}. Since the most unstable wavelength is relatively large \citep{Toomre1964}, gravitational instability will lead typically to giant planets at large orbital distances \citep{2023AASchib}. The mechanism needs massive gas disks and likely occurs during or just at the end of the infall phase when the protoplanetary disk is forming \citep{2021AASchib}. Its efficiency is still unknown, but it might be the formation mechanism of some distant massive planets \citep{2008ScienceMarois,2010NatureMarois}.

On the other hand, the bottom-up formation of planets starting out small and growing over time is seen as the standard pathway to form planets, both  gas giants and small rocky planets. In the following, we will review some of the key processes of planet formation via this core accretion paradigm which needs to connect micrometer-sized dust inherited from the interstellar medium to the final planets with sizes of several ten thousand kilometers.

While doing so, we follow the simplified picture of sequential growth shown in Fig. \ref{fig:schematic}, addressing physical processes that are relevant along the way. In the figure, the different stages are shown in the upper part, while in the lower part, physical processes relevant at these sizes are shown. Characteristic timescales are also given, but it should be noted that they are only very rough indicative values, and depend for example on the orbital distance that is considered.

We begin on the left of Fig. \ref{fig:schematic} with the protoplanetary gas disk \ref{sect:gasdisk} that sets the background structure in which planets form. The next section discusses the dynamics of the small building blocks, micrometer-sized dust and roughly cm-sized pebbles, where the latter form rapidly on a timescale of $10^4$ years through coagulation \ref{sect:dyndustpebbles}. At this stage, the radial drift through the disk is a key process.

The next step, that happens after $\sim10^5$ years, is the growth from pebbles to planetesimals as the first gravitationally bound objects which is addressed in Sect. \ref{sect:pebbles2planetesimals}. It likely involves the streaming instability as a way to concentrate pebbles to a point that a pebble cloud becomes self-gravitating and collapses to directly form planetesimals. 

For the next step, the way in which protoplanets (objects of a size of about 1000 km) grow further by the accretion of solid building blocks, several possibilities exist (Sect \ref{sec:solid_acc}), namely pebble accretion and planetesimal accretion. This is a process that occurs on a $\sim10^6$ year timescale. These protoplanets exchange angular momentum with the gas disk, leading to orbital migration. It is discussed in Sect. \ref{sec:orb_migration}. 

Some protoplanets grow massive enough ($\sim$ 10 $M_\oplus$) to be able to accrete nebular gas efficiently, leading to the formation of gas giants. The process of gas accretion is discussed in Sect. \ref{sec:gasaccretion} on a timescale of $\lesssim10^7$ years. 

After the dissipation of the gas disk, terrestrial planets form via giant impacts on a timescale of about 100 Myr, although some more recent models propose that substantial growth may already happen earlier \citep{2018AAOgihara,Broz2021,BatyginMorbidelli2023,Woo2023Icarus}.

Finally, in principle during the entire lifetime of a planetary system, the different components interact gravitationally and settle into a state that is stable over Gyr-timescales. On the individual planets, evolutionary processes like cooling and contraction, atmospheric escape, and various mechanisms like interior-atmosphere interactions (in/outgassing) and geophysical cycles take place.

\subsection{Gas disk}\label{sect:gasdisk}
On the modelling side, the structure of protoplanetary disks is simulated with 2 and 3D hydrodynamic numerical simulations that can include the effects of radiative energy transport and magnetic fields, where it is important for the latter to include non-ideal effects \citep[see][and references therein]{2014PPVITurner,2022PPVIILesur}. While yielding a detailed understanding of the structure of the disks, these simulations can only simulate short temporal intervals because of their computational costs. 

To simulate the full temporal evolution on a million year timescale, one dimensional (radial) axisymmetric models of gas disk evolution were developed by \citet{1974NMRASLyndenBellPringle,1981ARAAPringle} and for example reviewed in \citet{2019ArmitageSaasFee}. For our purposes, it is useful to recall the basic concept which will help to introduce the used quantities.

Protoplanetary disks are often assumed to rotate at the Keplerian orbital frequency $\Omega_{\rm K} = \sqrt{G M_{\star}/{r^3}}$, where the stellar mass $M_{\rm star}$, gravitational constant $G$, and distance to the star $r$ enter, although the gas is actually rotating slightly slower because of the radial pressure support as discussed in Sect. \ref{sect:dyndustpebbles}. Furthermore, if the disk is in vertical direction isothermal and in hydrostatic equilibrium, the vertical profile of the gas density is
\begin{equation}
\label{eq:density_z_hydro_eq}
\rho_{\rm g}(z) = \rho_{\rm g,0} \exp\left(-z^2/2H_{\rm g}^2\right)\,,
\end{equation}
where $\rho_{\rm g,0}$ is the gas density at the midplane, 
\beq
H_{\rm g} = \sqrt{\frac{k_B T}{\Omega_{\mathrm{K}}^2 \mu\, m_{\mathrm{H}}}} 
\eeq
is called the gas scale height, $T$ is the temperature, and $\mu m_{\mathrm{H}}$ is the mean molecular weight conventionally expressed in hydrogen atomic mass units $m_{\mathrm{H}}$.

Protoplanetary disks are observed to dissipate in about 1 to 10 Myr \citep{2001ApJHaisch,2022ApJPfalzner} (with a very large spread in lifetimes) and relatively high accretion rates  onto the host star $\sim 10^{-8}$ M$_{\odot}$/yr are observed \citep[e.g.][]{2017AAAlcala,2022PPVIIManara}. The initial mass of the disks is usually assumed to be a  few percents of the stellar mass but could also be an order of magnitude higher, in broad agreement with dust continuum measurements of young disks \citep{2020ApJTobinA}. Even under this optimistic assumption, the high accretion rates imply that most of the disk mass needs to be transported toward the star. Therefore, its angular momentum needs to be removed on the same timescale. The microscopic viscosity of the disk gas can not account for this, which motivated the introduction of a turbulent $\alpha$ viscosity $\nu = \alpha c_{sT} H_{\rm g}$ with the Shakura-Sunyaev $\alpha$ parameter \citep{1973A&AShakuraSunyaev}. Here, $c_{sT} =  \sqrt{\frac{k_B T}{\mu\, m_H}}$ is the speed of sound in an isothermal medium. The assumption of viscous shear transporting angular momentum and Keplerian orbital motion allows for deriving a single disk evolution equation which controls the radial motion of the gas disk and can be easily implemented in one dimensional (rotationally symmetric) models \citep{1952ZNatALust,1974NMRASLyndenBellPringle}
\begin{equation}
    \label{eq:disk_evo}
    \frac{d \Sigma}{d t} = \frac{3}{r}\frac{d }{d r} \left( r^{1/2} \frac{d}{d r}\left(r^{1/2} \nu\Sigma\right)\right)\,.
\end{equation}
This yields the evolution of the gas surface density $\Sigma$ as a function of time $t$ and distance from the star $r$.

In addition, a significant effect is expected from energetic radiation heating the molecules in the upper- or outermost regions of the disk. In addition to heating, this leads to ionization and dissociation of the molecules. From the absorption of the high-energy radiation the thermal energy of the lighter monoatomic hydrogen particles or hydrogen ions can exceed the local gravitational potential energy \citep{1994ApJHollenbach}. Thus, they become unbound and leave the disk as a wind. This photoevaporative process can be driven by radiation of the host star \citep{2009ApJGorti,2009ApJErcolano} or from external stars \citep{2016MNRASFacchini}, typically more massive O or B stars whose spectrum includes orders of magnitudes more high-energy photons. Depending on the source of the radiation and the wavelength, different prescriptions \citep{2019MNRASPicogna,2021MNRASPicogna,2021MNRASErcolano,2018MNRASHaworth} for an additional sink term $\dot{\Sigma}_{g, evap}$ in Equation \ref{eq:disk_evo} can be used. Recently, disk photoevaporation was reviewed in detail by \citet{2022EPJPWinterHaworth}.

Another sink or source term can be added in Equation \ref{eq:disk_evo} for those minor constituent gases other than H/He that can condense, respectively evaporate or sublimate from solid grains under nebular conditions. This happens in a narrow region at the icelines of the various volatile species, such as H$_2$O, CO, and CO$_2$. The transport of such gaseous species should be modelled individually using the transport equation for tracer fluids \citep[e.g.][]{Ciesla2011}. While in terms of mass, these species other than H/He are of negligible importance, they are of key interest for the resulting chemical composition of planets. More details on chemical modelling of protoplanetary disks are reviewed in Chapter 2 of this volume (Zhang \& Trapman, in press).

More recently, an alternative paradigm to the classic viscous evolution of the disk has become the focus of the research \citep[see review of][]{2022PPVIILesur} building upon the results of non-ideal magnetohydrodynamic simulations \citep[e.g.,][]{Bai2016,Bai2016a}. The magnetohydrodynamic instability which can drive significant turbulent viscosities as assumed in the classic picture is only active in regions where gas is sufficiently ionized \citep{Gammie1996}. This is the case at large temperatures where thermal ionization can take place, or at low surface densities where cosmic rays can ionize the full vertical extent of the disk. Between these two regions, a viscously 'dead' zone is expected where turbulence levels are low and viscous angular momentum transport is inhibited. In this region, angular momentum could be primarily removed by magnetically-driven winds from the surface layer of the disk. In this case, the evolution equation needs to be changed to include an advective term \citep{2016AASuzuki}, such as used in recent global disk evolution studies \citep{Tabone2022,Weder2023}.

\begin{figure*}[h]
	\centering
	\includegraphics[width=1\linewidth]{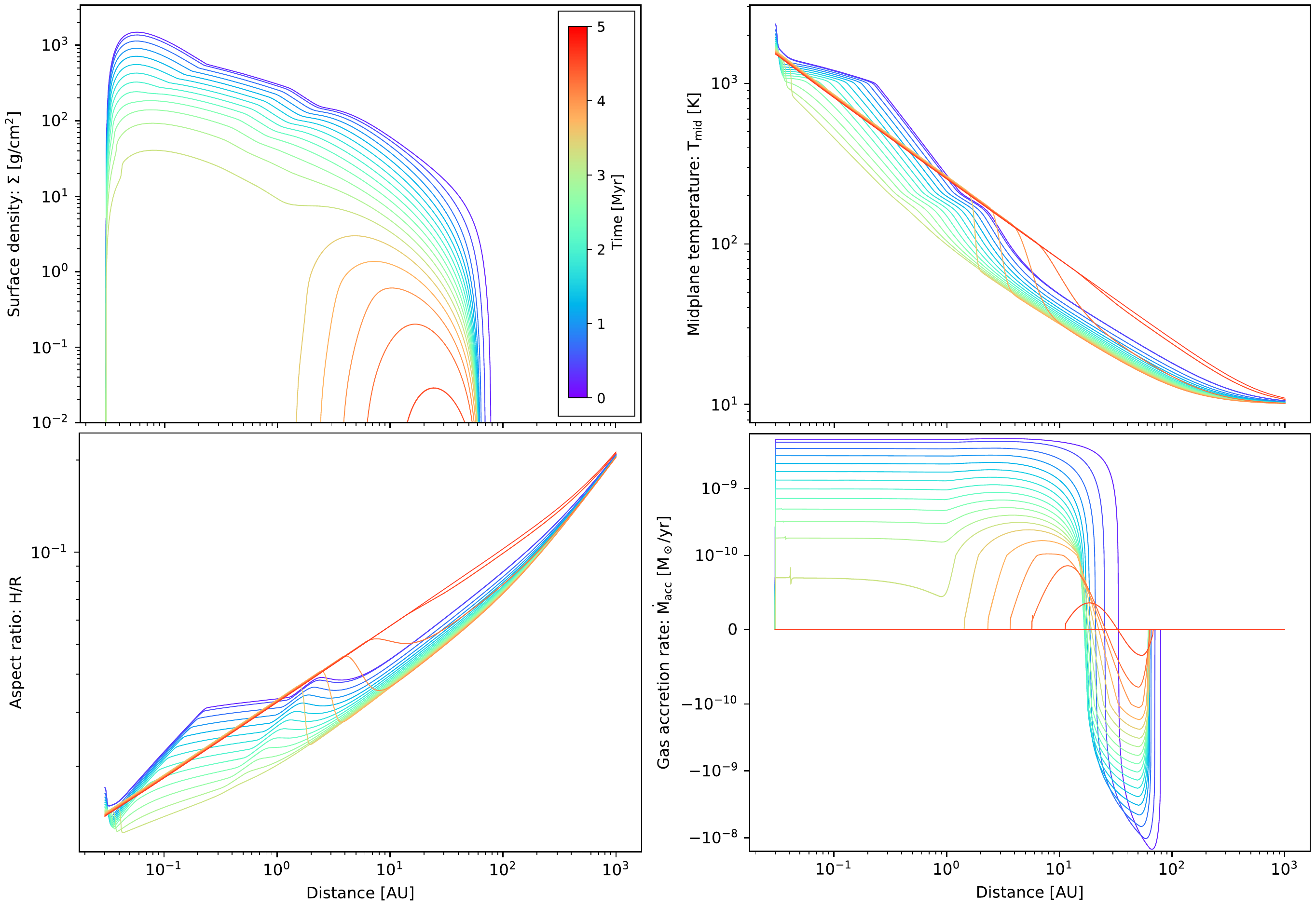}
	\caption{Evolution of a protoplanetary gas disk as a function of distance from the star and time in the classical turbulent $\alpha$-viscosity paradigm. Internal and external photoevaporation is also included. The panels show the gas surface density (top left), midplane temperature (top right), aspect ratio $H/r$ (bottom left), and accretion rate (bottom right). In the latter plot, positive values correspond to a gas flow towards the star (accretion) while negative values are outward spreading parts (decretion).}
	\label{fig:gasdisk}
\end{figure*}

Despite these developments, it is still helpful to study as a reference model the classical viscous picture with a constant $\alpha$ value, as shown in Figure \ref{fig:gasdisk}. The evolution of the overall gas surface density, the disk temperature and scale height, as well as the radial gas flow is shown. The  viscosity is assumed to heat the disk midplane due to viscous dissipation. This effect can be prominently seen in the temperature distribution interior of approximately 5\,au where the disk temperature slope is modified. In the viscously heated part, the temperature is sensitive to the opacity of the dust grains which in turn depend on the composition, here modelled following \citet{1994ApJBell}. In their opacity parameterization, a transition region at the water iceline is introduced smoothly from 150 to 170\,K. It manifests as a region with a shallow temperature slope which moves inwards as time evolves. Interior to the water opacity transition, the radial temperature gradient is large up to the silicate evaporation front at 1000\,K.

These temperature regimes leave an imprint visible in the bottom left panel on the aspect ratio since the scale height $H$ depends on temperature. In the same panel, we can see that the disk is flaring, that is, the aspect ratio increases with distance. In principle, this has an effect on the absorbed radiation via shadowing. Here, instead of a fully coupled treatment, a constant, but roughly consistent, flaring angle $d \ln H/d\ln r = 9/7$ \citep{1997ApJChiang} is assumed to calculate the absorbed irradiation from the star. This was used in several works \citep[e.g.][]{2005A&AHueso,2021AAEmsenhuberA} to obtain numerically stable solutions.

We note that although the temperatures in the disk can be sufficient to evaporate also refractory species, the region in which this occurs is limited to an annulus close to the star. Despite the increased surface density, little mass resides in this ring which is unlikely to leave an imprint in the mass budget of planets under standard assumptions. Therefore, the composition in refractory elements in planets is generally expected to be similar to that of the star \citep{2014AAThiabaud}.

As the viscosity depends on temperature, the steady-state surface density will also evolve to a profile which shows corresponding features. Since the viscous timescale  is shorter at smaller distances, such a steady-state accretion is reached over relatively short timescales. Therefore, the gas accretion rate (bottom right) is radially constant up to the transition at around 20\,au where the accretion switches to decretion. This point is known as the radius of maximum viscous couple or of velocity reversal. Such a transition is fundamental in solutions to any diffusion equation to conserve the total angular momentum \citep{1952ZNatALust,1974NMRASLyndenBellPringle}. The transition occurs where most of the mass reservoir is located from which the rest of the disk can draw.

The outer edge of the protoplanetary disk is given by the equilibrium between viscous spreading against external photoevaporation \citep[e.g.,][]{Coleman2024}. In Figure \ref{fig:gasdisk}, the initial profile is more extended than this equilibrium radius, thus the disk is initially shrinking.

The later stages of the disk evolution are dominated by photoevaporation since those sink terms are here assumed to be constant and independent on the local surface density. In the surface density evolution plot, it can be seen that an inner hole opens. This is a typical effect of internal photoevaporation, here following the prescription of \citet{2001MNRASClarke}. As the inner disk is cleared, radiation from the star can heat the disk through the midplane which leads to an increase in temperature to the equilibrium temperature in the absence of a disk. An increasing temperature front which propagates outwards can be seen in the top right panel for the latest timesteps (orange to red colors). 

\subsection{Dynamics of dust and pebbles}\label{sect:dyndustpebbles}
Like the interstellar medium, the protoplanetary disks which form together with the stars due to (partial) angular momentum conservation during infall do not only contain gas but also small particles. From multi-wavelength spectroscopic observations, we know that they range in size from \SI{0.01}{\micro\meter} to \SI{1}{\micro\meter} \citep{Mathis1977}. As for the gas, we refer to Chapter two in this issue or the literature \citep{Pollack1994,2014AAThiabaud} for a discussion of their composition.

How closely the particle dynamics is linked to the gas dynamics is expressed with the Stokes number $\rm{St}$ which corresponds to the dimensionless friction time. It relates the stopping time of a particle due to gas drag to a typical timescale for the gas motion, that is, in practice the orbital timescale. Thus, $\rm{St} = t_{\rm stop} \Omega_{\rm K}$, where we assume the gas orbits at a Keplerian frequency $\Omega_{\rm K}$. For small particles, which are in most of the disk in the Epstein drag regime \citep[see][for different drag regimes]{1977ApSSWeidenschilling}, the Stokes number at the midplane is 
\begin{equation}
    \rm{St} = \frac{\pi a \rho_{\rm s}}{2 \Sigma}\,
\end{equation}
for a given gas surface density $\Sigma$ and particle with size $a$ and bulk density $\rho_{\rm s}$ which is on the order of \SI{1}{\gram\per\centi\meter\cubed} but is expected to be higher if the local temperature led to evaporation of volatile species and lower if grains are porous or fractal which is currently only weakly constrained.

\subsubsection{Growth from Dust to Pebbles}
\label{sec:dust_to_pebbles}
For the initially micrometer-sized particles inherited from the interstellar medium, sometimes referred to as monomers, to grow, they need to collide with other particles. The collision rate between a spherical particle of size $a_t$ with particles of size $a_p$ is given by $\Gamma = \Delta v \sigma n$, where $\Delta v$ is the relative velocity, $\sigma = \pi (a_t + a_p)^2$ is the cross sectional area of the two spherical particles, and $n$ is the number density of particles in the gas.

To estimate the local number density it is important to account for particles settling towards the midplane which locally increases their number. This process is relatively fast and will reach an equilibrium where at regions close to the midplane, the particles of a given Stokes number can be described with a reduced scale height $H_{\rm d}$ compared to the gas \citep{2007IcarusYoudinLithwick}
\begin{equation}
    H_{\rm d} \approx H_{\rm g} \sqrt{\frac{\alpha}{\rm{St}}}\,.
    \label{eq:Hdust}
\end{equation}
At larger elevations ($z>H_{\rm g}$) a simple scale height approach is imprecise as discussed by \citet{Fromang2009}. It is further noteworthy that there exists a lower limit to the dust scale height as the shear between the settled layer of solids and the low-viscosity gas above it will trigger the Kelvin-Helmholtz instability which stirs-up the solids. Thus, the effect increases the local turbulent $\alpha$ \citep{Youdin2002,Johansen2006a}.

While the number density and size of the particles is usually constrained from disk models, the relative velocity among grains or pebbles depends on which process dominates the particle velocity \citep[see][for a review]{Birnstiel2016}. For a large portion of a turbulent disk and for small particles (low $\rm{St}$), the velocity due to turbulence is dominating the approach speed of two particles and is approximately given by \citep{Ormel2007}
\begin{equation}
    \Delta v \approx \sqrt{\frac{3 \alpha}{\rm{St}+\rm{St}^{-1}}} c_{\rm s}\,,
\end{equation}
where $\alpha$ is the \citet{1973A&AShakuraSunyaev} parameter for turbulent strength. We note that depending on the nature of the turbulence, the induced relative velocities might be lower because particles move in concert with each other. This is the case when the turn-over times of the large eddies is longer than the stopping time of the particles. Such a situation can be expected if turbulence is driven by the vertical shear instability \citep{Nelson2013,Lin2015}.

The relative velocities are not only used to estimate collision rates but are also key to determine the outcome of a collisional encounter. When the relative velocities are larger than a threshold velocity $v_{\rm frag}$, particles can break apart and prevent further growth. Laboratory experiments place $v_{\rm frag}$ in the range from \SI{1}{\meter\per\second} to \SI{10}{\meter\per\second} \citep{2008ARAABlumWurm,2018MNRASGundlach,2019ApJSteinpilz} depending on the composition. However, this is part of ongoing research and numerical simulation of porous grains hint at larger speeds giving rise to uninhibited growth to meter or kilometer-sized bodies, that is, planetesimals \citep{Okuzumi2012,Kobayashi2021}. However, on the way to planetesimals, other potential barriers need to be overcome. Instead of fragmenting, the collision of two grains could also lead to the impacting particle bouncing-off the target particle which also prevents sticking and growth \citep{Seizinger2013}. Recent work indicates that discrepancies \citep{Schrapler2022} between limits derived from laboratory and numerical approaches could be resolved by taking into account the size of the particle \citep{Arakawa2023} which should motivate the adoption of those limits in modern dust evolution models \citep[e.g.][]{Stammler2022}.

\subsubsection{Radial Drift}\label{sec:radialdrift}
A third barrier for growth is not related to collisions between particles but to their aerodynamic interaction with the gaseous material \citep{1977ApSSWeidenschilling}. It is a key element of today's dynamic picture of planet formation because of the mobility of building blocks discussed in Sect. \ref{sect:predictionsclassiccaltheorypardigmshifts}. 

Considering a parcel of gas, the pressure support of the gas interior to it reduces the orbital velocity relative to the Keplerian value to \citep{1976PThPhAdachi}
\begin{equation}
v_{\rm g,\phi} = v_{\rm K} \sqrt{1 - 2\eta}\,,
\end{equation}
where 
\begin{equation}
\eta  \equiv -\frac{r}{2 v_{\rm K}^2\rho} \frac{d P}{d r}\,,
\end{equation}
$v_{\rm K}$ is the Keplerian orbital velocity, and $\rho$ and $P$ are the density and pressure of the gas. Although $\eta$ is small for most disk profiles, the reduced gas speed introduces a headwind for dust particles orbiting at $v_{\rm K}$. At 1 AU, it is on the order of 100 m/s. While this is small relative to $_{\rm K}$, this is non-negligible in absolute terms.

Therefore, gas drag is acting and breaking the dust particle, removing its angular momentum and leading to a radial velocity component. Assuming the radial velocity to be in steady-state and small compared to the orbital velocity, a low dust-to-gas ratio, and a small value of $\eta$, it follows \citep{1986IcarusNakagawa}
\begin{equation}
    v_{\rm d,r} = \frac{-2 \eta v_\text{K} + \rm{St}^{-1} v_{g,r}}{\rm{St}^{-1}+\rm{St}}\,.
    \label{eq:radial_drift}
\end{equation}
We note that $\eta$ is typically negative due to the pressure decreasing with orbital distance, thus the particles indeed move -- or drift --towards the star. 

\begin{figure*}[h]
\begin{minipage}{0.55\textwidth}
\centering
\includegraphics[width=1.0\textwidth]{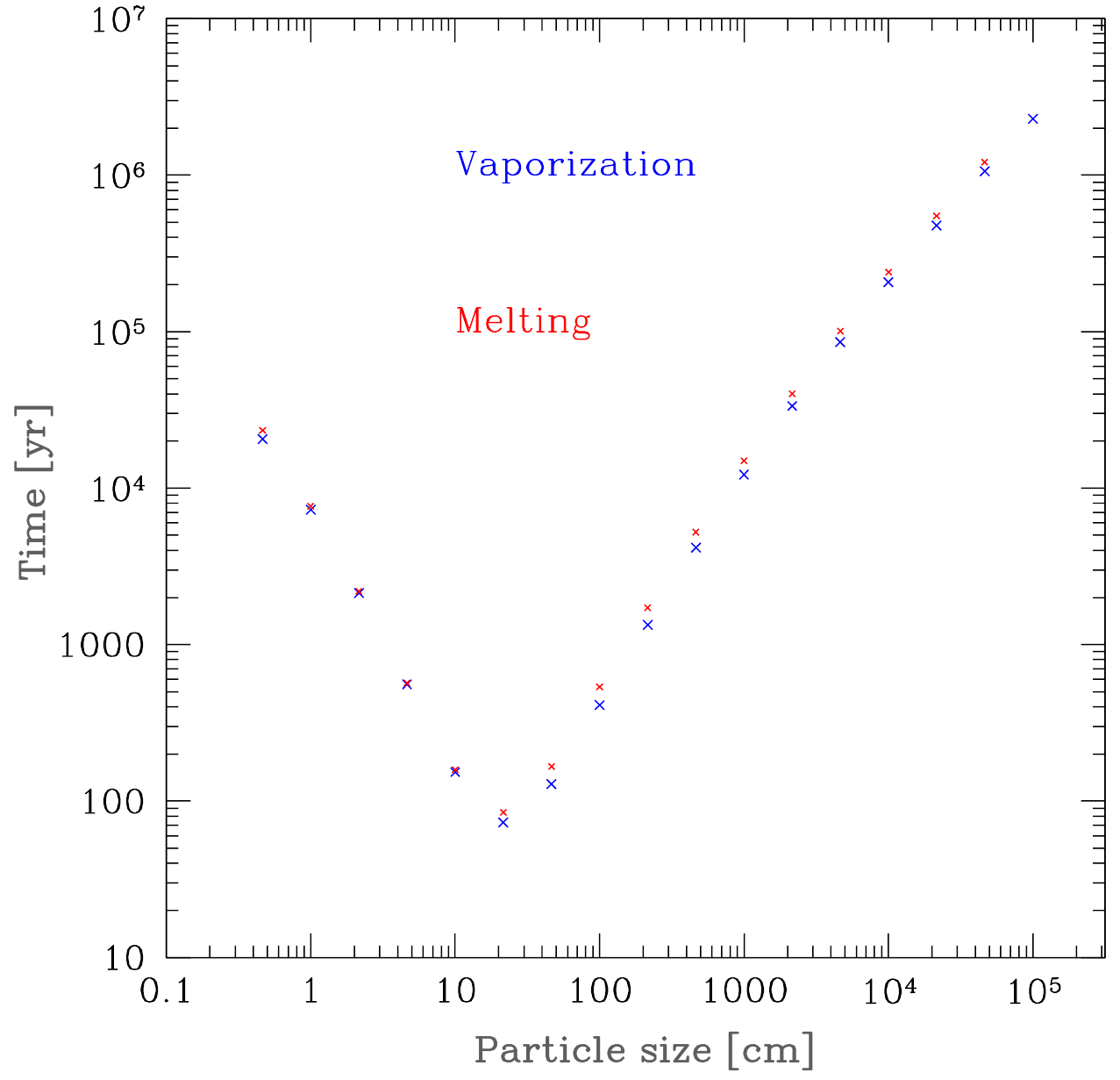}          
\end{minipage}
\hfill   
    \begin{minipage}{0.4\textwidth}  
    \caption{Radial drift resulting from the different rotation velocities of solid particles (pebbles, planetesimals) and nebular gas around the star. The figure shows the time for a rocky (silicate) body of different size to drift from 1 AU to its thermal destruction (either via vaporisation or melting) in the inner hot part of the disk. Destruction happens at the rockline, which corresponds to temperatures of about 1800 K usually found at orbital distances of a few 0.01 AU to 0.1 AU. In the size domain between 10 and 100 cm, bodies spiral to destruction extremely quickly (60-70 yrs). Smaller and larger bodies spiral in much slower. This plot emphasises the individual bodies' behaviour. The collective behaviour introduces new effects (Sect. \ref{sect:pebbles2planetesimals}).}
   \end{minipage}
    \label{fig:drift} 
\end{figure*}

From Eq. \ref{eq:radial_drift}, we see that particles with Stokes number close to unity will experience the strongest radial drift. As can be seen in Figure \ref{fig:drift}, this has the consequence that such  objects are removed extremely quickly from the inner parts of the Solar nebula (less than 100 years for roughly meter-sized bodies). For typical disk profiles, the radial drift timescale becomes shorter than dust growth timescales below a Stokes number of unity. Therefore, radial drift can act as a barrier which removes particles instead of allowing for growth. This fact was termed the meter-sized barrier and is a common issue discussed since the early works of \citet{Whipple1972}, \citet{1977ApSSWeidenschilling}, and \citet{1976PThPhAdachi}. 

However, if a local pressure maximum exists in the radial gas distribution, $\eta$ and the drift velocity can change sign. Thus, the solids accumulate at the pressure maximum and it will act as a dust trap. In these locations, the drift limit can be overcome. Symmetric dust features observed with the Atacama Large Millimeter/submm Array (ALMA) show observational evidence of this scenario \citep{ALMA-Partnership2015,2018ApJAndrewsA}.

However, today radial drift is no longer perceived as a process that is only detrimental to growth, but that on the contrary can critically help to form planetesimals and grow planets because of two key mechanisms: first, when considering the collective behaviour of drifting particles (instead of individual ones as in the previous considerations) and when including their back-reaction onto the gas disk, the following instability occurs:  spontaneously forming over-densities of drifting particles make the surrounding gas orbit a bit faster because of the back-reaction. This reduces the inward drift of this clump of particles. They are then catching other individual particles that drift faster, which further amplifies the process, making it run away. As discussed further in Sect. \ref{sect:pebbles2planetesimals}, this \textit{streaming instability} \citep{2005ApJYoudinGoodman} probably plays a key role for the formation of planetesimals.

Second, if the drifting particles can be accreted by planets \citep{2010AAOrmelKlahr}, they become a large reservoir for solid growth. At Stokes number of \SI{0.01}{} to \SI{0.1}{}, radial drift is efficient and since particles with these Stokes numbers correspond roughly to a few centimeters in size, the objects were termed pebbles and the accretion of these particles is known as \textit{pebble accretion} which will be discussed below in Sect. \ref{sec:solid_acc}.

\subsubsection{Entrainment of Grains in Photoevaporative Winds}
In contrast to the nebular gas, high energy radiation will not lead to heating of dust particles to thermal energies which would allow them to become unbound. However, small dust can be entrained in the gaseous photoevaporative wind which can affect the mass budget and composition and might provide direct observational clues \citep{2016MNRASFacchini,Hutchison2016,2020AAFranz,Sellek2020Entrainment}. For externally driven winds, it can be assumed that all grains are entrained with a size below a critical size \citep{Sellek2020Entrainment}
\begin{equation}
\label{eq:entrainment_size}
a_{\rm ent,ext} = \frac{v_{\rm th} \dot{\Sigma} r^2 }{\rho_{\rm s} \mathcal{F} G M_{\star}}\,,
\end{equation}
where $v_{\rm th}$ is the thermal velocity of the heated gas ($T\sim \SI{1000}{\kelvin}$), $\dot{\Sigma}$ is the gas surface density change due to photoevaporation, and $\mathcal{F} \approx H_{\rm g}/r$ is a geometrical factor. Equation \ref{eq:entrainment_size} is obtained by equating gravity to the gas drag from the escaping wind. For internally driven winds, the limiting size is most likely given by the particles which can be swept upwards by a vertical flow in the disk. For this case, \citet{2021MNRASBoothClarke} derived another limiting particles size which lies for typical disks close to the monomer size of the grains on the order of \SI{e-5}{\centi\meter} implying that the entrainment can be stalled completely for disks with weak photoevaporative winds. \citet{2022AABurnA} discuss the effects of entrainment on the solid mass budget for a range of parameters.

\subsection{Pebbles to planetesimals}\label{sect:pebbles2planetesimals}
To proceed on the way of converting the solids in disks from pebble size to eventually planet size, the bouncing, fragmentation, or drift limits discussed in the last section need to be overcome. This can be the case if particles spend little time as aggregates where they would drift or fragment and instead a process exists which directly bridges these critical size regime to at least several kilometer-sized objects.

Direct formation of large planetesimals overcomes several other threats which would affect intermediate-sized planetesimals. First, at tens of kilometer sizes, they would be sturdy enough to not fragment in mutual collisions \citep{1999IcarusBenzAsphaug}. Second, erosion of the planetesimal by interaction with the gas becomes inefficient and limited to a smaller region close to the star for larger planetesimals \citep{Demirci2020,Schonau2023}. Similarly, erosion due to collisions with pebbles are negligible at larger sizes \citep{2019AABurn}. Lastly, as visible in Fig. \ref{fig:drift}, large planetesimals are not affected by radial drift as their mass out-scales the surface-dependent drag force in the turbulent drag regime leading to negligible specific drag \citep{1977ApSSWeidenschilling,Birnstiel2016,2019AABurn}.

For these reasons, a scenario where planetesimals form relatively large is preferred. This can be achieved if dust is concentrated enough to trigger the gravitational collapse of a dust cloud. \citet{Goldreich1973} proposed this mechanism as a pathway for planetesimal formation where the required large dust-to-gas ratio is achieved by settling of dust to the midplane. However, for global dust-to-gas ratios on the order of a percent, including the aforementioned (Sect. \ref{sec:dust_to_pebbles}) lower limit due to self-induced turbulence of the dust layer, it was shown that this is impossible to realize \citep{Sekiya1998,Youdin2002}. Thus, the dust-to-gas ratio needs to be locally enhanced with a different process. To achieve this, \citet{Youdin2002} found factor $\sim$10 enhancements from radial drift only in a smooth disk which is a fringe case to trigger the collapse. Later, it was understood that the collective behavior of radially drifting dust can trigger what is now called the streaming instability via its back-reaction on the gas \citep{2005ApJYoudinGoodman,Squire2018}, as explained at the end of Sect. \ref{sec:radialdrift}. It has an automatic dust-concentration effect \citep{Johansen2007,Simon2016} aiding the potential gravitational collapse but it is also a source of turbulence which could act against the gravitational collapse \citep{2020ApJKlahr}. The growth of the streaming instability is only efficient at enhanced super-Solar pebble-to-gas ratios and particles with Stokes numbers $\rm{St}>0.01$ \citep{Bai2010,Carrera2015}. Furthermore, if a distribution of particle sizes \citep{Krapp2019} or a background turbulence \citep{Umurhan2020} is present, the range of conditions suitable for streaming instability shrinks. Thus, it might be a process which does not trigger planetesimal formation on its own. However, it is of great aid in the final clumping process. These considerations are discussed in the context of formulating a criterion for the formation of planetesimals as a function of disk conditions \citep{Gerbig2020}.

Therefore, to trigger planetesimal formation aided by the streaming instability concentrating pebbles in clouds eventually undergoing gravitational collapse, some other process could be required to reach the necessary dust concentration. It could be that planetesimals only form at preferential locations in the protoplanetary disk, such as the inner edge of the unionized and thus magnetically dead-zone \citep{Gammie1996,Drazkowska2016} or the iceline where water sublimates \citep{2017AADrazkowskaAlibert,Schoonenberg2017}. Alternatively, local concentrations of pebbles were assumed as put forward by \citet{2019ApJLenz} to construct a flux-regulated prescription for planetesimal formation useful for global planet formation models. Such an approach is motivated by a number of theoretically expected processes which can create  zonal flows and therefore local over-densities. The zonal flow can for exampl be caused by magnetic instabilities \citep{2013ApJDittrich,Bai2014,Bethune2016}. Particle rings could originate also from a dust-driven instability in the outer magnetically active regions \citep{Dullemond2018a}. Alternatively, particles could accumulate in forming vortices \citep{Raettig2015,Manger2018} caused by the vertical shear instability \citep{Nelson2013,Lin2015,Pfeil2019}. Most importantly in an observationally-driven field such as planet formation, such rings are observed with ALMA in many protoplanetary disks \citep[e.g.][]{ALMA-Partnership2015,2018ApJDullemond}.

While there is certainly a need for more research in determining the mechanism(s) of planetesimal formation, another important topic is the resulting size distribution. For the streaming instability, mass distributions of clumps of dust have been determined \citep{Simon2016,Simon2017,2017AASchaefer,Abod2019,Li2021}. However, these works are often missing the resolution to resolve the latest final gravitational collapse phase which might reduce the final planetesimal mass further \citep{Nesvorny2021,Polak2023}. Nevertheless, the order of magnitude in size of these first primordial planetesimals is likely larger than 100\,km. This size regime might be problematic for both pebble and planetesimal accretion presented below without further fragmenting the primordial bodies or making use of the concentration of solid \citep{Lorek2022}.

\subsection{Solid accretion mechanisms}\label{sec:solid_acc}
In this section, we discuss how protoplanets grow by accreting solids (as opposed to gas accretion). There are three different fundamental types, involving increasingly large bodies: pebble accretion, planetesimal accretion, and giant impacts. We first give an short overview describing how the  understanding of solid growth has changed over time and then address the physics of the mechanism separately. 

\subsubsection{Development of the concepts of solid accretion}
In the past fifty years, the concept of how protoplanets grow by accreting solids has evolved. At the beginning of modern theories still considered valid in a general sense stands the planetesimal accretion paradigm, proposed by Viktor Safronov in 1969 \citep{1969BookSafronov}. According to this concept, protoplanets (also called planetary embryos) grow through the gradual collisional accretion of small solid bodies, the planetesimals, which have sizes of about 1 to 100 km. This idea stems from the observation of the ubiquity of such bodies in the Solar System (asteroid and comets). 

Within the planetesimal paradigm, an important process is the one of planetesimal runaway growth \citep{1993IcarusWetherillStewart}. During planetesimal runaway accretion, the more massive bodies have a large gravitational focusing factor (the Savronov factor, see Eq. \ref{eq:mdotrho} below). This means that their effective collisional cross section is much larger than their physical size. In turn, the more massive bodies grow faster than less massive ones. This represents an unstable (runaway) situation that splits an originally monodispersed (same sized) planetesimal population into massive rapidly growing runaway bodies (protoplanets) and small slowly growing background planetesimals, i.e., a bimodal distribution. 

However, in 1993, \citet{1993IcarusIdaMakino} showed that the fast runaway growth regime cannot be sustained ad infinitum. Instead, once the large bodies are massive enough to dynamically heat the surrounding planetesimals (i.e., increase their eccentricities and inclination), the growth mode changes to the slower oligarchic  mode \citep{2010ApJOrmel}. In this regime, several neighbouring protoplanets (now called oligarchs) grow in lockstep, mutually separated by about 5 to 10 Hill spheres. The oligarchs still grow faster than the background planetesimals. In the oligarchic regime, the growth timescale becomes an increasing function of planet mass, opposite to the runaway regime.

However, in oligarchic growth there could be a timescale problem when it comes to the formation of giant planet cores in the outer regions of a planetary system, especially outside of about 10 AU. Given the high amounts of nebular gas that gas giants contain, the massive cores required to trigger gas runaway accretion (about 10 Earth masses) obviously need to form before the dissipation of the gas nebula i.e., within 3-10 Myr. \citet{2003IcarusThommes} highlighted this issue, indicating that the growth of these cores may take too long to occur within the observed lifetimes of  protoplanetary disks. The problem becomes more acute the further one moves away from the star since the growth timescale scales with the Keplerian frequency. While the in situ formation of Jupiter from 100 km planetesimals at 5.2 AU is still feasible in a nebula with about 2-4 times the surface density of the MMSN and modern models \citep{2020ApJPodolak}, it becomes very difficult to form in situ the ice giants or the giant extrasolar planets observed at distances of several tens of AUs. In this context it should be noted that the ice giants might have formed closer in (in the Nice model, they rather start forming at 6-8 AU, \citealt{2011NatureWalsh}), and the gravitational instability mechanism could have formed at least some of the massive distant extrasolar planets \citep{Boss2011}. An alternative explanation could be scattering events \citep{2019AAMarleau}.

To address the timescale problem, \citet{2004ApJIda1} suggested small planetesimals, around 1 km in size, as a solution. Planetesimal growth is the faster the smaller the planetesimals are, for two reasons: First, there is a stronger damping of eccentricities and inclinations by the gas drag of the surrounding  protoplanetary disk, leading to smaller random velocities of the planetesimals and thus a larger gravitational focusing factor. Second, the enhancement of the capture radius of the protoplanet because of drag in its gaseous envelope of the protoplanet is also increased for small planetesimals\footnote{When the effect of aerodynamic fragmentation of planetesimals in protoplanetary atmosphere is included besides drag, then  the capture radius effect is also efficient for large planetesimals, see \citealt{2020ApJPodolak}.}.

However, observations of the asteroid belt \citep{2009IcarusMorbidelli} and various streaming instability and planetesimal formation models \citep[e.g.,][]{Polak2023} rather indicated that planetesimals are born relatively large, around 100 km in size. This would pose a challenge to the idea of small planetesimals being the primary building blocks of planets, but it should be noted that contrasting evidence also exists (see \citealt{2021AAEmsenhuberB} for an extended discussion).

To reconcile these different perspectives, a new paradigm was introduced in a dedicated way by \citet{2010ApJOrmel} which proposes that  bodies much smaller than km-sized planetesimals are actually important planetary building blocks. This pebble accretion paradigm suggests that small pebbles (usually mm to dm-sized), can more efficiently accrete onto larger bodies, especially at larger orbital distances \citep{2012AALambrechtsJohansen}. The fundamental difference to planetesimal accretion lies in the fact that during the encounter, not only gravity is important but also dissipative drag forces \citep{2017ASSLOrmel}. This process is fast and can occur even in the outer regions of a planetary system, as the capture radius scales with the Hill sphere which in turn scales with the semi-major axis. The physical size of the core  which is more relevant for planetesimal accretion does in contrast not increase with distance. 

However, pebble accretion in the outer system requires a large enough starting seed because pebbles cannot be accreted by another pebble. This can be problematic if the streaming instability does not lead to the formation of such bodies \citep{Lorek2022}. This would bring us back to the need for planetesimals during the intermediate growth stage (from the largest mass formed directly out of the planetesimal formation mechanism to the lowest mass where efficient pebble accretion starts), which would in turn bring back the aforementioned timescale problem in the outer disk.

One potential solution to this new challenge is the formation of protoplanets in a structured disk that contains rings \citep{2022AALau,JiangOrmel2023}, or more accurately pressure maxima where drifting pebbles accumulate. The high concentration of pebbles and the low headwind velocity inside the ring renders pebble accretion very efficient. These studies suggest that the presence of pressure maxima can facilitate the formation of massive protoplanets even at large orbital distances (50-100 AU). Disk structures can furthermore act as traps for the orbital migration of protoplanets, further enhancing the efficiency of planet formation \citep[e.g.,][]{Coleman2016}. 

In summary, solid accretion involves various paradigms and models, including the planetesimal paradigm with runaway and oligarchic growth stages, pebble accretion and giant impacts. The latter is thought to be the final stage in the growth of terrestrial planets. The role of pebbles, planetesimal, and impacts in different regions and phases of planetary system growth are central subjects of ongoing research and hybrid models start to be developed \citep{Coleman2021,2022AAVoelkel}.

\subsubsection{Basic concepts of planetesimal accretion}
A key concept in collisional growth via planetesimals is the one of the gravitational focusing. In this section, we address this effect in a more educative fashion than processes addressed in the other parts. More realistic state-of-the-art calculations are considerably more complex \citep{2001IcarusInaba}, but the simple picture presented here illustrates the basic concepts. 

\begin{figure*}[h]
	\centering
	\includegraphics[width=1\linewidth]{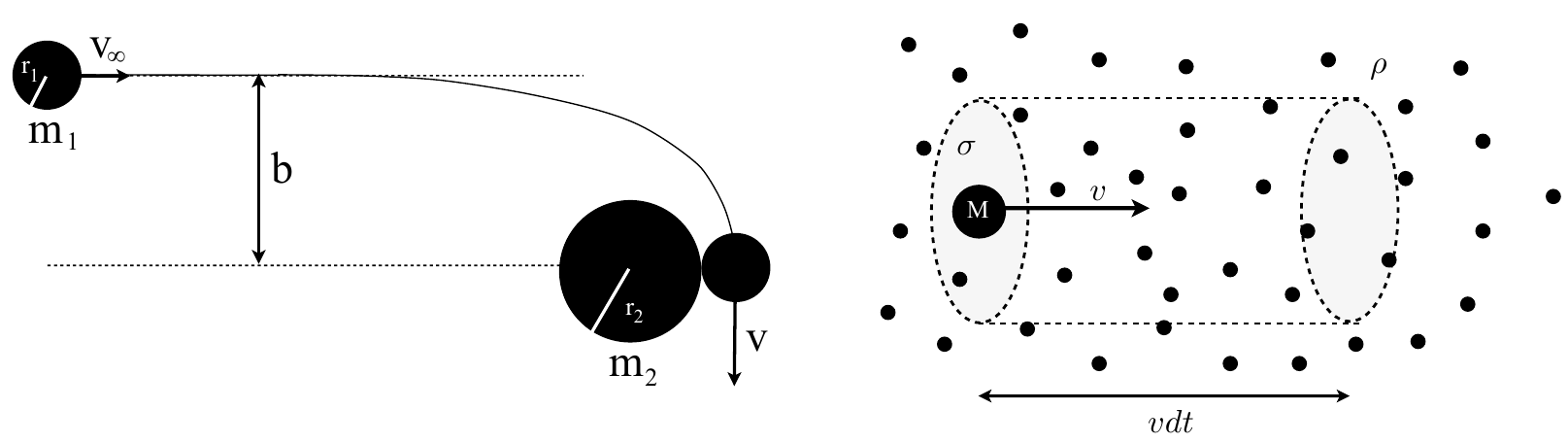}
	\caption{Illustration of the approach of two gravitating bodies (left) and the cylinder of accreted mass onto a moving body (right). The cylinder (dotted) is visualized for a particle with mass $M$ moving at velocity $v$ through a medium with density $\rho$ for a time $dt$ at a certain interaction cross-section $\sigma$. In the left schematic, particle 1 with mass $m_1$ and radius $r_1$ is drawn at two different times on its path (arrow), once at infinite separation from particle 2, where it has a velocity $v_{\infty}$, and at its closest approach to particle 2 where it has a differing velocity $v$ due to the gravitational attraction.}
	\label{fig:planaccschematic}
\end{figure*}

Consider two bodies of masses $m_1$ and $m_2$ with radii $r_1$ and $r_2$ approach each other at an initial relative velocity $v_\infty$ and with an impact parameter $b$ (Figure \ref{fig:planaccschematic}). The velocity is formally the one at infinity, or in practice when the planets are still very far from each other. 

In a billiard game, the collisional cross sections of two bodies is simply given by the geometrical cross sections 
\begin{equation}
    \sigma=\sigma_{\rm geo}=\pi(r_1+r_2)^2
\end{equation}
However, the attracting nature of gravity leads for planetary growth to an increase of the collisional cross section over the geometrical one. This is called gravitational focusing. Energy and angular momentum conservation allows one to calculate the collisional cross section for two arbitrary sized, gravitating (spherical) bodies, neglecting the influence of the Sun (two body approximation).

Energy conservation yields (with values at infinity on the left and values at the closest approach on the right):
\begin{equation}
\frac{1}{2}\mu v_\infty^2=\frac{1}{2}\mu v^2-G\frac{m_1 m_2}{r_1+r_2}
\end{equation}
with $\mu$ the reduced mass $(m_1 m_2)/(m_1+m_2)$. The conservation of angular momentum yields
\begin{equation}
\mu b v_\infty=\mu (r_1+r_2)v
\end{equation}
which can be solved to yield the velocity at the closest approach, $v=b v_\infty/(r_1+r_2)$. Together with the definition of the escape velocity
\begin{equation}
v_{\rm esc}=\sqrt{\frac{2 G (m_1+m_2)}{r_1+r_2}}
\end{equation}
combining gives
\begin{equation}
    b^2=(r_1+r_2)^2\left(1+\frac{v^2_{\rm esc}}{v^2_{\infty}}\right)
\end{equation}
This yields
\begin{equation}
    \sigma=\pi b^2=\pi(r_1+r_2)^2\left(1+\frac{v^2_{\rm esc}}{v^2_{\infty}}\right)=\sigma_{\rm geo}(1+\Theta)
\end{equation}
where we have used the so-called Safronov number $\Theta=v^2_{\rm esc}/v^2_{\infty}$ in honour of Viktor Safronov who was the first to develop this collisional accretion scenario. We thus find that the collisional cross section is increased over the geometrical cross section by a factor which is proportional to the square of the escape velocity to the random velocity. Clearly, this ratio can become quite big thus increasing strongly the collision probability.  

The accretion of mass per time is now a simple geometrical task. As visualized in the right panel of Fig. \ref{fig:planaccschematic}, a particle with an interaction cross-section of $\sigma$ moving through a medium of density $\rho$ will accrete over a time $dt$ the mass in the visualized cylinder
\begin{equation}
\dot{M} dt = \sigma \rho v dt\,,
\end{equation}
where we introduced the mass accretion rate $\dot{M}$. Thus, for our derived cross-section, we get
\begin{equation}
\label{eq:mdotrho}
\dot{M} = \rho v_{\infty} \pi(r_1+r_2)^2\left(1+\frac{v^2_{\rm esc}}{v^2_{\infty}}\right)\,,
\end{equation}
where the velocity with which the particle under consideration (with mass $M$) can be set to $v_{\infty}$ which should be interpreted as the average relative velocities between the considered particle and all other particles at large distance.

For applications, we need to describe the density $\rho$ used in the above formula. Assuming the medium to be composed of planetesimals, that is, particles well decoupled from the gas, $rho$ can be estimated using the eccentricity $e$ and inclination $i$ distribution of the particles. Under these assumptions, the particles move on Keplerian orbits. Therefore, their velocity is on the order of $v \approx \sqrt{ e^2 + i^2} v_{\rm K}$ \citep{Lissauer1993}.

This velocity can also be used to describe the planetesimals as a fluid and derive a planetesimal scale height $H_{\rm pla} = v/\Omega_{\mathrm{K}} = a \sqrt{e^2 + i^2}$ with a corresponding approximate planetesimal midplane density of \citep{1969BookSafronov}
\begin{equation}
\rho_{\rm pla} \approx \frac{\Sigma_{\rm pla}}{\sqrt{2\pi (e^2+i^2)}}\,.
\end{equation}
Combined with Eq. \ref{eq:mdotrho}, this yields an easy-to-use estimate of the planetesimal accretion rate
\begin{equation}
\label{eq:mdotRunaway}
\dot{M}_{\rm pla} \approx \Sigma_{\rm pla} v_{\rm K} (r_1+r_2)^2\left(1+\frac{v^2_{\rm esc}}{(e^2+i^2) v_{\rm K}^2}\right)\,,
\end{equation}
where we omitted factors of order unity to not overstate the precision of the approach. We see that the more massive a growing planet, the larger the escape velocity and the more rapid its growth. Therefore, if the other factors are kept constant this accretion formula describes relatively well the \textit{runaway planetesimal accretion}.

As seen in this derivation, the complexity emerges when trying to estimate the exact approach speed of two bodies. For non-circular orbits, it depends on the location on the orbit of both the target and the impactor. The integration then becomes non-trivial and when additionally accounting for approaches which do not neglect the gravity of the central star, the problem is no longer solvable analytically. Therefore, numerical studies were conducted \citep{1990IcarusGreenzweigLissauer,1992IcarusGreenzweigLissauer,2001IcarusInaba} which obtained much more precise prescriptions for the accretion rate of planetesimals.

As the aforementioned insights were gained, it also became clear that the eccentricities and inclinations of the planetesimals are key parameters which are not independent of the planetary mass. In the vicinity of a relatively massive growing protoplanet, the eccentricities and inclinations of the planetesimals are affected due to gravitational stirring \cite{1990IcarusIda,Ohtsuki1999} and thus the velocities and accretion probabilities are affected. This reduces the fast, runaway accretion via the denominator in Eq \ref{eq:mdotRunaway} and leads to a transition the slower so-called \textit{oligarchic} accretion. Prescriptions to model the gravitational stirring were formulated and applied in a number of works \citep{Ohtsuki1999,Chambers2006,2013A&AFortier}.

The particles in the gas disk are not only subject to gravitational stirring of the planet but also experience a similar effect from gas turbulence \citep{Ormel2012,2016ApJKobayashi} and from each other \citep{2002IcarusOhtsuki}. The stirring effects are compensated by aerodynamic gas drag which slows down the planetesimals as long as the gaseous disk is present \citep{1976PThPhAdachi}. Since the drag force scales with the surface of the planetesimals while the other processes mainly depend on the mass, gas drag is more efficient for small planetesimals. Thus, planetesimal accretion becomes more efficient if planetsimals are born small (of order km in radius) or fragments are produced frequently.

In addition, another aerodynamic effect emerges from the gaseous envelope of the planet. As the gas density is increased in the vicinity of the planet over the the density in the disk, the aerodynamic drag increases as particles get close to the planet. Instead of passing the planet, it can then be accreted \citep{2003A&AInaba,2006ConfMordasini}.

While one might be tempted to assume that this is an effect which takes only place close to the planet when the particles are anyways on collision-course, one needs to consider that during the gas disk stage, the planetary envelope can be very extended and have a size comparable to the Hill sphere (see Sect. \ref{sec:gasaccretion}).

\subsubsection{Pebble accretion}
The pebble accretion process builds on the aforementioned gas-drag enhanced capture radius of the planet. This effect was studied with focus on larger planetesimals for decades \citep{1988IcarusPodolak,2003A&AInaba}. For even smaller particles, which are experiencing the strongest drag effects (i.e. Stokes numbers close to unity), the cross section of the planet is even more significantly increased over the physical cross-section due to aerodynamic drag.

The idea to accrete the small, millimeter to centimeter sized particles was brought forward by \citet{Klahr2006a}, although it was applied to giant planets forming in a vortex instead of classical core-accretion, which was only done in \citet{2010AAOrmelKlahr}. The advantages over planetesimal accretion are faster growth and a larger reservoir of solids, in principle all solids exterior to the growing planets location. This motivated the full development of an alternative paradigm of solid accretion termed pebble accretion \citep[see][for in-depth reviews]{2017ASSLOrmel,Johansen2017}.

Here, we briefly review the fundamentals of the approach. As for planetesimals, an accretion rate of solids can be expressed as the flux of particles through the cross sectional area of the planet
\begin{equation}
\dot{M}_{\rm peb} = \pi R_{\rm acc, peb}^2 \rho_{\rm peb} v_{\rm rel, peb}\,.
\label{eq:mdotpeb}
\end{equation}
Since the particles drift through the protoplanetary disk, the relative velocity $v_{\rm rel, peb}$ contains a significant contribution from the radial motion. Approximately, the sum of the difference between Keplerian orbits and the radial velocity can be used for the relative velocity \citep{2017ASSLOrmel}
\begin{equation}
v_{\rm rel, peb} = v_{\rm d,r} + \frac{3}{2} \Omega_{\mathrm{K}} R_{\rm acc, peb}\,.
\end{equation}
This expression neglects flow patterns of the gas due to the planet or turbulence which can become relevant for low Stokes numbers.

On the other hand, eccentricities and inclinations of pebbles are damped due to efficient gas drag to a degree where they can be neglected. As for planetesimals, the relevant mass density of pebbles distributed in space (not to be confused with the bulk density of an individual pebble)  is a key quantity to determine. In steady-state, it can be approximately from the reduced scale height of pebbles or dust (Equation \ref{eq:Hdust}) using the definition of a surface density $\rho_{\rm peb} \approx \Sigma_{\rm peb}/(\sqrt{2\pi} H_{\rm peb})$. This expression is accurate if the planet's radius of influence $R_{\rm acc,peb}$ (see below), is small with respect to the scale height of the pebbles. However, once the planet becomes vertically more extended than the scale height, it transitions from this so-called 3D regime to a 2D case, where the full vertical extent of the pebble distribution can be accreted, resulting in $\dot{M}_{\rm peb} = 2 R_{\rm acc, peb} v_{\rm rel, peb}$ instead of Equation \ref{eq:mdotpeb}.

The final quantity to be determined is the accretion radius $R_{\rm acc, peb}$. It can be estimated by comparing the timescale that particles require to settle toward the planetary core against the timescale they reside within the planets' sphere of influence. The latter time is either determined from the pebble radially drifting toward the star (headwind regime) or passing azimuthally due to the difference in Keplerian orbital velocities (shear regime). Following \citet{2017ASSLOrmel}, this leads to an expression for the headwind regime 
\beq
R_{\rm acc,peb,hw} \sim \sqrt{ \frac{2 G M \mathrm{St} \Omega_{\mathrm{K}}}{v_{\rm d,r}}}
\eeq
while in the shear regime it evaluates to $R_{\rm acc,peb,hw} \sim \mathrm{St}^{1/3} R_{\rm H}$ , where the Hill radius 
\begin{equation}
R_{\rm H} = \left(\frac{ M }{3 M_{\star}}\right)^{1/3} a
\end{equation}
is used where $a$ is the semimajor axis of the planet.

This approximate treatment can be used for pebbles with Stokes numbers of order 0.01 to 1. For lower Stokes number, that is, for dust, particles follow the gas streamlines around the planet without accretion \citep{2017ASSLOrmel}. For larger Stokes number, that is, entering the regime of planetesimals, the assumptions of the Epstein drag regime and zero inclinations and eccentricities break.

At larger planetary masses, pebble accretion terminates once the planet significantly perturbs the gas disk (see Section \ref{sec:orb_migration}) to create a local pressure maximum where pebble drift is stopped \citep{2014AALambrechts,2018AAAtaiee}. While this effect depends on the gas disk properties, such as viscosity and scale height, the typical pebble isolation mass is on the order of 10 to 40 \,M$_{\oplus}$.

Recently, hybrid models combining planetesimal and pebble accretion have started to be constructed \citep[e.g.,][]{2018NatAsAlibert,2022AAVoelkel,KesslerAlibert2023A&A}. They find a highly complex interplay with phases where one or the other mechanism is dominant resulting in multiple generations of protoplanets, where the additional planet luminosity resulting from planetesimal accretion can delay gas accretion relative to the pure pebble case. They also find that giant planet formation alone from 100 km planetesimals is efficient in forming giant planets with final orbital distances out of about 1 AU, but not further away \citep{2020AAVoelkel}. This could indicate a distance dependency of the importance of pebbles (larger distances) and planetesimals (smaller distances). Together with the effect of planetesimal fragmentation \citep{KaufmannAlibert2023}, that can link the two scenarios, more work is warranted to study the relative roles of planetesimal and pebble accretion and their interplay.

\subsection{Gas accretion}\label{sec:gasaccretion}
The process of gas accretion can be separated into two stages, an initial cooling-limited 
stage at early times and lower planet masses. During this first stage, the planet's envelope is smoothly attached to the surrounding protoplanetary disk. This phase can be followed by a later disk-limited stage for sufficiently high planet masses where the planet's surface is detached from the disk. 

\subsubsection{Cooling-limited, attached stage}
In the initial stage, as a protoplanet grows by solid accretion, its gravitational pull will increasingly attract H/He gas from the protoplanetary disk until it is balanced by the pressure support of already attracted gas. The pressure support is sustained by the luminosity in the planet's envelope, which is in turn generated by the envelope gas itself or by the accretion of solids. Before the advent of 3D hydrodynamic simulations, this situation was treated as a fully static 1D spherically symmetric problem where the cooling and contraction of the already accreted gas allows new nebular gas to flow into the planet's sphere of influence, regulating thereby the gas accretion rate \citep{1986IcarusBodenheimerPollack,2000ApJIkoma}. Originally, this sphere of influence was taken to be simply the smaller of the Bondi $R_{\rm B}$ and the Hill sphere radius $R_{\rm H}$ \citep{1986IcarusBodenheimerPollack}. However, hydrodynamic simulations show that only about the inner 0.25 $R_{\rm H}$ are actually bound to the planet, while the outer layers participate in the disk's surrounding shear flow \citep{2009IcarusLissauer}. 

In this inner part, the relevant equations are similar to the stellar structure equations. Spherical symmetry is assumed. Then, mass conservation, momentum conservation in the form of hydrostatic equilibrium , energy conservation, and energy transport can be written using the radius from the planet's center $r$ as coordinate \citep{1986IcarusBodenheimerPollack}:
\begin{align}\label{eq:internalstruct}
\frac{\partial m}{\partial r}&=4 \pi r^{2} \rho  &\quad  \quad \frac{\partial P}{\partial r}&=-\frac{G m}{r^{2}}\rho    \\
\frac{\partial l}{\partial r}&=4 \pi r^{2} \rho\left(\varepsilon -P \frac{\partial V}{\partial t} -\frac{\partial u}{\partial t}\right)             & \frac{ \partial T}{\partial r}&=\frac{T}{P}\frac{\partial P}{\partial r}\nabla(T,P)
\end{align}
where $G$ is the gravitational constant, $m$ the enclosed mass at radius $r$, and $P$, $T$, and $\rho$ are the pressure, temperature, and density in a layer. The gradient $\nabla$ depends on the process by which the energy is more efficiently transported (radiation or convection) as judged by the Schwarzschild criterion. In the case of convection, tabulated adiabatic gradients from specialised non-ideal equations of state are used which also yield the density as a function of pressure and temperature. Assuming radiative diffusion, the radiative gradient  is given by
\begin{equation}
\nabla_\mathrm{rad}=\frac{3 \kappa l}{64 \pi \sigma G m T^3},
\end{equation}
where $\kappa$ is the Rossland mean opacity and $\sigma$ is the Stefan-Boltzmann constant.

We notice that the energy equation is the only equation which explicitly depends on time $t$. This implies that it drives the temporal evolution. In this equation, $l$ is the luminosity (energy flux), $\varepsilon$ an extra energy source such as impacts or radiogenic heating (for stars it would be nuclear fusion), $V=1/\rho$ is the specific volume, and $u$ is the specific internal energy. Using the first law of thermodynamics, one can also consider the temporal change of the gas' entropy instead of the change of the volume and internal energy separately. This is particularly useful if (most of) the envelope is adiabatic and thus characterised by one entropy. 

These equations can be solved numerically \citep{1986IcarusBodenheimerPollack,2000ApJIkoma,2014ApJPisoYoudin,2012A&AMordasiniB,Kimura2020,2021AAEmsenhuberA} given boundary conditions, opacities, and an equation of state to close the system of equations. In the initial stage, the outer boundary conditions are given by the disk background conditions modified by the effects of the circulating flow \citep{2020MNRASali-dib}. Specialised equations of state (EOS) that cover the required regime of pressures and temperatures in giant planet interiors deviate strongly from the ideal gas case but must include also the degenerate limit \citep{1995ApJSSaumon,2021ApJChabrierDebras}. Such EOS are specially developed for the interior of planets and brown dwarfs \citep[see][for a review]{2020Helled}. Modern equations of state can to some extent account for varying helium fractions and metallicities $Z$, but many questions still remain \citep{2023HowardGuillot}. 

For the early stage in which solid accretion is the main source of luminosity for the planet, a useful simplification is to assume a constant luminosity throughout the structure of the planet and that for sufficiently large impactors (planetesimals), energy is deposited at the envelope to core boundary \citep{2012A&AMordasiniB}. This is a good approximation when the gaseous envelope is thin enough to allow for planetesimals to reach the solid core \citep{1988IcarusPodolak,2006ConfMordasini}. For small initial relative velocities between the accreted material and the planet, the luminosity of accreted material is given by the released potential energy in the gravitational field of the planet,
\begin{equation}
    L_{\rm solid} = G \frac{\dot{M}_{\rm acc,solid} M_{\rm c}}{R_{\rm c}}\,,
\end{equation} 
where $\dot{M}_{\rm acc,solid}$ is the accretion rate  of solids, and $M_{\rm c}$ and $R_{\rm c}$ are the core mass and core radius. This establishes a link between gas and solid accretion (Section \ref{sec:solid_acc}). Once the envelope becomes ``opaque'' to the incoming bodies, the liberated energy depends on the radius at which the particles are mainly stopped $R_{\rm stop}$ and  the corresponding encompassed mass at this radius $M_{\rm stop}$ \citep{1988IcarusPodolak}.

In the initial cooling-limited stage, the temporal evolution is controlled by the timescale of the Kelvin-Helmholtz cooling of the envelope $\tau_{\rm KH}$, therefore the characteristic timescale for the accretion of gas for a planet of mass $M$ can be estimated as
\beq
\dot{M}_{\rm gas}=\frac{M}{\tau_{\rm KH}}.
\eeq
The solution of the structure equations shows that under simplifying assumptions (for example, no solid accretion) the Kelvin-Helmholtz timescale $\tau_{\rm KH}$ can be parameterised as \citep{2000ApJIkoma,2014AAMordasiniA} 
\begin{equation}\label{eq:tKH}
\tau_{\rm KH}=10^{p_{\rm KH}} {\ \rm yr} \left(\frac{M}{M_\oplus }\right)^{q_{\rm KH}}\left(\frac{\kappa}{1 {\rm \ g \ cm}^{-2}}\right).
\end{equation}
In this equation, the parameters $p_{\rm KH}$ and $q_{\rm KH}$ are found  by fitting the accretion rate found by solving the internal structure equations \citep[e.g.,][]{2004ApJIda1,2011MNRASMiguel,2014AAMordasiniA}. \citet{2014AAMordasiniA} for example derive $p_{\rm KH}=10.4$, $q_{\rm KH}=-1.5$, and $\kappa=3\times10^{-3}$ cm$^{2}$/g. Once can see that the gas accretion rate is thus a increasing function of planet mass. Once the planet mass is sufficient high (about 5-10 Earth masses), $\tau_{\rm KH}$ becomes comparable to the disk lifetime, implying that gas accretion becomes important.
\subsubsection{Disk-limited, detached stage}
As the planet's mass increases further (where its envelope mass becomes increasingly important relative to the core mass), the  gas accretion process speeds up further, meaning that runaway gas accretion sets in \citep{1974IcarusPerriCameron,1980PThPhMizuno,1982PSSStevenson}. The exact critical mass at which this occurs depends on several factors including the opacity in the envelope \citep[given mainly by solid grain opacity][]{2014AAMordasiniB,ormel2014,Kimura2020} but also the solid accretion rate. At some point, core and envelope mass become equal, which is known as the crossover mass \citep{1996IcarusPollack}. In addition to the solid accretion luminosity, contributions from the gas' cooling and contraction become relevant and eventually dominate. 

At some point in the runaway phase, the outer radius of the planet contracts so rapidly (but still quasi-statically, \citealt{1986IcarusBodenheimerPollack}) that the gas disk can no longer supply nebular gas at a rate sufficient to fill the rapidly emptying shell -- the planet's surface thus detaches from the protoplanetary disk \citep{2000IcarusBodenheimer,2012A&AMordasiniB} and the growth enters the disk-limited accretion stage. The planet's radius now shrinks to a value that is much smaller than the Hills sphere of about 1.5 - 5 Jovian radii depending on the entropy of the gas in the interior \citep{2012A&AMordasiniC,2017ApJMarleau}. This is also the moment when the structure changes from an approximately spherically symmetric shape during the attached stage to a flattened one \citep{2012MNRASAyliffeBate,2017MNRASSzulagyiMordasini} with a circumplanetary disk surrounding the growing gas giant \citep{2022ApJAdamsBatygin}. The outer boundary conditions of the structure equations are also modified as gas now falls with high velocity on the surface of the planet where it shocks 
\citep{2012A&AMordasiniB,Marleau2023}. The resulting emission in the H-$\alpha$ line has been observed recently for forming extrasolar gas giants \citep{Haffert2019}.

In this later stage of massive planets ($M\gtrsim100 M_\oplus$), the gas accretion rate is no longer limited by the cooling of the envelope but by the supply rate of gas from the protoplanetary disk. The exact limit is influenced by the three dimensional structure of the gas around the planet. A first estimate yields the classical Bondi/Hill accretion rate \citep{2008ApJDAngeloLubow,2012A&AMordasiniB}, i.e., the rate at which the planet sweeps nebular gas given relative velocities caused by the Keplerian shear 
\beq\label{eq:mdotbondi}
\dot{M}_{\rm e, Bondi}\approx\frac{\Sigma}{H}\left(\frac{R_{\rm H}}{3}\right)^{3}\Omega.
\eeq
In this equation, $\Sigma$, $H$, $R_{\rm H}$, and $\Omega$ are the gas surface density averaged over the planet's feeding zone, the vertical scale height of the disk, the Hill sphere radius, and the Keplerian orbital frequency at the planet's position. More accurate rates can be derived from 2 and 3D hydrodynamic simulations \citep[e.g.,][]{Machida2010,2008ApJDAngeloLubow,2013ApJBodenheimer,Choksi2023} that can be used to calibrate the 1D approaches \citep{2022AASchib}.

Three-dimensional hydrodynamic simulations have in recent years also revealed that envelope gas can be recycled back to the protoplanetary disk with potential influence on the energy budget of the planet \citep{ormel2015,2021AAMoldenhauer,2022AAMoldenhauer}. Especially at low planet masses and small orbital distances, protoplanetary envelopes are  not closed  hydrostatic 1D systems, but dynamically exchange gas with the surrounding disk. The resulting advection of high entropy material can delay gas accretion \citep{Cimermankuiper2017}. The basic picture that the KH-contraction of the inner bound region ultimately regulates growth remains however valid, but the outer layers need to take into account the multi-dimensional hydrodynamical effects \citep{Bailey2023}. Regarding the composition, further high-resolution investigations are required to assess to what extent this effect could potentially reset the atmospheric composition of a migrated gas-rich planet to a more local composition. 

\begin{figure}[h]
	\centering
	\includegraphics[width=1\linewidth]{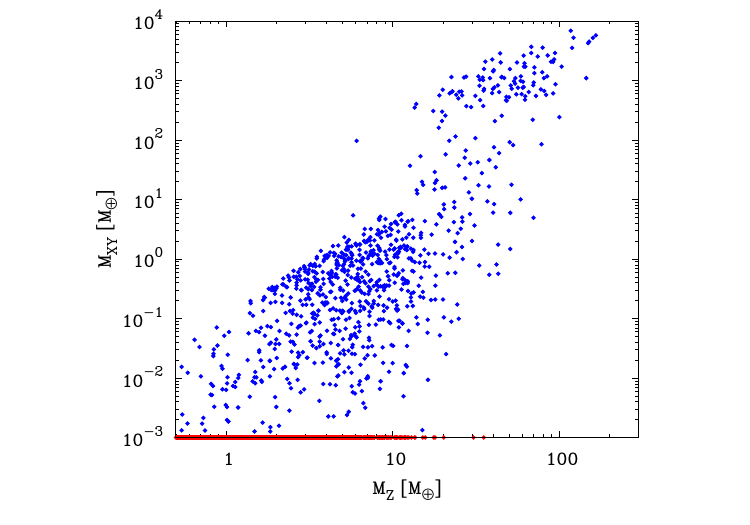}
	\caption{Illustration of the effect of envelope accretion (and loss). The blue points show the mass of hydrogen and helium $M_{\rm XY}$ as a function of the mass of heavy elements $M_{\rm Z}$ (iron, silicate, potentially water) of synthetic planets predicted by the global planet formation model of \citet{Burn2024}. This model also includes the processes of envelope loss via impact stripping and hydrodynamic atmospheric escape driven by high-energy photons. Red points are planets without H/He. The age of the planets is 5 Gyr.}
	\label{fig:mzmhhe}
\end{figure}

Figure \ref{fig:mzmhhe} shows the mass of H/He as a function of the heavy element (core) mass predicted by the Generation III Bern global planet formation model \citep{2021AAEmsenhuberA} that is based on the core accretion paradigm. The initial conditions of the model were varied according to observed properties of protoplanetary disks \citep{2021AAEmsenhuberB} to synthesise a population of model planets. The host star mass is 1 $M_\odot$ and the age of the synthetic planets is 5 Gyr. 

The envelope masses were derived by solving the internal structure equations in 1D. The opacity caused by grains in the protoplanetary atmospheres was assumed to be a factor 0.003 reduced \citep{2014AAMordasiniA} relative to ISM grain opacities \citep{1994ApJBell}, meaning that the KH-contraction of the envelopes is relatively efficient. This has the consequence that even relatively low-mass planets can accrete some H/He which is then reflected in the planetary mass-radius relation \citep{2012A&AMordasiniB}. The disk-limited gas accretion rate was found in a way similar to  Eq. \ref{eq:mdotbondi}. 

At low masses, the envelope mass increases with increasing heavy element mass. This is expected from the general scaling of the KH-timescale with mass (Eq. \ref{eq:tKH}). There is, however, a large spread. This is on one hand caused by different formation pathways which result in different efficiencies of gas accretion during the nebular phase. This diversity can  not be fully captured by simpler semi-analytical expressions for gas accretion \citep{2019AAAlibertVenturini}. On the other hand, additional effects modify (reduce) the envelope mass post-formation: during the long-term evolutionary phase, i.e., from the end of the lifetime of the disk to 5 Gyr, the planets can lose their H/He via impact stripping \citep{2020MNRASDenman} or atmospheric escape \citep[e.g.,][]{2014ApJJin,Owen2019AREPS,Affolter2023}. Planets without H/He are shown by red points.

Then, at higher heavy element / core masses of about 10-20 $M_\oplus$, runaway  gas accretion sets in. In the plot, this shifts the planets almost vertically upwards in the figure to high H/He masses, whereby they become gas giants. 

At intermediate masses of about 20-200 $M_\oplus$, there are fewer planets, known as the ``planetary desert'' \citep{2004ApJIda1}. It is caused by the relatively high gas accretion rates in the runaway and initial disk-limited phases of about a few $10^{-4}$ to $10^{-3}$ $M_\oplus$/yr \citep{2017AAMordasini}. This makes it less likely that the gas disk dissipates at just the moment of intermediate masses. Whether this ``smoking gun'' of core accretion is actually visible in the observed planetary mass function is a subject of debate. While some observational studies support the existence of the desert \citep{2011MayorArxiv,Bertaux2022}, others do not \citep{2018ApJSuzuki,Bennett2021}.

\subsection{Orbital migration}
\label{sec:orb_migration}
The mentioned orbital migration of planets within the gaseous disk is likely affecting the overall population of exoplanets which leave observable imprints on the planetary population level \citep{Burn2024}. Thus, we briefly review the process. More detailed reviews were published in \citet{2016SSRvBaruteau,Paardekooper2023}.

The physical reason for orbital migration lies in the perturbation of the disk gas in the vicinity of the planet which in turn exerts a torque on the planet. The leading order contribution is due to the mass accumulating at the Lindblad resonance, which is both trailing behind the planet exterior to the planetary orbit as well as leading in front of the planet just interior to the planetary orbit. Due to the geometry of the problem, the outer over-density typically exerts the larger torque on the planet and therefore decelerates it. This leads to a radial motion towards the star \citep{1979ApJGoldreichTremaine,1997IcarusWard,2002ApJTanaka}.
In detailed studies, the process is usually discussed with reference to a contribution from the Lindblad torque
\begin{equation}
\Gamma_{0}  =\left(\frac{q}{h}\right)^{2}\Sigma_{\rm g} a_{\rm p}^{4}\Omega_{\rm K}^{2 }\,,
\end{equation}
where the Keplerian orbital frequency $\Omega_{\rm K}$ needs to be evaluated at the planet location $a_{\rm p}$, $q$ is the planet-to-star mass ratio and $h$ is the local aspect ratio in the disk $H/a_{\rm p}$. A torque $\Gamma_{\rm tot}$ exerted on the planet on a circular orbit will lead to a change in semi-major axis of
\begin{equation}
     \frac{d a}{dt}=2 a_{\rm p} \frac{\Gamma_{\rm tot}}{J}\,
\end{equation}
where $J=M\sqrt{G M_{\star} a_{\rm p}}$. In addition to the Lindblad torque, there are several other contributions to $\Gamma_{\rm tot}$, which can be categorized into four regimes based on planetary mass and disk viscosity.

\begin{figure}
    \centering
    \includegraphics[width=\linewidth]{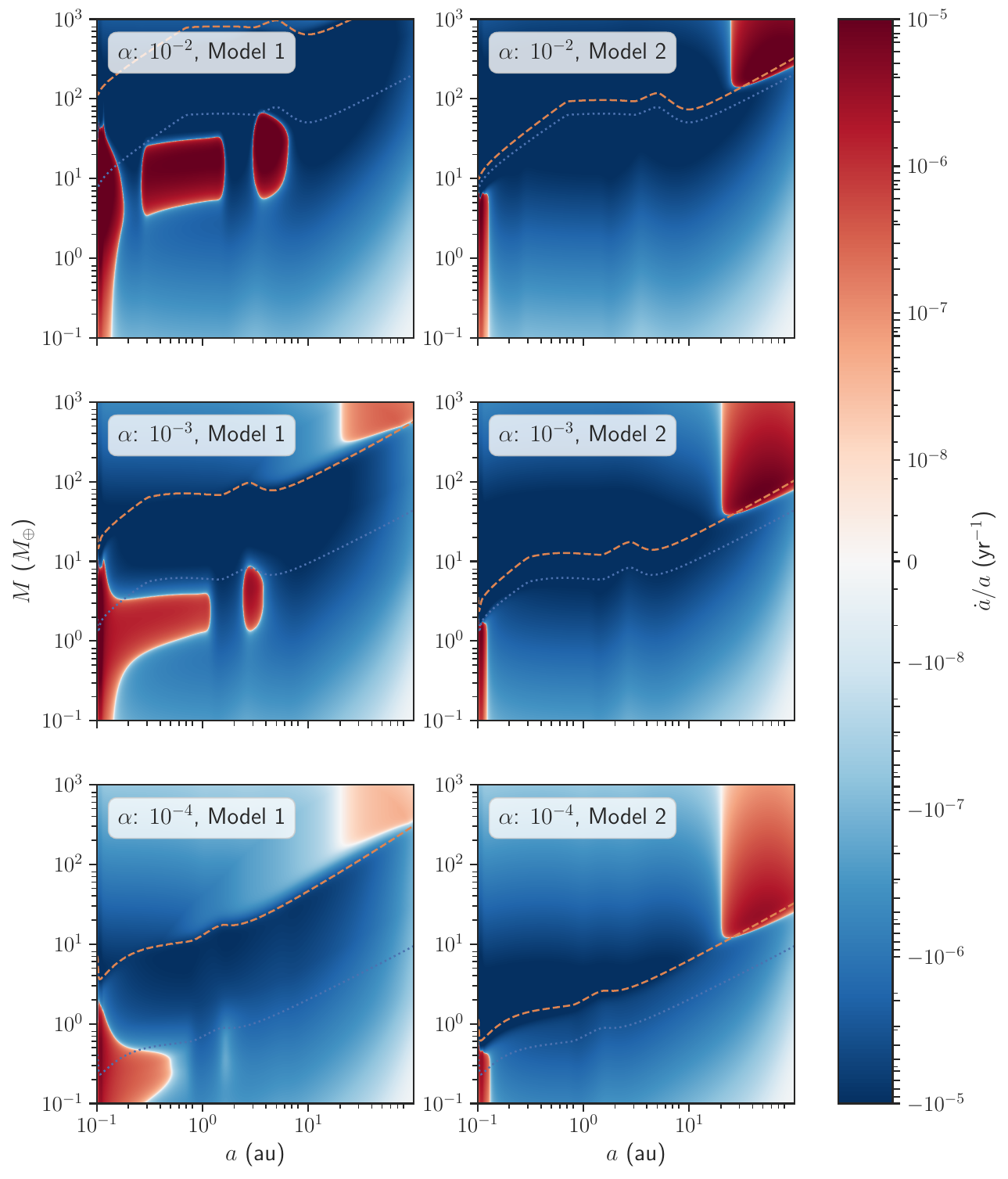}
    \caption{Migration rate of planetary embryos as a function of distance and planetary mass in a protoplanetary disk with different viscosities after 0.1\,Myr of evolution. The rate is normalized by the distance (thus showing the inverse timescale). Red regions indicate outward migration. $\alpha$ is used here for migration, disk evolution, and viscous heating. The rates for Model 1 (left column) follow \citet{2011MNRASPaardekooper} for viscous, low-mass planets (type I) and \citet{2014A&ADittkrist} for larger planets (type II). The transition between the regimes is marked with a dashed line and corresponds to the gap opening mass following Equations \ref{eq:gap_opening_crida} for Model 1 and the used Equation \ref{eq:gap_opening_kanag} for Model 2 which further uses \citet{2017MNRASJimenezMasset} type I torques. Saturation masses (Equation \ref{eq:msat}) are marked with a dotted line. They provide a useful estimate of the upper mass limit of outward migration regions for Model 1.}
    \label{fig:migration_map}
\end{figure}

\paragraph{In the high viscosity, low planetary mass regime,} the contributions are well-studied and analytical expressions are available. This regime is the classical viscous type I regime \citep{1997IcarusWard,2002ApJTanaka}. In addition to the Lindblad torque, the co-rotation torque can contribute significantly and even invert the typically inwards direction. The co-rotation torque originates from material orbiting the star in close vicinity to the planet. It is sensitive to the gradient in thermal properties over the region where parcels of gas describe a U-turn motion in proximity of the planet \citep{2010MNRASPaardekooper,2011MNRASPaardekooper,2012ARAAKleyNelson,2017MNRASJimenezMasset}, that is, over a region of width
\begin{equation}
r_\mathrm{co} \approx  \frac{a_{\rm p}}{\gamma^{1/4}} \sqrt{\frac{M a_{\rm p}}{M_\star h}}\,,
\end{equation}
where $\gamma$ is the adiabatic index. For typical disk properties, the co-rotation torque alone would lead to an outward motion of the planet. However, it can be weaker than the Lindblad torque and it further saturates at a certain planetary mass. This is the case when gradients in the gas properties are not re-established by material flowing into the relevant region due to viscous diffusion. The timescale of viscous diffusion over the corotation region is $\tau_\nu = r_{\mathrm co}^2 /\nu$. By comparing this timescale to the time it takes the gas to complete a full libration \citep{2012MNRASHellaryNelson} 
\beq
t_\mathrm{lib}=\frac{8\pi a_{\rm p}}{3\Omega_\mathrm{K} r_\mathrm{co}},
\eeq
and with inclusion of more precise factors of order unity, \citet{2023EPJPEmsenhuber} derived a useful critical planetary mass for saturation of the co-rotation torque
called the saturation mass for planetesimal accretion
\begin{equation}
     M_\mathrm{sat} = \left(\frac{8 \pi \alpha }{3}\right)^{2/3} \sqrt{\frac{\gamma}{C_{\rm HS}^4}} M_\star \left(\frac{H}{a_{\rm p}}\right)^{7/3}\,,
     \label{eq:msat}
\end{equation}
where $C_{\rm HS} \approx 1.1$ is based on numerical experiments \citep{2010MNRASPaardekooper}. The direct dependence on $\alpha$ is visible in the middle and top left panels of Figure \ref{fig:migration_map}, where saturation is reached at the upper mass end of the red regions caused by the corotation torque.

\paragraph{In the high viscosity, high planetary mass regime,} the torques on the planet decrease because the region around the planet is emptied and a gap forms. Formulas for gap emergence and width were obtained by \citet{2006IcarusCrida} and \citet{2017PASJKanagawa}.  In Figure \ref{fig:migration_map}, the transition is smoothed \citep[following][]{2021AAEmsenhuberA} and a gap is assumed to have emerged when \citep{2006IcarusCrida}
\begin{equation}
\label{eq:gap_opening_crida}
\frac{3 H_g}{4 R_H} + \frac{50 \nu M_\star}{M a_{\rm p}^2 \Omega_\mathrm{K}} \leq 1\,.
\end{equation}
A more simple criterion was found by \citet{2018ApJKanagawa}. In their work, they estimate a critical gap-opening mass of
\begin{equation}
    \label{eq:gap_opening_kanag}
    M_{\rm gap} = 8\times 10^{-5} M_{\star} \sqrt{\frac{\alpha}{10^{-3}}} \left( \frac{H_{\rm g}/a_{\rm p}}{0.05}\right)^{2.5}\,.
\end{equation}

In that case, migration is found to be suppressed. \citet{2015AADurmannKley} and \citet{2018ApJKanagawa} suggest to use the type I torque (including the co-rotation torque) and reduce it linearly with the reduction of the surface density at the bottom of the gap. Prior works suggested a more pronounced transition of regimes where the planet migration would be linked to the radial velocity of the gas \citep{2014A&ADittkrist} but with a reduction once the planet mass dominates over the local disk mass \citep{2009ApJAlexanderArmitage}. This latter approach is used in Figure \ref{fig:migration_map}, which causes an outward directed type II migration at the locations where the gas flow is also directed outwards.

\paragraph{For low viscosity disks ($\alpha \lesssim 10^{-4}$),} the field is currently determining accurate migration rates \citep{McNally2019,Lega2022}. Of large importance could be the dynamical co-rotation torque \citep{Paardekooper2014,Pierens2015}. It emerges in the same fashion as the co-rotation torque in a viscous disk but for cases where the viscous co-rotation torque would saturate. When a planet grows to significant mass in such a disk, due to the lack of viscous diffusion, gas is trapped in the horseshoe region and migrates together with the planet, effectively slowing down its Lindblad-torque-driven migration. To quantify it, tracking the history of the planet is required.

At larger planetary mass, there is s till research to be done to find a conclusive migration rate. Vortices and magnetic field lines penetrating into the planetary gap \citep{Aoyama2023,Wafflard-Fernandez2023} make this regime particularly challenging to explore \citep{Paardekooper2023}.

Finally, under certain circumstances and not limited to low-viscosity disks, further processes can exert a considerable torque, such as the torque from planetesimals \citep{levisonthommes2010,ormelida2012} or dust \citep{Benitez-Llambay2018} in the vicinity of the planet as well as heating effects \citep{Lega2014,benitez-llambaymasset2015,Masset2017}.

\section{Putting the pieces together: global models and planetary population synthesis}
\label{sec:results_phys}
In this section, we aim to showcase how global models of planet formation that combine several of the physical processes discussed in the last section bridge the gap from protoplanetary disks to the final planetary systems. As an example, we will focus on the results obtained with the Bern model of planet formation \citep{2005A&AAlibert,2012A&AMordasiniB,2013A&AAlibert,2021AAEmsenhuberA}. Other global models were developed by several authors \citep{2004ApJIda1,2014MNRASColemanNelson,2015AABitschB,2019AALambrechts,2018ApJChambers,2022MNRASAlessiPudritz,KimuraIkoma2022NatAs} and are  reviewed in  detail in \citet{2018BookMordasini,2022PPVIIDrazkowska}.

\begin{figure*}[htb]
    \centering
 \begin{minipage}{0.6\textwidth}
 \centering

    \includegraphics[width=\linewidth]{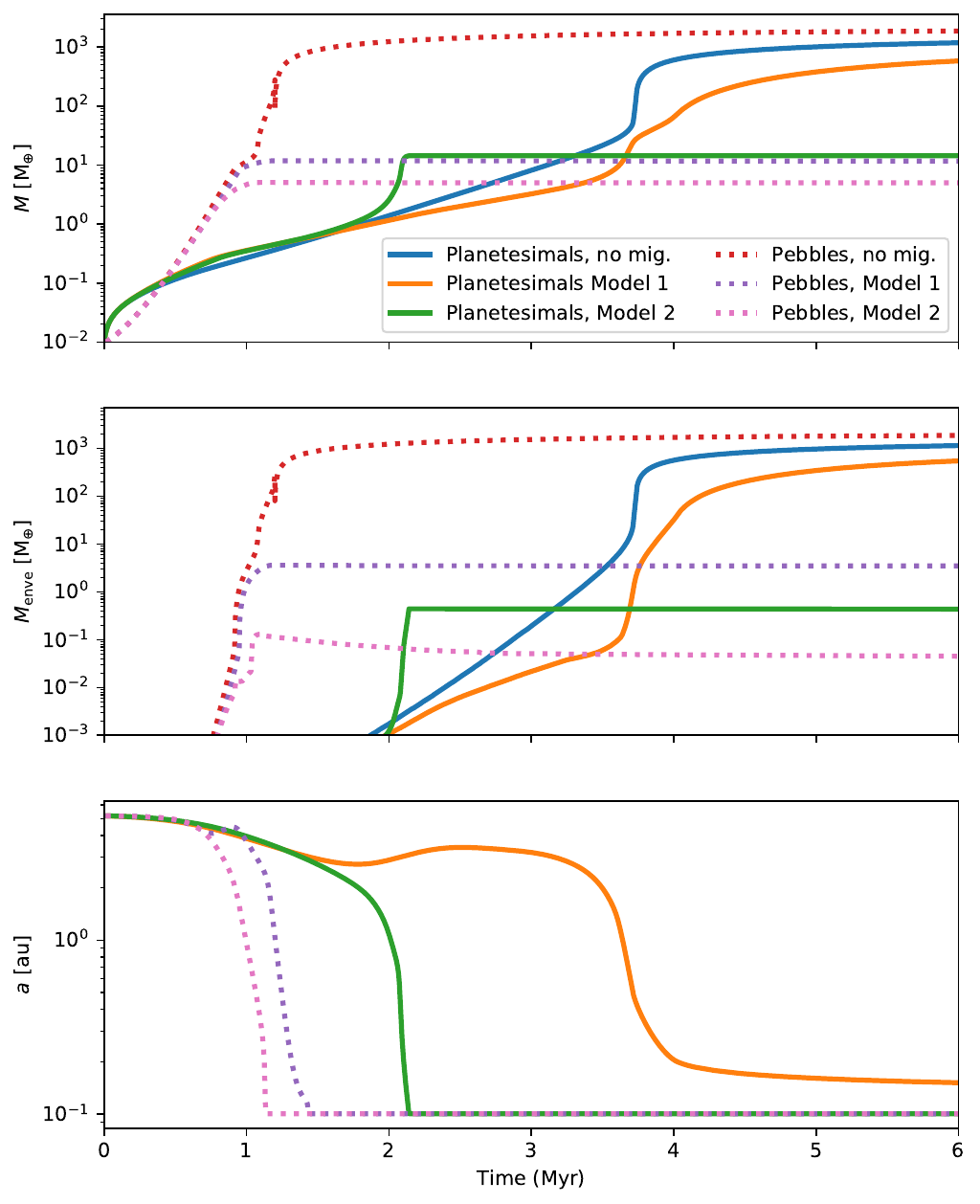}
      \end{minipage}
 \hfill  
     \begin{minipage}{0.35\textwidth}  
    \caption{Time evolution of the formation of planets for different models. A protoplanetary embryo with a mass of 0.01\,M$_{\oplus}$ is injected into the same gas disk and its migration is calculated following Model 1 or Model 2 (Section \ref{sec:orb_migration}). Accretion of solids follows the presented pebble- or planetesimal-accretion recipes.}
    \label{fig:tracks_t_M}
    \end{minipage}
\end{figure*}

\subsection{Coupled planet formation models}
To highlight the interplay of different processes, we show in Figure \ref{fig:tracks_t_M} the time evolution of a planet modelled using different assumptions. The protoplanetary disk evolution is calculated and follows a gas disk with identical parameters but twice the mass of the one shown in Section \ref{sec:orb_migration} (initial mass of 0.043\,M$_{\odot}$). The initial solid to gas mass ratio is 0.05 (715\,M$_\oplus$), which is a massive and solid-rich, but not unrealistic case compared to the population of observed disks \citep{2020ApJTobinA}. The solids are either distributed as pebbles or as already formed planetesimals with 100\,km diameter. The protoplanet is injected at 5.2\,au and grows by solid accretion of planetesimals in the oligarchic regime \citep{2013A&AFortier} or pebbles \citep{2017ASSLOrmel} and consistent gas accretion given the solid luminosity and later limited by the gas disk supply \citet{2013ApJBodenheimer}. Its migration follows \citet{2017MNRASJimenezMasset} and \citet{2018ApJKanagawa} (Model 2) or \citet{2011MNRASPaardekooper} and \citet{2014A&ADittkrist} (Model 1).

From the results, we can first see that migration often limits growth to larger masses. To form a giant planet, the critical timescale to overcome is typically the Type I migration timescale. This is particularly true if no outward migration region forms. Such an outward-migration zone keeps the planet growing with planetesimal accretion and Model 1 migration at a larger distance until it reaches several Earth masses after 3.5\,Myr. It then proceeds to become a hot Jupiter aided by accretion of planetesimals as the planet migrates, a scenario which is, however, not likely when multiple embryos would be considered since larger bodies are expected to emerge on shorter timescales in the inner system \citep{2021AAEmsenhuberA}.

If migration is turned off, cold giant planets can form under a wide range of parameters. Although this case is chosen for illustrative purposes here, it might be a realistic scenario if the disk is structured in a  configuration resulting in migration traps even before the first planets emerge \citep{Matsumura2007,2010ApJLyra,Hasegawa2013,Coleman2016}.

The rest of the model runs with migration produce hot sub-Neptunes which is a typical outcome of planet formation models and might explain the large number of observed sub-Neptunes and potentially super Earths.

Solid accretion in the pebble scenario is proceeding faster than planetesimal accretion at the chosen initial distance of 5.2\,au. At larger distances, the differences grow even larger with pebble accretion as a likely scenario for giant planet emergence from regions exterior to 10\,au while planetesimal accretion becomes inefficient at these separations, unless the planetesimal are km-sized. These differences were also highlighted by \citet{2020AABrugger,2020AAVoelkel}. Here, we note that the high efficiency of pebble accretion can also be detrimental to planets growing to become giants because there is a large disk mass at early times which leads to shorter migration timescales. After pebble accretion terminates at the pebble isolation mass, gas accretion proceeds at these early stages more slowly than type I migration. However, this depends on the chosen viscosity ($\alpha = 10^{-3}$ here).

We note that the timing of the emergence of an embryo at the chosen location is in principle not unconstrained. Here we chose time zero, but in reality it would take considerable time to assemble an object with a mass of 0.01\,M$_{\oplus}$. \citet{2021AAVoelkel,2022AAVoelkel} create an analytic model to realistically inject these objects.

\subsection{Planetary population synthesis}
To gain a more general understanding of planet formation, the example given above is far from sufficient. As hinted, the chosen disk conditions are tuned towards giant planet formation and the chosen starting location was further aiding in promoting growth.

To address the influence of these initial conditions, it is required to sample from realistic distributions and run the global planet formation model many times. This is exactly the idea of planetary population synthesis. Disk initial conditions are nowadays observed for large statistical samples. This allows for drawing initial conditions as Monte Carlo variables from the distribution of observed disks. The key distributions are disk gas mass, disk size, disk solid content, and lifetime \citep[see e.g.][for reviews]{2014PPVIBenz,2018BookMordasini,2023EPJPEmsenhuber}.

The population synthesis approach was recently used to generate a set of new populations yielding many quantities that can be directly compared with observations \citep{2021AAEmsenhuberB,2021AASchleckerA}. The underlying Generation III Bern Model uses the aforementioned Model 1 for migration and planetesimal accretion as reviewed in \citet{2021AAEmsenhuberA}. An important change was to use the distribution of younger Class I objects to start the simulations \citep{2018ApJSTychoniec}. The initial conditions were further tested against observations of more evolved disks \citep{Emsenhuber2023}.

\begin{figure}
    \centering
    \includegraphics[width=\linewidth]{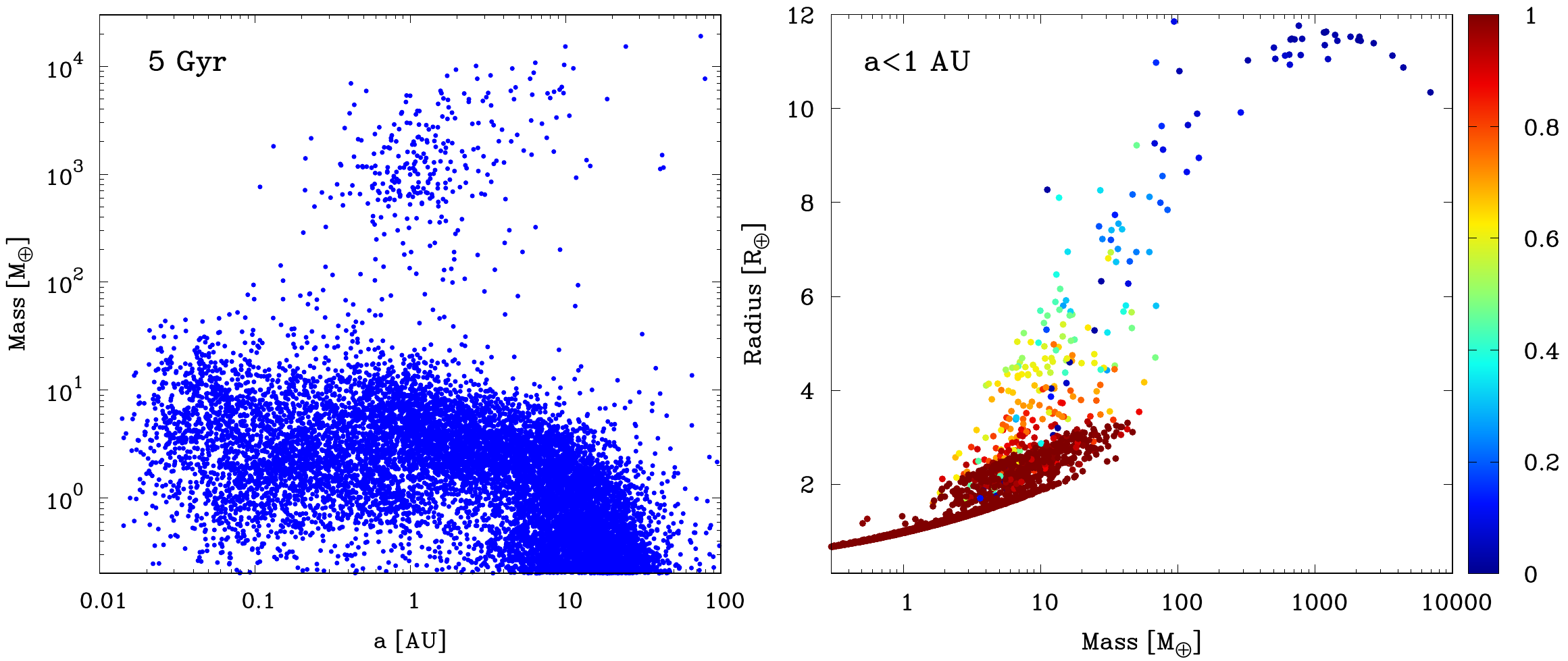}
    \caption{Mass-distance diagram (left) and mass-radius relation in the nominal planetary population of 1 $M_\odot$ stars obtained from planetary population synthesis with the Generation III Bern Model \citep{2021AAEmsenhuberA,Burn2024}. The age is 5 Gyr. The population consists of 1000 planetary system which each is initially seeded with 100 lunar-mass planetary embryos. On the right, only planets with $a<1$ AU are included as a zero-order representation of the observational bias of the transit method. The color code gives the envelope metallicity $Z$, where zero would correspond to a pure H-He envelope.  }
    \label{fig:NG76_evo27}
\end{figure}

The resulting population of planets shares a number of important features with the observed exoplanetary population \citep{2021AAEmsenhuberB}. As an illustration, Figure \ref{fig:NG76_evo27} shows the synthetic mass-distance diagram and the mass-radius diagram at an age of the population of 5 Gyr (compare with the observed versions in Figures \ref{fig:amobserved} and Fig. \ref{fig:exoplanetconstraints}). One can identify a number of features that are also seen in the observed population, like the increase of the frequency of giant planets outside of about 1 AU and the large population of close-in lower mass planets. Furthermore, thanks to including N-body interactions, the regime of low-mass planets down to Earth mass can be realistically explored where the radius valley \citep{Burn2024} and the peas-in-a pod trend \citep{2018AJWeiss} are reproduced \citep{2021AAMishra}. In particular, collisions resulting from N-body interactions erase to a large degree the imprint of the solid accretion timescale and limits, discussed in Sections \ref{sec:solid_acc} and \ref{sect:predictionsclassiccaltheorypardigmshifts}, on the final population as the inner system. They consist early in the history of  system of several low-mass planets that gets dynamically excited and polluted by migrated planets. An extension toward lower stellar masses \citep{2021AABurn} was also tested against available radial velocity surveys \citep{2022AASchlecker}, where important observed feature can be reproduced.

However, some rare massive planets around low-mass stars \citep[e.g.][]{2019ScienceMorales} are not explained by this population synthesis model. Furthermore, the observed trend of increasing planet occurrence with decreasing stellar mass from transit surveys \citep{2015ApJMuldersB} is not reproduced at the same quantitative level \citep{2021AABurn}. Further challenges remain the eccentricity distribution of giant planets \citep[potentially requiring a revision of the planet-disk interplay,][]{2020AABitsch} and the production of hot Jupiters with the same model that can also create cold Jupiters. This is the case because, typically, a change in migration and maximum gas accretion rate tilts the outcome to favor only one of the two categories. For a more comprehensive review of recent population synthesis calculations, we refer to \citet{2018BookMordasini,2023EPJPEmsenhuber}.

\section{The compositional opportunity}\label{sec:results_compo}

\subsection{Chemical and compositional clues to the origin and evolution of planets}
As planet formation occurs in a protoplanetary disk, the chemical inventory thereof is inherited as a starting point for a planet's composition. When trying to model the process (see \citealp{Oberg2023}, for a recent review), the initial conditions are key. However, the initial disk composition is still disputed with cold and low-density regions not allowing for chemical evolution to reach an equilibrium state. Instead, the composition in the outer disk is likely inherited from the star-forming region \citep{Eistrup2016}. More recently, the static picture, which was for simplicity assumed in the seminal work of \citet{2011ApJObergB}, was substantially revised by works that consider the time evolution of the protoplanetary disk composition \cite{Eistrup2018,2019MNRASBooth,Krijt2020}. 

 \begin{figure*}[htb!]
    \centering
    \includegraphics[width=0.8\linewidth]{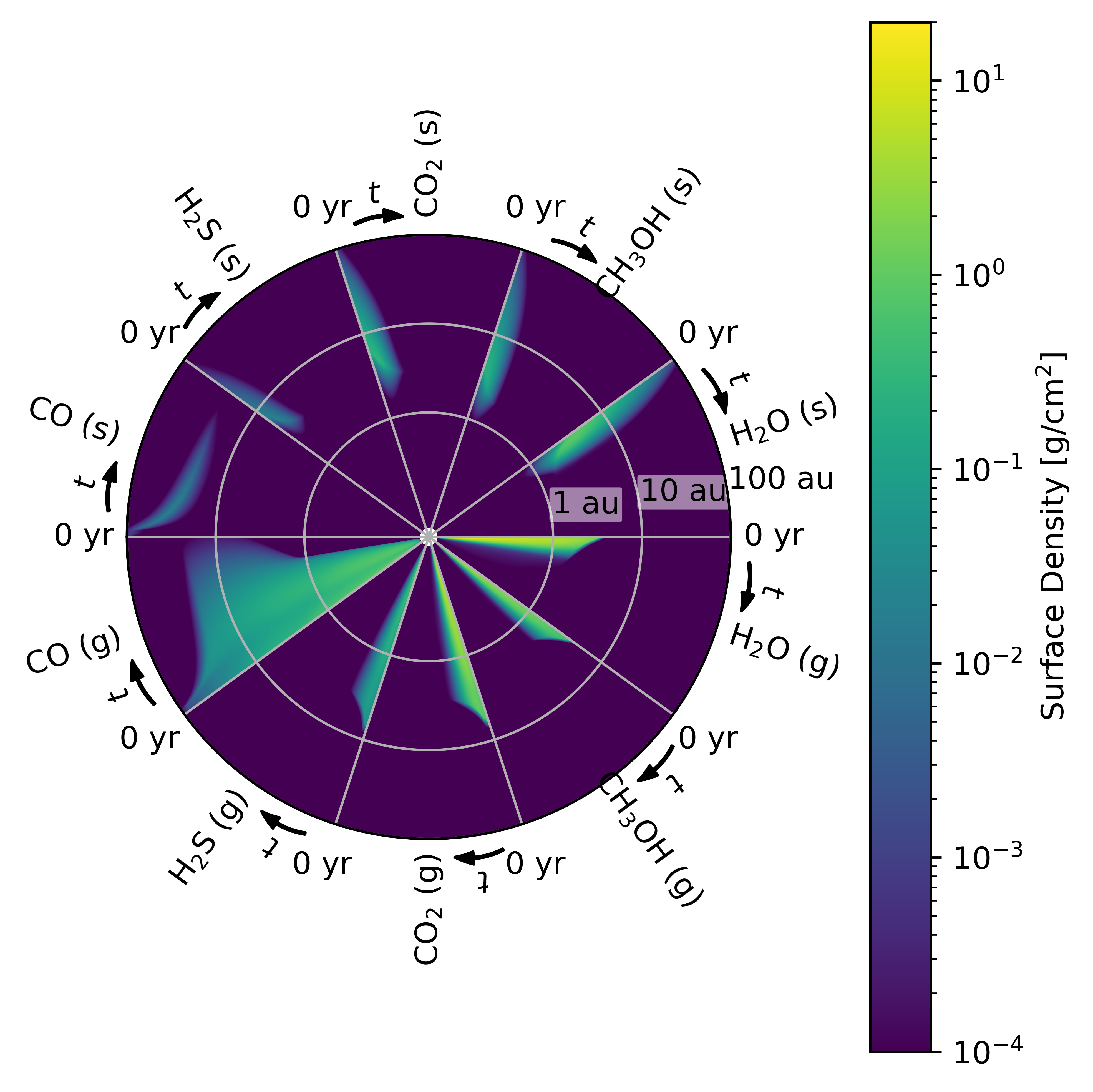}
    \caption{Evolution over 15\,Myr of the surface density of solid and gaseous species in a protoplanetary disk around a Solar-mass star. In each sector, the linear time evolution is shown in clockwise orientation and the surface density is color-coded. Distance to the star is shown in logarithmic spacing with an inner boundary at the disk edge at 0.04\,au. The upper half of the plot shows the volatile species in the solid phase where an inner boundary corresponds to the species iceline. The lower half shows the same species in the gaseous phase.}
    \label{fig:disk_compo_evolution}
\end{figure*}

To illustrate this, we present in Fig. \ref{fig:disk_compo_evolution} the time-evolution of the composition of a protoplanetary disk using the Bern model of planet formation \citep{2021AAEmsenhuberA} extended with solid evolution and entrainment in photoevaporative winds \citep{2020AAVoelkel,2022AABurnA} as well as compositional tracking of the pebbles and gas as discussed in Section \ref{sect:gasdisk}. The evaporation and recondensation of the volatile species use the saturation pressure curves collected by \citet{Fray2009}. For numerical consistency, a diffusion flux limiter is used and the process of evaporation is slowed down to decrease the abundance of water by a factor of $e$ over a gas scale height $H_{\rm g}$. A complete description of this model is currently in preparation (Burn et al.).

Figure \ref{fig:disk_compo_evolution} shows the time evolution of a standard disk with initial mass of 0.038\,M$_{\odot}$ and an exponential cut-off radius of 30\,au. Several stages of evolution take place, before the composition evolves significantly, dust grows to pebble mass, then pebbles drift and transport material to the inner system. At the same stage, the disk cools down and the snowlines move closer to the star. Pebble drift becomes inefficent once all dust has grown and drifted towards the star. As the supply of fresh material subsides, the inner disk gas becomes depleted of volatiles except CO, which has a snowline exterior of where the bulk of the disk mass resides. Finally, internal photoevaporation opens a gap (visible in the CO (g) sector in Figure \ref{fig:disk_compo_evolution}) and an outer ring gradually disperses. 

For CO, we see a ring of solids which is made of condensed CO ice. These rings exist as long as the pebble supply is present for all species but are not as prominent in the other cases. The exact feedback and amount of condensation depends on the viscosity of the gas but also on the numerical smoothing as icelines are in principle a 2D surface in radial-vertical extent instead of a sharp line. Re-condensation of gas is prominent for CO where the iceline lies in the outer decreting part of the disk which leads to a persistent condensation front until the end of the gas disk lifetime. Other species accrete on the star in a timeframe dominated by pebble drift which is efficient under reasonable assumptions of moderate turbulence ($\alpha$: $10^{-3}$) and sturdy grains ($v_{\rm frag}$: 2\,m/s).

It was shown by \citet{2019MNRASBooth} that chemical evolution plays a role over Myr timescales in the inner disk, which was not included in the simulations here. A simple reaction network was used in that work to keep the computation time short. More extensive chemical modeling is required to resolve the lower-density regions and for comparison to observations \citep{Woitke2009,Semenov2011,Walsh2015}. On the other hand, by excluding radial transport of ices, \citet{Cridland2016,Cridland2017} could use a more extended chemical network when modelling planet formation and obtaining planetary compositions. Another effect which was so far not coupled to other dynamical processes affecting the disk composition is that of radioactive heating leading to outgassing of water from 100\,km-sized planetesimals \citep{Lichtenberg2019}. These shortcomings show that there is still need of theoretical work to fully do justice to all relevant effect within one model.

With the advent of the James Webb Space Telescope (JWST), we have the opportunity to test models of protoplanetary disk composition using infrared spectroscopy. The technique can probe molecular lines in the warm ($T\gtrapprox 300$\,K) gas as well as the continuum emission which is typically dominated by the silicate grain features. First results \citep[see][for an overview]{vanDishoeck2023} highlight a surprising diversity with some water-rich \citep{Banzatti2023,Perotti2023}, over CO$_2$-rich \citep{Grant2023}, as well as a silicate- and water-poor, carbon-rich \citep{Tabone2023} inner disks. Several large JWST programs are being conducted which aim to resolve the compositional variability and potential correlations with stellar type, large-scale disk structure, or age. In particular, the synergy with constraints from ALMA that can probe species in the outer disk \citep{Oberg2021} might be fruitful. The ALMA data on its own are more challenging to link to planetary compositions as fewer planets form in the outer disk (at least in the conventional core accretion scenario).

\subsection{Disk carbon to oxygen content}
\begin{figure}[htb]
    \centering
    \includegraphics[width=0.7\linewidth]{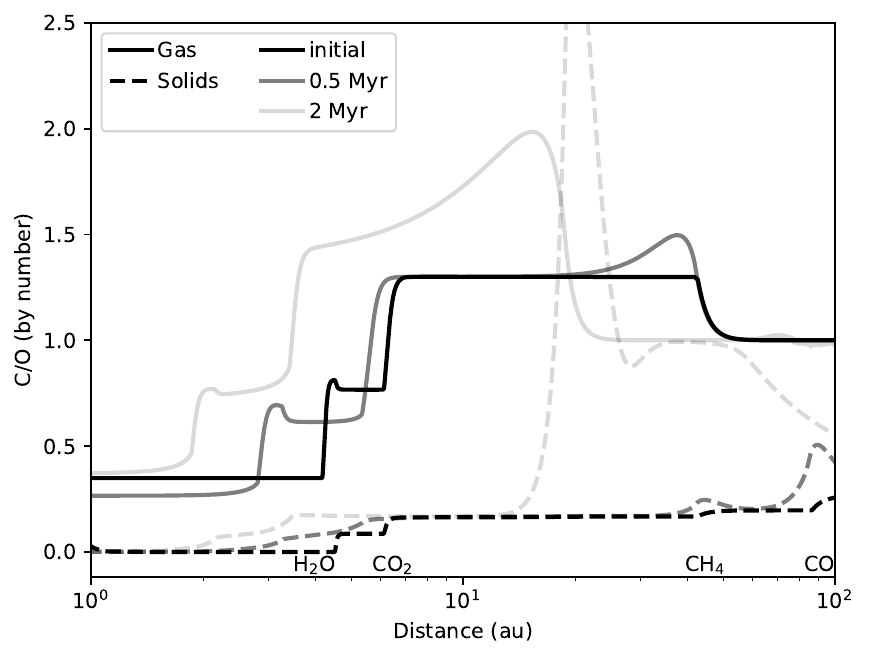}
    \caption{Carbon to oxygen number ratio as a function of distance (``Öberg plot'', following the work of \citealp{2011ApJObergB}) in a protoplanetary disk around a Solar mass star. In contrast to \citet{2011ApJObergB}, we used the initialization of \citet{2014AAMarboeufA} and did not add refractory carbon to the solids (C/O of refractories is zero), which is why the solid C/O drops to zero interior to the methanol and water icelines in the inner disk. The same disk evolution with the species shown in Figure \ref{fig:disk_compo_evolution} is used to calculate C/O ratios. The CH$_4$ and CO snowlines significantly alter the local composition at later times. In the inner disk, the short viscous timescale (for $\alpha = 10^{-3}$) leads to a quick equilibration instead of a local accumulation of material. We note that there is a minor impact by the methanol line which is not labelled.}
    \label{fig:disk_compo_c_over_o}
\end{figure}

Despite the challenging nature of the subject \citep{2022ApJMolliere}, the protoplanetary disk composition has been linked to the composition of gas giants \citep[e.g.,][]{Madhusudhan2014,2015A&AThiabaud,2016ApJMordasini,Schneider2021,Bitsch2022}. To illustrate the link, it is useful to consider here the pioneering work of \citet{2011ApJObergB} who discussed the carbon to oxygen ratio in the protoplanetary disk gas and solids. The C/O ratio changes with distance to the star at the respective snowlines. This is shown in Figure \ref{fig:disk_compo_c_over_o} for the same disk for which the evolution was discussed above. For example, exterior to the CO$_2$ snowline, the gas phase contains only CH$_4$ and CO as carbon- or oxygen-bearing species. This means moving interior to the CO$_2$ snowline, a jump towards a C/O of 0.5 (from CO$_2$) occurs. Even more oxygen is present interior to the water snowline, further reducing C/O.

 The basic idea is that once the C/O in the atmosphere of a giant planet can be measured, its formation location can be identified. The formation location is a key information for planet formation. Originally, it was thought that the disk gas phase is the determining factor (solid line in Fig. \ref{fig:disk_compo_c_over_o}). However, \citep{2016ApJMordasini} showed that for typical enrichment levels of Hot Jupiters, rather the accreted solids (the dashed line in Fig. \ref{fig:disk_compo_c_over_o}) determine the atmospheric C/O if atmosphere and interior mix. The reason is that nebular gas usually contains only few C and O molecules compared to the solids in which these elements are major components (for example oxygen in silicate rocks or C in carbon-bearing ices). As discussed in this section, the picture has become even more complex with effects of the disk's temporal evolution, pebble drift and evaporation, inhomogeneous planetary interiors and other factors.   

Figure \ref{fig:disk_compo_c_over_o} also shows the time evolution of the modelled protoplanetary disk. In this case, we see a change of the inner protoplanetary gas disk C/O to a quasi-static value of approximately 0.3. This value is sensitive to a number of assumptions on the molecular inventory of the disk, such as the water to CO and CO$_2$ ratio \citep{2014AAMarboeufA}. More recently, it was also speculated whether hydrocarbons could be present and evaporate at the soot line which would be located in the hot inner disk ($\sim$500\,K, \citealp{Li2021b}). This is one of the hypothesis to obtain an observable, carbon-rich composition \citep{Tabone2023}. Alternatively, \citet{Mah2023} proposed that the carbon-rich disk could be in a later evolutionary stage in the disk. The process proposed in that work is the same as what develops in Figure \ref{fig:disk_compo_c_over_o}: At later stages, all the material with high condensation temperatures, such as water, has accreted onto the star due to, first, radial drift of solids followed by viscous accretion of gas. Instead, what remains in the inner disk is carbon-rich material from the outer disk (2 Myr line). It is the species which exist primarily in the gas phase (CO and CH$_4$) which are not as rapidly depleted as the volatiles in the rapidly drifting solid phase. Qualitatively this scenario is a common outcome for disks with rapid pebble drift ($v_{\rm frag}\gtrapprox 1\,$m/s, \citealp{Mah2023}). The prediction of this hypothesis is that the disk is strongly depleted in overall solids. These novel results motivate more detailed studies of the microphysics at the snowlines and the their efficiency to trap and cycle volatiles from the gas phase back to the solid phase.

\subsection{Bulk and atmospheric composition}
\begin{figure*}[h]
	\centering
\includegraphics[width=1\linewidth]{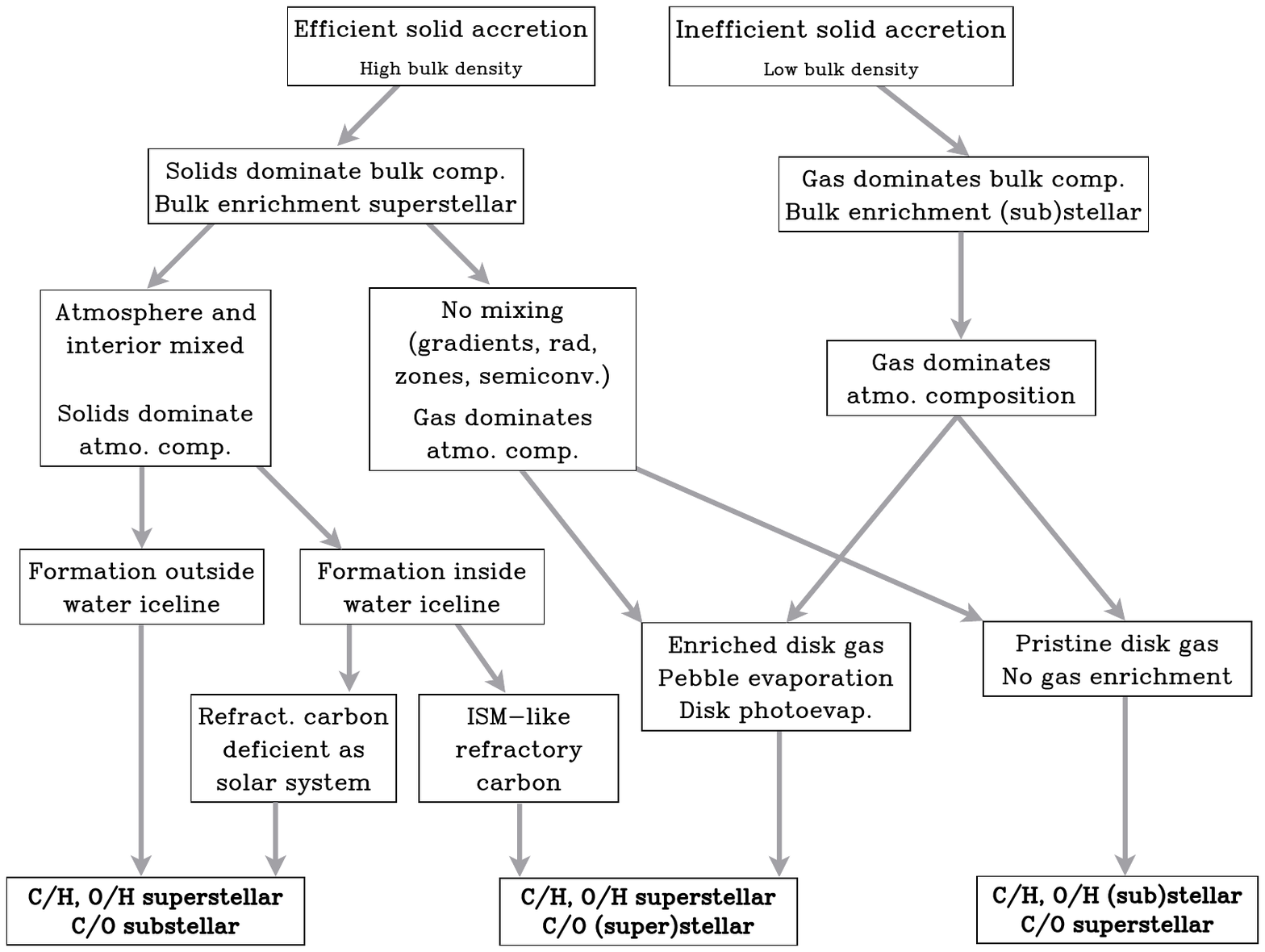}
	\caption{Formation pathways and physical processes leading to different atmospheric composition in terms of C/H, O/H, and C/O for Hot Jupiters. These can be probed with atmospheric spectroscopy. The bulk enrichment is also noted as another observational constraint accessible via the planet's mean density. Updated with additional physical processes based on  \citet{2016ApJMordasini}.}
\label{fig:atmocompohotjupiters}
\end{figure*}

While the disk composition is key to determine the bulk composition of planets, the partitioning of elements to different reservoirs in the planet as well as the chemical form in which they present themselves to observers is not trivial. Figure \ref{fig:atmocompohotjupiters} shows various potential pathways to different atmospheric compositions for hot Jupiters in terms of their C/H, O/H, and C/O, which are most easily observed. 

The plot is based on \citet{2016ApJMordasini} but updated to include several new aspects like pebble evaporation \citep{Booth2017MNRAS}. In addition to the disk-related aspects (formation location with respect to icelines), the efficiency of solid accretion and the mixing of species within the atmosphere are key aspects. In particular, if mixing is not efficient (which seems to be the case at least for our Jupiter, see Sect. \ref{sec:obs_constraints}), the observable composition is sensitive to the latest formation stage where the planet has already obtained its bulk mass but might not stop solid accretion completely \citep{Shibata2023}.

Going from left to right and from top to bottom in Fig. \ref{fig:atmocompohotjupiters}, we first consider that solid accretion is efficient. The Hot Jupiter will then have a comparatively high mean density and solids will determine the bulk  enrichment which will be at a superstellar level. If interior and atmosphere are well mixed, the composition of the accreted solids will also dominate the atmospheric enrichment. Accreted solids will be the prime source of heavy elements in the planetary envelope and atmosphere, dominating over the heavy elements accreted together with the H/He gas. If the planet formed outside of the water iceline or the CO$_2$ line, the accreted solids will contain oxygen (and carbon), but the composition of icy solids containing water ice is overall oxygen dominated. This can be seen with the dashed line in Fig. \ref{fig:disk_compo_evolution} with C/O values of about 0.1 to 0.3. This finally results in an atmosphere with a superstellar O/H, C/H and a low substellar C/O.

Next we consider still the case that the bulk and atmospheric enrichment is dominated by the accreted solids, but for an accretion of the solids inside of the water iceline. Here it depends on the composition of the refractory solids that are accreted. If these solids are carbon depleted relative to the ISM as it is the case in the inner Solar System \citep{Bergin2015PNAS} and as might be generally the case as suggested by most polluted white dwarfs \citep{XuBonsor2021Eleme}, then again a superstellar C/H, O/H and substellar C/O will result, because of the oxygen in the accreted silicates, at least if the silicates do not condense out again in the deeper layers \citep{Kitzmann2024MNRAS}. On the other hand, if the solids inside of the iceline contain large quantities of refractory carbon similar to the ISM, a planet with a superstellar C/H, O/H and a stellar or superstellar C/O will result.  

The right part of Figure \ref{fig:atmocompohotjupiters} discusses the outcome if solid accretion was comparatively inefficient and thus the atmospheric C and O is dominated by the one of the accreted gas. This applies also to the case that atmosphere and interior are poorly mixed because of compositional gradients and inefficient semi-convection \citep{2013MNRASVazan}, even if solid accretion was efficient overall. This case further branches into sub-scenarios depending on whether the gas has a pristine background composition or whether effects like disk photoevaporation \citep{GuillotHueso2006} and pebble drift and evaporation \citep{Booth2017MNRAS} enrich the nebular gas at the place of the planet. In the former case, a planetary atmosphere with a (sub)stellar O/H and C/H and a superstellar C/O would result, while the latter would lead to enrichment levels of O/H, C/H and a C/O ratio which are (super)stellar. Despite being strongly simplified, these different pathways already give a hint of the complexity of linking formation and planetary atmospheres. To address this, other elements like N, S or refractories can be used to add further complementary constraints \citep{Polman2023A&A,Crossfield2023ApJ,OhnoFortney2023}. 

Additionally, for lower planetary mass, the solid core of a planet can be a substantial reservoir for volatile elements. The core can then control the atmospheric composition over long timescales. But whenever a primordial atmosphere of significant mass is present, both reservoirs need to be included and modelled consistently \citep{Bower2019,Bower2022,Lichtenberg2023}. The self-consistent coupling between outer atmosphere and the deeper envelope (which can, for example, be water-dominated) is also important \citep{Guzman2022MNRAS} and allows to combine spectroscopic constraints on the atmospheric composition with observational constraints on the planet's mean density and thus its bulk composition. Sub-Neptunes fall in the regime where large quantities of  volatile elements different from hydrogen and helium might make-up most of the envelope of the planet \citep{Burn2024} which could even mix with rocky material for hot planets in a supercritical phase \citep{Vazan2022}. More interdisciplinary research connecting geophysics with planet formation and astronomical constraints is required to determine the exact partitioning of elements and the interior evolution \citep[e.g.,][]{kite2019}.

\section{Summary and conclusions}\label{sec:conclusion}
In this chapter, we have reviewed the theory of planet formation. In the first part, key observational constraints  from the Solar System and from the exoplanet population were discussed. Then, some of the physical mechanisms governing planetary system formation were introduced. We then addressed how these mechanisms can be put together to form global planet formation models that can be used in planetary population synthesis. Population synthesis makes quantitative statistical comparisons possible between formation theory and exoplanet observations. Finally, we showed how the compositional links from protoplanetary disks to planetary atmospheres put novel constraints on planet formation theory.  

Triggered mainly by observational progress, the understanding how planets form has evolved from a static picture inspired solely by the Solar System to a highly dynamic one which is reflected in the diversity of the  population of extrasolar planets. This evolution is accompanied by a number of ongoing shifts of paradigms: first, from in situ formation to mobility of the building blocks both on the dust/pebble scale (drift) and on the scale of protoplanets (orbital migration). Second, from solid accretion only via planetesimals to a multifaceted process were pebbles, planetesimals, and giant impacts play a role. Third, from a formation in smooth viscous disks to structured and potentially MHD wind-driven disks. At the moment, it is not yet possible to say to what extent these new paradigms will replace the older ones or if in the end a synthesis of older and newer views will result. 

In order to make progress, it will be important to address a number of key questions and frontiers:
\begin{enumerate}
\item Are the structures in protoplanetary disks cause or consequence of planet formation? What is the formation mechanism of the protoplanets that seem to emerge at large orbital distances already at early times as suggested by observed disk structures?
\item Is the evolution of protoplanetary disks driven by viscous accretion or by MHD-winds and what are the consequences for planet formation, like potentially reduced orbital migration?
\item What is the importance of pebbles, planetesimals, and giant impacts as mechanisms of solid growth? Is there a distance dependency? How does this influence the planetary composition and structure?
\item How large are planetary gas accretion rates in the runaway and disk-limited phase and which mechanisms control it? Is there a planetary desert in the mass function? 
\item How did the Solar System form, capitalizing on the special constraints we only have in our system like the meteoritic record and the minor bodies? Was there a grand tack? How much material passed Jupiter? How special is its architecture?
\item How can the complexity of the interiors of the giant planets in our Solar System (especially Jupiter) be translated to exoplanets for which typically simple fully adiabatic interiors are assumed? 
\item What is the origin of the very abundant population of close-in low-mass/small exoplanets? Are some of these planets ocean worlds that have migrated from beyond the iceline to their current position? Are there hints from architectural patterns? 
\item What is the role of gravitational instability as a giant planet formation mechanism and/or potentially as a seed for core accretion?
\item What is the impact of the cluster/stellar environment on planet formation via late infall, external photoevaporation of disks, and stellar encounters?
\item What can be learned from observations of ongoing planet formation? Is it possible to derive much more direct constraints on processes like formation timescales, formation location, gas accretion, migration, and circumplanetary disks? Can we identify an evolutionary sequence like for stars?
\item Do planets keep an atmospheric memory about their origin? Can it be accessed remotely via spectroscopic observations? How are bulk and observable atmospheric composition linked?
\item What is the diversity of low-mass planets in the habitable zone in terms of water content, bulk and atmospheric composition, or remaining primordial H/He envelope? How does this depend on host star properties like stellar mass and on the architecture of the host planetary system? Which fraction is really Earth-like and habitable?
\end{enumerate} 

Fortunately, chances are high that significant advancements will be possible in the coming years for this dozen points because the number of observational constraints will continue to grow rapidly. 

The JWST is a first already active example. Despite its recent launch, it already leads to important insights and triggers the development of more precise models for the composition of the inner disk, including effects like a soot line or late carbon-rich disk gas. This showcases the potential of linking the outer reservoir of the disk to the inner disk composition and finally the planetary atmospheres, at least if they can be related to the bulk of the material accreted during formation.  

In the Solar System, space missions like JUICE, BepiColombo, or the Uranus mission but also ultra-precise laboratory measurements will help to better understand our own planetary system, while ALMA, ANDES, or RISTRETTO will probe the formation environment, the atmospheric composition, and ongoing planet formation from the ground. In space, GAIA, PLATO, and the Roman Space Telescope will critically extend the parameter space of the known exoplanet demographics in the high- and  low-mass regime to areas currently unknown. ARIEL will add via atmospheric spectroscopy statistical constraints on planetary composition for a larger sample of planets which is important given the complexity when linking formation and atmospheres. Finally, over longer timescales, the LIFE mission and the Habitable Worlds Observatory should be able to observationally address the last question listed above about the diversity of potentially habitable planets and maybe even life. 

Taken together, these observational efforts will yield very rich data to build a set of many multifaceted constraints for planetary formation theory. Theoretical results can be confronted to them, improving eventually our understanding of the origin of planets.
\\
\\
\small{Acknowledgements: We thank Willy Benz, Alexandre Emsenhuber, Benjamin Fulton, Erik Petigura, Roy van Boekel, and Jesse Weder for important input. C.M. acknowledges the support from the Swiss National Science Foundation under grant 200021\_204847 ``PlanetsInTime''. R.B. acknowledges financial support from the German Excellence Strategy via the Heidelberg Cluster of Excellence (EXC 2181 - 390900948) ``STRUCTURES'' under Exploratory Project 8.4. Parts of this work has been carried out within the framework of the NCCR PlanetS supported by the Swiss National Science Foundation under grants 51NF40\_182901 and 51NF40\_205606.}

\bibliographystyle{aasjournal}
\bibliography{manu,remo}

\begin{thebibliography}{}
\expandafter\ifx\csname natexlab\endcsname\relax\def\natexlab#1{#1}\fi
\providecommand{\url}[1]{\href{#1}{#1}}
\providecommand{\dodoi}[1]{doi:~\href{http://doi.org/#1}{\nolinkurl{#1}}}
\providecommand{\doeprint}[1]{\href{http://ascl.net/#1}{\nolinkurl{http://ascl.net/#1}}}
\providecommand{\doarXiv}[1]{\href{https://arxiv.org/abs/#1}{\nolinkurl{https://arxiv.org/abs/#1}}}

\bibitem[{Abod {et~al.}(2019)Abod, Simon, Li, Armitage, Youdin, \&
  Kretke}]{Abod2019}
Abod, C.~P., Simon, J.~B., Li, R., {et~al.} 2019, The Astrophysical Journal,
  883, 192, \dodoi{10.3847/1538-4357/ab40a3}

\bibitem[{{Adachi} {et~al.}(1976){Adachi}, {Hayashi}, \&
  {Nakazawa}}]{1976PThPhAdachi}
{Adachi}, I., {Hayashi}, C., \& {Nakazawa}, K. 1976, Progress of Theoretical
  Physics, 56, 1756, \dodoi{10.1143/PTP.56.1756}

\bibitem[{{Adams} \& {Batygin}(2022)}]{2022ApJAdamsBatygin}
{Adams}, F.~C., \& {Batygin}, K. 2022, \apj, 934, 111,
  \dodoi{10.3847/1538-4357/ac7a3e}

\bibitem[{{Adams} {et~al.}(2020){Adams}, {Batygin}, {Bloch}, \&
  {Laughlin}}]{Adams2020}
{Adams}, F.~C., {Batygin}, K., {Bloch}, A.~M., \& {Laughlin}, G. 2020, \mnras,
  493, 5520, \dodoi{10.1093/mnras/staa624}

\bibitem[{{Affolter} {et~al.}(2023){Affolter}, {Mordasini}, {Oza},
  {Kubyshkina}, \& {Fossati}}]{Affolter2023}
{Affolter}, L., {Mordasini}, C., {Oza}, A.~V., {Kubyshkina}, D., \& {Fossati},
  L. 2023, \aap, 676, A119, \dodoi{10.1051/0004-6361/202142205}

\bibitem[{{Alcal{\'a}} {et~al.}(2017){Alcal{\'a}}, {Manara}, {Natta}, {Frasca},
  {Testi}, {Nisini}, {Stelzer}, {Williams}, {Antoniucci}, {Biazzo}, {Covino},
  {Esposito}, {Getman}, \& {Rigliaco}}]{2017AAAlcala}
{Alcal{\'a}}, J.~M., {Manara}, C.~F., {Natta}, A., {et~al.} 2017, \aap, 600,
  A20, \dodoi{10.1051/0004-6361/201629929}

\bibitem[{{Alessi} \& {Pudritz}(2022)}]{2022MNRASAlessiPudritz}
{Alessi}, M., \& {Pudritz}, R.~E. 2022, \mnras, 515, 2548,
  \dodoi{10.1093/mnras/stac1782}

\bibitem[{{Alexander} \& {Armitage}(2009)}]{2009ApJAlexanderArmitage}
{Alexander}, R.~D., \& {Armitage}, P.~J. 2009, \apj, 704, 989,
  \dodoi{10.1088/0004-637X/704/2/989}

\bibitem[{{Ali-Dib} {et~al.}(2020){Ali-Dib}, {Cumming}, \&
  {Lin}}]{2020MNRASali-dib}
{Ali-Dib}, M., {Cumming}, A., \& {Lin}, D. N.~C. 2020, \mnras, 494, 2440,
  \dodoi{10.1093/mnras/staa914}

\bibitem[{{Alibert}(2019)}]{Alibert2019}
{Alibert}, Y. 2019, \aap, 624, A45, \dodoi{10.1051/0004-6361/201834592}

\bibitem[{{Alibert} {et~al.}(2013){Alibert}, {Carron}, {Fortier}, {Pfyffer},
  {Benz}, {Mordasini}, \& {Swoboda}}]{2013A&AAlibert}
{Alibert}, Y., {Carron}, F., {Fortier}, A., {et~al.} 2013, \aap, 558, A109,
  \dodoi{10.1051/0004-6361/201321690}

\bibitem[{{Alibert} {et~al.}(2005){Alibert}, {Mordasini}, {Benz}, \&
  {Winisdoerffer}}]{2005A&AAlibert}
{Alibert}, Y., {Mordasini}, C., {Benz}, W., \& {Winisdoerffer}, C. 2005, \aap,
  434, 343, \dodoi{10.1051/0004-6361:20042032}

\bibitem[{{Alibert} \& {Venturini}(2019)}]{2019AAAlibertVenturini}
{Alibert}, Y., \& {Venturini}, J. 2019, \aap, 626, A21,
  \dodoi{10.1051/0004-6361/201834942}

\bibitem[{{Alibert} {et~al.}(2018){Alibert}, {Venturini}, {Helled}, {Ataiee},
  {Burn}, {Senecal}, {Benz}, {Mayer}, {Mordasini}, {Quanz}, \&
  {Sch{\"o}nb{\"a}chler}}]{2018NatAsAlibert}
{Alibert}, Y., {Venturini}, J., {Helled}, R., {et~al.} 2018, \natas, 2, 873,
  \dodoi{10.1038/s41550-018-0557-2}

\bibitem[{{ALMA-Partnership} {et~al.}(2015){ALMA-Partnership}, Brogan,
  P{\'e}rez, Hunter, Dent, Hales, Hills, Corder, Fomalont, Vlahakis, Asaki,
  Barkats, Hirota, Hodge, Impellizzeri, Kneissl, Liuzzo, Lucas, Marcelino,
  Matsushita, Nakanishi, Phillips, Richards, Toledo, Aladro, Broguiere, Cortes,
  Cortes, Espada, Galarza, {Garcia-Appadoo}, {Guzman-Ramirez}, Humphreys, Jung,
  Kameno, Laing, Leon, Marconi, Mignano, Nikolic, Nyman, Radiszcz, Remijan,
  Rod{\'o}n, Sawada, Takahashi, Tilanus, Vila~Vilaro, Watson, Wiklind, Akiyama,
  Chapillon, {De Gregorio-Monsalvo}, Di~Francesco, Gueth, Kawamura, Lee,
  Nguyen~Luong, Mangum, Pietu, Sanhueza, Saigo, Takakuwa, Ubach, Van~Kempen,
  Wootten, {Castro-Carrizo}, Francke, Gallardo, Garcia, Gonzalez, Hill,
  Kaminski, Kurono, Liu, Lopez, Morales, Plarre, Schieven, Testi, Videla,
  Villard, Andreani, Hibbard, \& Tatematsu}]{ALMA-Partnership2015}
{ALMA-Partnership}, Brogan, C.~L., P{\'e}rez, L.~M., {et~al.} 2015,
  Astrophysical Journal Letters, 808, \dodoi{10.1088/2041-8205/808/1/L3}

\bibitem[{{Altwegg} {et~al.}(2015){Altwegg}, {Balsiger}, {Bar-Nun},
  {Berthelier}, {Bieler}, {Bochsler}, {Briois}, {Calmonte}, {Combi}, {De
  Keyser}, {Eberhardt}, {Fiethe}, {Fuselier}, {Gasc}, {Gombosi}, {Hansen},
  {H{\"a}ssig}, {J{\"a}ckel}, {Kopp}, {Korth}, {LeRoy}, {Mall}, {Marty},
  {Mousis}, {Neefs}, {Owen}, {R{\`e}me}, {Rubin}, {S{\'e}mon}, {Tzou}, {Waite},
  \& {Wurz}}]{Altwegg2015}
{Altwegg}, K., {Balsiger}, H., {Bar-Nun}, A., {et~al.} 2015, Science, 347,
  1261952, \dodoi{10.1126/science.1261952}

\bibitem[{{Anand} \& {Mezger}(2023)}]{AnandMezger2023}
{Anand}, A., \& {Mezger}, K. 2023, Chemie der Erde / Geochemistry, 83, 126004,
  \dodoi{10.1016/j.chemer.2023.126004}

\bibitem[{{Andrews}(2020)}]{2020ARAAAndrews}
{Andrews}, S.~M. 2020, \araa, 58, 483,
  \dodoi{10.1146/annurev-astro-031220-010302}

\bibitem[{{Andrews} {et~al.}(2018{\natexlab{a}}){Andrews}, {Terrell},
  {Tripathi}, {Ansdell}, {Williams}, \& {Wilner}}]{2018ApJAndrewsA}
{Andrews}, S.~M., {Terrell}, M., {Tripathi}, A., {et~al.} 2018{\natexlab{a}},
  \apj, 865, 157, \dodoi{10.3847/1538-4357/aadd9f}

\bibitem[{{Andrews} {et~al.}(2009){Andrews}, {Wilner}, {Hughes}, {Qi}, \&
  {Dullemond}}]{2009ApJAndrews}
{Andrews}, S.~M., {Wilner}, D.~J., {Hughes}, A.~M., {Qi}, C., \& {Dullemond},
  C.~P. 2009, \apj, 700, 1502, \dodoi{10.1088/0004-637X/700/2/1502}

\bibitem[{{Andrews} {et~al.}(2018{\natexlab{b}}){Andrews}, {Huang},
  {P{\'e}rez}, {Isella}, {Dullemond}, {Kurtovic}, {Guzm{\'a}n}, {Carpenter},
  {Wilner}, {Zhang}, {Zhu}, {Birnstiel}, {Bai}, {Benisty}, {Hughes},
  {{\"O}berg}, \& {Ricci}}]{2018ApJAndrewsB}
{Andrews}, S.~M., {Huang}, J., {P{\'e}rez}, L.~M., {et~al.} 2018{\natexlab{b}},
  \apjl, 869, L41, \dodoi{10.3847/2041-8213/aaf741}

\bibitem[{Aoyama \& Bai(2023)}]{Aoyama2023}
Aoyama, Y., \& Bai, X.-N. 2023, The Astrophysical Journal, 946, 5,
  \dodoi{10.3847/1538-4357/acb81f}

\bibitem[{{Arakawa} {et~al.}(2023){Arakawa}, {Okuzumi}, {Tatsuuma}, {Tanaka},
  {Kokubo}, {Nishiura}, {Furuichi}, \& {Nakamoto}}]{Arakawa2023}
{Arakawa}, S., {Okuzumi}, S., {Tatsuuma}, M., {et~al.} 2023, \apjl, 951, L16,
  \dodoi{10.3847/2041-8213/acdb5f}

\bibitem[{{Armitage} \& {Kley}(2019)}]{2019ArmitageSaasFee}
{Armitage}, P.~J., \& {Kley}, W. 2019, {From Protoplanetary Disks to Planet
  Formation}, \dodoi{10.1007/978-3-662-58687-7}

\bibitem[{{Ataiee} {et~al.}(2018){Ataiee}, {Baruteau}, {Alibert}, \&
  {Benz}}]{2018AAAtaiee}
{Ataiee}, S., {Baruteau}, C., {Alibert}, Y., \& {Benz}, W. 2018, \aap, 615,
  A110, \dodoi{10.1051/0004-6361/201732026}

\bibitem[{{Ayliffe} \& {Bate}(2012)}]{2012MNRASAyliffeBate}
{Ayliffe}, B.~A., \& {Bate}, M.~R. 2012, \mnras, 427, 2597,
  \dodoi{10.1111/j.1365-2966.2012.21979.x}

\bibitem[{{Bae} {et~al.}(2023){Bae}, {Isella}, {Zhu}, {Martin}, {Okuzumi}, \&
  {Suriano}}]{2022PPVIIBae}
{Bae}, J., {Isella}, A., {Zhu}, Z., {et~al.} 2023, in Astronomical Society of
  the Pacific Conference Series, Vol. 534, Protostars and Planets VII, ed.
  S.~{Inutsuka}, Y.~{Aikawa}, T.~{Muto}, K.~{Tomida}, \& M.~{Tamura}, 423,
  \dodoi{10.48550/arXiv.2210.13314}

\bibitem[{Bai(2016)}]{Bai2016a}
Bai, X.-N. 2016, The Astrophysical Journal, 821, 80,
  \dodoi{10.3847/0004-637X/821/2/80}

\bibitem[{{Bai} \& {Stone}(2010)}]{Bai2010}
{Bai}, X.-N., \& {Stone}, J.~M. 2010, \apj, 722, 1437,
  \dodoi{10.1088/0004-637X/722/2/1437}

\bibitem[{Bai \& Stone(2014)}]{Bai2014}
Bai, X.-N., \& Stone, J.~M. 2014, The Astrophysical Journal, 796, 31,
  \dodoi{10.1088/0004-637X/796/1/31}

\bibitem[{Bai {et~al.}(2016)Bai, Ye, Goodman, \& Yuan}]{Bai2016}
Bai, X.-N., Ye, J., Goodman, J., \& Yuan, F. 2016, The Astrophysical Journal,
  818, 152, \dodoi{10.3847/0004-637x/818/2/152}

\bibitem[{{Bailey} \& {Zhu}(2023)}]{Bailey2023}
{Bailey}, A., \& {Zhu}, Z. 2023, arXiv e-prints, arXiv:2310.03117,
  \dodoi{10.48550/arXiv.2310.03117}

\bibitem[{Banzatti {et~al.}(2023)Banzatti, Pontoppidan, Carr, Jellison,
  Pascucci, Najita, {Mu{\~n}oz-Romero}, {\"O}berg, Kalyaan, Pinilla, Krijt,
  Long, Lambrechts, Rosotti, Herczeg, Salyk, Zhang, Bergin, Ballering, Meyer,
  Bruderer, \& Collaboration}]{Banzatti2023}
Banzatti, A., Pontoppidan, K.~M., Carr, J.~S., {et~al.} 2023, The Astrophysical
  Journal Letters, 957, L22, \dodoi{10.3847/2041-8213/acf5ec}

\bibitem[{{Baruteau} {et~al.}(2016){Baruteau}, {Bai}, {Mordasini}, \&
  {Molli{\`e}re}}]{2016SSRvBaruteau}
{Baruteau}, C., {Bai}, X., {Mordasini}, C., \& {Molli{\`e}re}, P. 2016, \ssr,
  205, 77, \dodoi{10.1007/s11214-016-0258-z}

\bibitem[{{Baruteau} {et~al.}(2014){Baruteau}, {Crida}, {Paardekooper},
  {Masset}, {Guilet}, {Bitsch}, {Nelson}, {Kley}, \&
  {Papaloizou}}]{2014PPVIBaruteau}
{Baruteau}, C., {Crida}, A., {Paardekooper}, S.~J., {et~al.} 2014, in
  Protostars and Planets VI, ed. H.~{Beuther}, R.~S. {Klessen}, C.~P.
  {Dullemond}, \& T.~{Henning}, 667,
  \dodoi{10.2458/azu_uapress_9780816531240-ch029}

\bibitem[{{Bashi} \& {Zucker}(2021)}]{Bashi2021}
{Bashi}, D., \& {Zucker}, S. 2021, \aap, 651, A61,
  \dodoi{10.1051/0004-6361/202140699}

\bibitem[{{Batygin} \& {Morbidelli}(2023)}]{BatyginMorbidelli2023}
{Batygin}, K., \& {Morbidelli}, A. 2023, Nature Astronomy, 7, 330,
  \dodoi{10.1038/s41550-022-01850-5}

\bibitem[{{Bekaert} {et~al.}(2020){Bekaert}, {Broadley}, \&
  {Marty}}]{Bekaert2020}
{Bekaert}, D.~V., {Broadley}, M.~W., \& {Marty}, B. 2020, Scientific Reports,
  10, 5796, \dodoi{10.1038/s41598-020-62650-3}

\bibitem[{{Bell} \& {Lin}(1994)}]{1994ApJBell}
{Bell}, K.~R., \& {Lin}, D.~N.~C. 1994, \apj, 427, 987, \dodoi{10.1086/174206}

\bibitem[{{Benisty} {et~al.}(2021){Benisty}, {Bae}, {Facchini}, {Keppler},
  {Teague}, {Isella}, {Kurtovic}, {P{\'e}rez}, {Sierra}, {Andrews},
  {Carpenter}, {Czekala}, {Dominik}, {Henning}, {Menard}, {Pinilla}, \&
  {Zurlo}}]{Benisty2021}
{Benisty}, M., {Bae}, J., {Facchini}, S., {et~al.} 2021, \apjl, 916, L2,
  \dodoi{10.3847/2041-8213/ac0f83}

\bibitem[{{Ben{\'{\i}}tez-Llambay} {et~al.}(2015){Ben{\'{\i}}tez-Llambay},
  {Masset}, {Koenigsberger}, \& {Szul{\'a}gyi}}]{benitez-llambaymasset2015}
{Ben{\'{\i}}tez-Llambay}, P., {Masset}, F., {Koenigsberger}, G., \&
  {Szul{\'a}gyi}, J. 2015, \nat, 520, 63, \dodoi{10.1038/nature14277}

\bibitem[{{Ben{\'i}tez-Llambay} \& Pessah(2018)}]{Benitez-Llambay2018}
{Ben{\'i}tez-Llambay}, P., \& Pessah, M.~E. 2018, The Astrophysical Journal,
  855, L28, \dodoi{10.3847/2041-8213/aab2ae}

\bibitem[{{Bennett} {et~al.}(2021){Bennett}, {Ranc}, \&
  {Fernandes}}]{Bennett2021}
{Bennett}, D.~P., {Ranc}, C., \& {Fernandes}, R.~B. 2021, \aj, 162, 243,
  \dodoi{10.3847/1538-3881/ac2a2b}

\bibitem[{{Benz} \& {Asphaug}(1999)}]{1999IcarusBenzAsphaug}
{Benz}, W., \& {Asphaug}, E. 1999, \icarus, 142, 5,
  \dodoi{10.1006/icar.1999.6204}

\bibitem[{{Benz} {et~al.}(1989){Benz}, {Cameron}, \& {Melosh}}]{1989IcarusBenz}
{Benz}, W., {Cameron}, A.~G.~W., \& {Melosh}, H.~J. 1989, \icarus, 81, 113,
  \dodoi{10.1016/0019-1035(89)90129-2}

\bibitem[{{Benz} {et~al.}(2014){Benz}, {Ida}, {Alibert}, {Lin}, \&
  {Mordasini}}]{2014PPVIBenz}
{Benz}, W., {Ida}, S., {Alibert}, Y., {Lin}, D., \& {Mordasini}, C. 2014, in
  Protostars and Planets VI, ed. H.~{Beuther}, R.~S. {Klessen}, C.~P.
  {Dullemond}, \& T.~{Henning}, 691,
  \dodoi{10.2458/azu_uapress_9780816531240-ch030}

\bibitem[{{Berger} \& {Loutre}(1991)}]{Berger1991}
{Berger}, A., \& {Loutre}, M.~F. 1991, Quaternary Science Reviews, 10, 297,
  \dodoi{10.1016/0277-3791(91)90033-Q}

\bibitem[{{Bergin} {et~al.}(2015){Bergin}, {Blake}, {Ciesla}, {Hirschmann}, \&
  {Li}}]{Bergin2015PNAS}
{Bergin}, E.~A., {Blake}, G.~A., {Ciesla}, F., {Hirschmann}, M.~M., \& {Li}, J.
  2015, Proceedings of the National Academy of Science, 112, 8965,
  \dodoi{10.1073/pnas.1500954112}

\bibitem[{{Bergsten} {et~al.}(2023){Bergsten}, {Pascucci}, {Hardegree-Ullman},
  {Fernandes}, {Christiansen}, \& {Mulders}}]{Bergsten2023}
{Bergsten}, G.~J., {Pascucci}, I., {Hardegree-Ullman}, K.~K., {et~al.} 2023,
  \aj, 166, 234, \dodoi{10.3847/1538-3881/ad03ea}

\bibitem[{{Bertaux} \& {Ivanova}(2022)}]{Bertaux2022}
{Bertaux}, J.-L., \& {Ivanova}, A. 2022, \mnras, 512, 5552,
  \dodoi{10.1093/mnras/stac777}

\bibitem[{B{\'e}thune {et~al.}(2016)B{\'e}thune, Lesur, \&
  Ferreira}]{Bethune2016}
B{\'e}thune, W., Lesur, G., \& Ferreira, J. 2016, Astronomy and Astrophysics,
  589, A87, \dodoi{10.1051/0004-6361/201527874}

\bibitem[{Birnstiel {et~al.}(2016)Birnstiel, Fang, \& Johansen}]{Birnstiel2016}
Birnstiel, T., Fang, M., \& Johansen, A. 2016, Space Science Reviews, 205, 41,
  \dodoi{10.1007/s11214-016-0256-1}

\bibitem[{{Birnstiel} {et~al.}(2012){Birnstiel}, {Klahr}, \&
  {Ercolano}}]{2012A&ABirnstiel}
{Birnstiel}, T., {Klahr}, H., \& {Ercolano}, B. 2012, \aap, 539, A148,
  \dodoi{10.1051/0004-6361/201118136}

\bibitem[{{Bitsch} {et~al.}(2015){Bitsch}, {Lambrechts}, \&
  {Johansen}}]{2015AABitschB}
{Bitsch}, B., {Lambrechts}, M., \& {Johansen}, A. 2015, \aap, 582, A112,
  \dodoi{10.1051/0004-6361/201526463}

\bibitem[{Bitsch {et~al.}(2022)Bitsch, Schneider, \& Kreidberg}]{Bitsch2022}
Bitsch, B., Schneider, A.~D., \& Kreidberg, L. 2022, Astronomy \& Astrophysics,
  665, A138, \dodoi{10.1051/0004-6361/202243345}

\bibitem[{{Bitsch} {et~al.}(2020){Bitsch}, {Trifonov}, \&
  {Izidoro}}]{2020AABitsch}
{Bitsch}, B., {Trifonov}, T., \& {Izidoro}, A. 2020, \aap, 643, A66,
  \dodoi{10.1051/0004-6361/202038856}

\bibitem[{{Blum} \& {Wurm}(2008)}]{2008ARAABlumWurm}
{Blum}, J., \& {Wurm}, G. 2008, \araa, 46, 21,
  \dodoi{10.1146/annurev.astro.46.060407.145152}

\bibitem[{{Bodenheimer} {et~al.}(2013){Bodenheimer}, {D'Angelo}, {Lissauer},
  {Fortney}, \& {Saumon}}]{2013ApJBodenheimer}
{Bodenheimer}, P., {D'Angelo}, G., {Lissauer}, J.~J., {Fortney}, J.~J., \&
  {Saumon}, D. 2013, \apj, 770, 120, \dodoi{10.1088/0004-637X/770/2/120}

\bibitem[{{Bodenheimer} {et~al.}(2000){Bodenheimer}, {Hubickyj}, \&
  {Lissauer}}]{2000IcarusBodenheimer}
{Bodenheimer}, P., {Hubickyj}, O., \& {Lissauer}, J.~J. 2000, \icarus, 143, 2,
  \dodoi{10.1006/icar.1999.6246}

\bibitem[{{Bodenheimer} \& {Pollack}(1986)}]{1986IcarusBodenheimerPollack}
{Bodenheimer}, P., \& {Pollack}, J.~B. 1986, \icarus, 67, 391,
  \dodoi{10.1016/0019-1035(86)90122-3}

\bibitem[{{Bonfils} {et~al.}(2013){Bonfils}, {Delfosse}, {Udry}, {Forveille},
  {Mayor}, {Perrier}, {Bouchy}, {Gillon}, {Lovis}, {Pepe}, {Queloz}, {Santos},
  {S{\'e}gransan}, \& {Bertaux}}]{2013AABonfils}
{Bonfils}, X., {Delfosse}, X., {Udry}, S., {et~al.} 2013, \aap, 549, A109,
  \dodoi{10.1051/0004-6361/201014704}

\bibitem[{{Booth} \& {Clarke}(2021)}]{2021MNRASBoothClarke}
{Booth}, R.~A., \& {Clarke}, C.~J. 2021, \mnras, 502, 1569,
  \dodoi{10.1093/mnras/stab090}

\bibitem[{{Booth} {et~al.}(2017){Booth}, {Clarke}, {Madhusudhan}, \&
  {Ilee}}]{Booth2017MNRAS}
{Booth}, R.~A., {Clarke}, C.~J., {Madhusudhan}, N., \& {Ilee}, J.~D. 2017,
  \mnras, 469, 3994, \dodoi{10.1093/mnras/stx1103}

\bibitem[{{Booth} \& {Ilee}(2019)}]{2019MNRASBooth}
{Booth}, R.~A., \& {Ilee}, J.~D. 2019, \mnras, 487, 3998,
  \dodoi{10.1093/mnras/stz1488}

\bibitem[{{Borucki} {et~al.}(2010){Borucki}, {Koch}, {Basri}, {Batalha},
  {Brown}, {Caldwell}, {Caldwell}, {Christensen-Dalsgaard}, {Cochran},
  {DeVore}, {Dunham}, {Dupree}, {Gautier}, {Geary}, {Gilliland}, {Gould},
  {Howell}, {Jenkins}, {Kondo}, {Latham}, {Marcy}, {Meibom}, {Kjeldsen},
  {Lissauer}, {Monet}, {Morrison}, {Sasselov}, {Tarter}, {Boss}, {Brownlee},
  {Owen}, {Buzasi}, {Charbonneau}, {Doyle}, {Fortney}, {Ford}, {Holman},
  {Seager}, {Steffen}, {Welsh}, {Rowe}, {Anderson}, {Buchhave}, {Ciardi},
  {Walkowicz}, {Sherry}, {Horch}, {Isaacson}, {Everett}, {Fischer}, {Torres},
  {Johnson}, {Endl}, {MacQueen}, {Bryson}, {Dotson}, {Haas}, {Kolodziejczak},
  {Van Cleve}, {Chandrasekaran}, {Twicken}, {Quintana}, {Clarke}, {Allen},
  {Li}, {Wu}, {Tenenbaum}, {Verner}, {Bruhweiler}, {Barnes}, \&
  {Prsa}}]{2010ScienceBorucki}
{Borucki}, W.~J., {Koch}, D., {Basri}, G., {et~al.} 2010, Science, 327, 977,
  \dodoi{10.1126/science.1185402}

\bibitem[{{Boss}(1997)}]{1997ScienceBoss}
{Boss}, A.~P. 1997, \sci, 276, 1836, \dodoi{10.1126/science.276.5320.1836}

\bibitem[{{Boss}(2011)}]{Boss2011}
---. 2011, \apj, 731, 74, \dodoi{10.1088/0004-637X/731/1/74}

\bibitem[{{Bottke} \& {Norman}(2017)}]{BottkeNorman2017}
{Bottke}, W.~F., \& {Norman}, M.~D. 2017, Annual Review of Earth and Planetary
  Sciences, 45, 619, \dodoi{10.1146/annurev-earth-063016-020131}

\bibitem[{Bouma {et~al.}(2022)Bouma, Kerr, Curtis, Isaacson, Hillenbrand,
  Howard, Kraus, Bieryla, Latham, Petigura, \& Huber}]{bouma2022}
Bouma, L.~G., Kerr, R., Curtis, J.~L., {et~al.} 2022, The Astronomical Journal,
  164, 215, \dodoi{10.3847/1538-3881/ac93ff}

\bibitem[{Bower {et~al.}(2022)Bower, Hakim, Sossi, \& Sanan}]{Bower2022}
Bower, D.~J., Hakim, K., Sossi, P.~A., \& Sanan, P. 2022, The Planetary Science
  Journal, 3, 93, \dodoi{10.3847/PSJ/ac5fb1}

\bibitem[{Bower {et~al.}(2019)Bower, Kitzmann, Wolf, Sanan, Dorn, \&
  Oza}]{Bower2019}
Bower, D.~J., Kitzmann, D., Wolf, A.~S., {et~al.} 2019, Astronomy and
  Astrophysics, 631, A103, \dodoi{10.1051/0004-6361/201935710}

\bibitem[{{Bowler}(2016)}]{2016PASPBowler}
{Bowler}, B.~P. 2016, \pasp, 128, 102001,
  \dodoi{10.1088/1538-3873/128/968/102001}

\bibitem[{{Bro{\v{z}}} {et~al.}(2021){Bro{\v{z}}}, {Chrenko}, {Nesvorn{\'y}},
  \& {Dauphas}}]{Broz2021}
{Bro{\v{z}}}, M., {Chrenko}, O., {Nesvorn{\'y}}, D., \& {Dauphas}, N. 2021,
  Nature Astronomy, 5, 898, \dodoi{10.1038/s41550-021-01383-3}

\bibitem[{{Br{\"u}gger} {et~al.}(2020){Br{\"u}gger}, {Burn}, {Coleman},
  {Alibert}, \& {Benz}}]{2020AABrugger}
{Br{\"u}gger}, N., {Burn}, R., {Coleman}, G.~A.~L., {Alibert}, Y., \& {Benz},
  W. 2020, \aap, 640, A21, \dodoi{10.1051/0004-6361/202038042}

\bibitem[{{Bryan} {et~al.}(2016){Bryan}, {Knutson}, {Howard}, {Ngo}, {Batygin},
  {Crepp}, {Fulton}, {Hinkley}, {Isaacson}, {Johnson}, {Marcy}, \&
  {Wright}}]{2016ApJBryan}
{Bryan}, M.~L., {Knutson}, H.~A., {Howard}, A.~W., {et~al.} 2016, \apj, 821,
  89, \dodoi{10.3847/0004-637X/821/2/89}

\bibitem[{{Budde} {et~al.}(2019){Budde}, {Burkhardt}, \& {Kleine}}]{Budde2019}
{Budde}, G., {Burkhardt}, C., \& {Kleine}, T. 2019, Nature Astronomy, 3, 736,
  \dodoi{10.1038/s41550-019-0779-y}

\bibitem[{{Burn} {et~al.}(2022){Burn}, {Emsenhuber}, {Weder}, {V{\"o}lkel},
  {Klahr}, {Birnstiel}, {Ercolano}, \& {Mordasini}}]{2022AABurnA}
{Burn}, R., {Emsenhuber}, A., {Weder}, J., {et~al.} 2022, \aap, 666, A73,
  \dodoi{10.1051/0004-6361/202243262}

\bibitem[{{Burn} {et~al.}(2019){Burn}, {Marboeuf}, {Alibert}, \&
  {Benz}}]{2019AABurn}
{Burn}, R., {Marboeuf}, U., {Alibert}, Y., \& {Benz}, W. 2019, \aap, 629, A64,
  \dodoi{10.1051/0004-6361/201935780}

\bibitem[{{Burn} {et~al.}(2024){Burn}, {Mordasini}, {Mishra}, {Haldemann},
  {Venturini}, {Emsenhuber}, \& {Henning}}]{Burn2024}
{Burn}, R., {Mordasini}, C., {Mishra}, L., {et~al.} 2024, Nature Astronomy,
  Advanced Online Publication, \dodoi{10.1038/s41550-023-02183-7}

\bibitem[{{Burn} {et~al.}(2021){Burn}, {Schlecker}, {Mordasini}, {Emsenhuber},
  {Alibert}, {Henning}, {Klahr}, \& {Benz}}]{2021AABurn}
{Burn}, R., {Schlecker}, M., {Mordasini}, C., {et~al.} 2021, \aap, 656, A72,
  \dodoi{10.1051/0004-6361/202140390}

\bibitem[{{Cameron}(1988)}]{cameron1988}
{Cameron}, A.~G.~W. 1988, \araa, 26, 441,
  \dodoi{10.1146/annurev.aa.26.090188.002301}

\bibitem[{{Canup} \& {Asphaug}(2001)}]{2001NatureCanup}
{Canup}, R.~M., \& {Asphaug}, E. 2001, \nat, 412, 708, \dodoi{10.1038/35089010}

\bibitem[{{Capistrant} {et~al.}(2024){Capistrant}, {Soares-Furtado},
  {Vanderburg}, {Jankowski}, {Mann}, {Ross}, {Srdoc}, {Hinkel}, {Becker},
  {Magliano}, {Limbach}, {Stephan}, {Nine}, {Tofflemire}, {Kraus}, {Giacalone},
  {Winn}, {Bieryla}, {Bouma}, {Ciardi}, {Collins}, {Covone}, {de Beurs},
  {Huang}, {Jenkins}, {Kreidberg}, {Latham}, {Quinn}, {Seager}, {Shporer},
  {Twicken}, {Wohler}, {Vanderspek}, {Yarza}, \& {Ziegler}}]{Capistrant2024}
{Capistrant}, B.~K., {Soares-Furtado}, M., {Vanderburg}, A., {et~al.} 2024,
  \aj, 167, 54, \dodoi{10.3847/1538-3881/ad1039}

\bibitem[{{Carrera} {et~al.}(2015){Carrera}, {Johansen}, \&
  {Davies}}]{Carrera2015}
{Carrera}, D., {Johansen}, A., \& {Davies}, M.~B. 2015, \aap, 579, A43,
  \dodoi{10.1051/0004-6361/201425120}

\bibitem[{{Cassan} {et~al.}(2012){Cassan}, {Kubas}, {Beaulieu}, {Dominik},
  {Horne}, {Greenhill}, {Wambsganss}, {Menzies}, {Williams}, {J{\o}rgensen},
  {Udalski}, {Bennett}, {Albrow}, {Batista}, {Brillant}, {Caldwell}, {Cole},
  {Coutures}, {Cook}, {Dieters}, {Dominis Prester}, {Donatowicz}, {Fouqu{\'e}},
  {Hill}, {Kains}, {Kane}, {Marquette}, {Martin}, {Pollard}, {Sahu}, {Vinter},
  {Warren}, {Watson}, {Zub}, {Sumi}, {Szyma{\'n}ski}, {Kubiak}, {Poleski},
  {Soszynski}, {Ulaczyk}, {Pietrzy{\'n}ski}, \&
  {Wyrzykowski}}]{CassanKubas2012}
{Cassan}, A., {Kubas}, D., {Beaulieu}, J.~P., {et~al.} 2012, \nat, 481, 167,
  \dodoi{10.1038/nature10684}

\bibitem[{{Chabrier} \& {Debras}(2021)}]{2021ApJChabrierDebras}
{Chabrier}, G., \& {Debras}, F. 2021, \apj, 917, 4,
  \dodoi{10.3847/1538-4357/abfc48}

\bibitem[{{Chambers}(2018)}]{2018ApJChambers}
{Chambers}, J. 2018, \apj, 865, 30, \dodoi{10.3847/1538-4357/aada09}

\bibitem[{Chambers(2006)}]{Chambers2006}
Chambers, J.~E. 2006, Icarus, 180, 496, \dodoi{10.1016/j.icarus.2005.10.017}

\bibitem[{{Chiang} \& {Goldreich}(1997)}]{1997ApJChiang}
{Chiang}, E.~I., \& {Goldreich}, P. 1997, \apj, 490, 368,
  \dodoi{10.1086/304869}

\bibitem[{{Choksi} {et~al.}(2023){Choksi}, {Chiang}, {Fung}, \&
  {Zhu}}]{Choksi2023}
{Choksi}, N., {Chiang}, E., {Fung}, J., \& {Zhu}, Z. 2023, \mnras, 525, 2806,
  \dodoi{10.1093/mnras/stad2269}

\bibitem[{Ciesla(2011)}]{Ciesla2011}
Ciesla, F.~J. 2011, ApJ, 740, 9, \dodoi{10.1088/0004-637X/740/1/9}

\bibitem[{{Cimerman} {et~al.}(2017){Cimerman}, {Kuiper}, \&
  {Ormel}}]{Cimermankuiper2017}
{Cimerman}, N.~P., {Kuiper}, R., \& {Ormel}, C.~W. 2017, \mnras, 471, 4662,
  \dodoi{10.1093/mnras/stx1924}

\bibitem[{{Clarke} {et~al.}(2001){Clarke}, {Gendrin}, \&
  {Sotomayor}}]{2001MNRASClarke}
{Clarke}, C.~J., {Gendrin}, A., \& {Sotomayor}, M. 2001, \mnras, 328, 485,
  \dodoi{10.1046/j.1365-8711.2001.04891.x}

\bibitem[{{Coleman}(2021)}]{Coleman2021}
{Coleman}, G. A.~L. 2021, \mnras, 506, 3596, \dodoi{10.1093/mnras/stab1904}

\bibitem[{{Coleman} {et~al.}(2024){Coleman}, {Mroueh}, \&
  {Haworth}}]{Coleman2024}
{Coleman}, G. A.~L., {Mroueh}, J.~K., \& {Haworth}, T.~J. 2024, \mnras, 527,
  7588, \dodoi{10.1093/mnras/stad3692}

\bibitem[{{Coleman} \& {Nelson}(2014)}]{2014MNRASColemanNelson}
{Coleman}, G. A.~L., \& {Nelson}, R.~P. 2014, \mnras, 445, 479,
  \dodoi{10.1093/mnras/stu1715}

\bibitem[{{Coleman} \& {Nelson}(2016)}]{Coleman2016}
---. 2016, \mnras, 460, 2779, \dodoi{10.1093/mnras/stw1177}

\bibitem[{{Crida} {et~al.}(2006){Crida}, {Morbidelli}, \&
  {Masset}}]{2006IcarusCrida}
{Crida}, A., {Morbidelli}, A., \& {Masset}, F. 2006, \icarus, 181, 587,
  \dodoi{10.1016/j.icarus.2005.10.007}

\bibitem[{{Cridland} {et~al.}(2016){Cridland}, {Pudritz}, \&
  {Alessi}}]{Cridland2016}
{Cridland}, A.~J., {Pudritz}, R.~E., \& {Alessi}, M. 2016, \mnras, 461, 3274,
  \dodoi{10.1093/mnras/stw1511}

\bibitem[{{Cridland} {et~al.}(2017){Cridland}, {Pudritz}, {Birnstiel},
  {Cleeves}, \& {Bergin}}]{Cridland2017}
{Cridland}, A.~J., {Pudritz}, R.~E., {Birnstiel}, T., {Cleeves}, L.~I., \&
  {Bergin}, E.~A. 2017, \mnras, 469, 3910, \dodoi{10.1093/mnras/stx1069}

\bibitem[{{Crossfield}(2023)}]{Crossfield2023ApJ}
{Crossfield}, I. J.~M. 2023, \apjl, 952, L18, \dodoi{10.3847/2041-8213/ace35f}

\bibitem[{{Cumming} {et~al.}(2008){Cumming}, {Butler}, {Marcy}, {Vogt},
  {Wright}, \& {Fischer}}]{2008PASPCumming}
{Cumming}, A., {Butler}, R.~P., {Marcy}, G.~W., {et~al.} 2008, \pasp, 120, 531,
  \dodoi{10.1086/588487}

\bibitem[{{Dai} {et~al.}(2023){Dai}, {Liu}, {Yang}, \& {Zhou}}]{Dai2023}
{Dai}, Y.-Z., {Liu}, H.-G., {Yang}, J.-Y., \& {Zhou}, J.-L. 2023, \aj, 166,
  219, \dodoi{10.3847/1538-3881/acff67}

\bibitem[{{D'Angelo} \& {Lubow}(2008)}]{2008ApJDAngeloLubow}
{D'Angelo}, G., \& {Lubow}, S.~H. 2008, \apj, 685, 560, \dodoi{10.1086/590904}

\bibitem[{{Dauphas} \& {Pourmand}(2011)}]{DauphasPourmand2011}
{Dauphas}, N., \& {Pourmand}, A. 2011, \nat, 473, 489,
  \dodoi{10.1038/nature10077}

\bibitem[{{David} {et~al.}(2016){David}, {Hillenbrand}, {Petigura},
  {Carpenter}, {Crossfield}, {Hinkley}, {Ciardi}, {Howard}, {Isaacson}, {Cody},
  {Schlieder}, {Beichman}, \& {Barenfeld}}]{David2016Nature}
{David}, T.~J., {Hillenbrand}, L.~A., {Petigura}, E.~A., {et~al.} 2016, \nat,
  534, 658, \dodoi{10.1038/nature18293}

\bibitem[{{Dawson} \& {Johnson}(2018)}]{2018ARAADawson}
{Dawson}, R.~I., \& {Johnson}, J.~A. 2018, \araa, 56, 175,
  \dodoi{10.1146/annurev-astro-081817-051853}

\bibitem[{{Debras} \& {Chabrier}(2019)}]{DebrasChabrier2019}
{Debras}, F., \& {Chabrier}, G. 2019, \apj, 872, 100,
  \dodoi{10.3847/1538-4357/aaff65}

\bibitem[{{DeMeo} \& {Carry}(2014)}]{DeMeoCarry2014}
{DeMeo}, F.~E., \& {Carry}, B. 2014, \nat, 505, 629,
  \dodoi{10.1038/nature12908}

\bibitem[{{Demirci} {et~al.}(2020){Demirci}, {Schneider}, {Teiser}, \&
  {Wurm}}]{Demirci2020}
{Demirci}, T., {Schneider}, N., {Teiser}, J., \& {Wurm}, G. 2020, \aap, 644,
  A20, \dodoi{10.1051/0004-6361/202039312}

\bibitem[{{Denman} {et~al.}(2020){Denman}, {Leinhardt}, {Carter}, \&
  {Mordasini}}]{2020MNRASDenman}
{Denman}, T.~R., {Leinhardt}, Z.~M., {Carter}, P.~J., \& {Mordasini}, C. 2020,
  \mnras, 496, 1166, \dodoi{10.1093/mnras/staa1623}

\bibitem[{{Dittkrist} {et~al.}(2014){Dittkrist}, {Mordasini}, {Klahr},
  {Alibert}, \& {Henning}}]{2014A&ADittkrist}
{Dittkrist}, K.~M., {Mordasini}, C., {Klahr}, H., {Alibert}, Y., \& {Henning},
  T. 2014, \aap, 567, A121, \dodoi{10.1051/0004-6361/201322506}

\bibitem[{{Dittrich} {et~al.}(2013){Dittrich}, {Klahr}, \&
  {Johansen}}]{2013ApJDittrich}
{Dittrich}, K., {Klahr}, H., \& {Johansen}, A. 2013, \apj, 763, 117,
  \dodoi{10.1088/0004-637X/763/2/117}

\bibitem[{{Dole}(1970)}]{Dole1970}
{Dole}, S.~H. 1970, \icarus, 13, 494, \dodoi{10.1016/0019-1035(70)90095-3}

\bibitem[{{Donati} {et~al.}(2016){Donati}, {Moutou}, {Malo}, {Baruteau}, {Yu},
  {H{\'e}brard}, {Hussain}, {Alencar}, {M{\'e}nard}, {Bouvier}, {Petit},
  {Takami}, {Doyon}, \& {Collier Cameron}}]{Donati2016Nature}
{Donati}, J.~F., {Moutou}, C., {Malo}, L., {et~al.} 2016, \nat, 534, 662,
  \dodoi{10.1038/nature18305}

\bibitem[{{Dong} {et~al.}(2018){Dong}, {Xie}, {Zhou}, {Zheng}, \&
  {Luo}}]{DongXie2018}
{Dong}, S., {Xie}, J.-W., {Zhou}, J.-L., {Zheng}, Z., \& {Luo}, A. 2018,
  Proceedings of the National Academy of Science, 115, 266,
  \dodoi{10.1073/pnas.1711406115}

\bibitem[{{Drazkowska} {et~al.}(2022){Drazkowska}, {Bitsch}, {Lambrechts},
  {Mulders}, {Harsono}, {Vazan}, {Liu}, {Ormel}, {Kretke}, \&
  {Morbidelli}}]{2022PPVIIDrazkowska}
{Drazkowska}, J., {Bitsch}, B., {Lambrechts}, M., {et~al.} 2022, arXiv
  e-prints, arXiv:2203.09759.
\newblock \doarXiv{2203.09759}

\bibitem[{{Dr{\k{a}}{\.z}kowska} \& {Alibert}(2017)}]{2017AADrazkowskaAlibert}
{Dr{\k{a}}{\.z}kowska}, J., \& {Alibert}, Y. 2017, \aap, 608, A92,
  \dodoi{10.1051/0004-6361/201731491}

\bibitem[{{Dr{\k{a}}{\.z}kowska} {et~al.}(2016){Dr{\k{a}}{\.z}kowska},
  {Alibert}, \& {Moore}}]{Drazkowska2016}
{Dr{\k{a}}{\.z}kowska}, J., {Alibert}, Y., \& {Moore}, B. 2016, \aap, 594,
  A105, \dodoi{10.1051/0004-6361/201628983}

\bibitem[{Dullemond \& Penzlin(2018)}]{Dullemond2018a}
Dullemond, C.~P., \& Penzlin, A. B.~T. 2018, Astronomy and Astrophysics, 609,
  A50, \dodoi{10.1051/0004-6361/201731878}

\bibitem[{{Dullemond} {et~al.}(2018){Dullemond}, {Birnstiel}, {Huang},
  {Kurtovic}, {Andrews}, {Guzm{\'a}n}, {P{\'e}rez}, {Isella}, {Zhu}, {Benisty},
  {Wilner}, {Bai}, {Carpenter}, {Zhang}, \& {Ricci}}]{2018ApJDullemond}
{Dullemond}, C.~P., {Birnstiel}, T., {Huang}, J., {et~al.} 2018, \apjl, 869,
  L46, \dodoi{10.3847/2041-8213/aaf742}

\bibitem[{Dunlop {et~al.}(2013)Dunlop, Encrenaz, Bibring, Blanc, Barucci,
  Roques, \& Zarka}]{dunlop2013solar}
Dunlop, S., Encrenaz, T., Bibring, J., {et~al.} 2013, The Solar System,
  Astronomy and Astrophysics Library (Springer Berlin Heidelberg).
\newblock \url{https://books.google.ch/books?id=fWfyCAAAQBAJ}

\bibitem[{{D{\"u}rmann} \& {Kley}(2015)}]{2015AADurmannKley}
{D{\"u}rmann}, C., \& {Kley}, W. 2015, \aap, 574, A52,
  \dodoi{10.1051/0004-6361/201424837}

\bibitem[{Eistrup {et~al.}(2016)Eistrup, Walsh, \& {van
  Dishoeck}}]{Eistrup2016}
Eistrup, C., Walsh, C., \& {van Dishoeck}, E.~F. 2016, Astronomy \&
  Astrophysics, 595, A83, \dodoi{10.1051/0004-6361/201628509}

\bibitem[{Eistrup {et~al.}(2018)Eistrup, Walsh, \& {van
  Dishoeck}}]{Eistrup2018}
---. 2018, Astronomy \& Astrophysics, 613, A14,
  \dodoi{10.1051/0004-6361/201731302}

\bibitem[{Emsenhuber {et~al.}(2023)Emsenhuber, Burn, Weder, Monsch, Picogna,
  Ercolano, \& Preibisch}]{Emsenhuber2023}
Emsenhuber, A., Burn, R., Weder, J., {et~al.} 2023, Astronomy and Astrophysics,
  673, A78, \dodoi{10.1051/0004-6361/202244767}

\bibitem[{{Emsenhuber} {et~al.}(2023){Emsenhuber}, {Mordasini}, \&
  {Burn}}]{2023EPJPEmsenhuber}
{Emsenhuber}, A., {Mordasini}, C., \& {Burn}, R. 2023, European Physical
  Journal Plus, 138, 181, \dodoi{10.1140/epjp/s13360-023-03784-x}

\bibitem[{{Emsenhuber} {et~al.}(2021{\natexlab{a}}){Emsenhuber}, {Mordasini},
  {Burn}, {Alibert}, {Benz}, \& {Asphaug}}]{2021AAEmsenhuberB}
{Emsenhuber}, A., {Mordasini}, C., {Burn}, R., {et~al.} 2021{\natexlab{a}},
  \aap, 656, A70, \dodoi{10.1051/0004-6361/202038863}

\bibitem[{{Emsenhuber} {et~al.}(2021{\natexlab{b}}){Emsenhuber}, {Mordasini},
  {Burn}, {Alibert}, {Benz}, \& {Asphaug}}]{2021AAEmsenhuberA}
---. 2021{\natexlab{b}}, \aap, 656, A69, \dodoi{10.1051/0004-6361/202038553}

\bibitem[{Encrenaz \& Lequeux(2021)}]{encrenaz2021solar}
Encrenaz, T., \& Lequeux, J. 2021, The Solar System 1: Telluric and Giant
  Planets, Interplanetary Medium and Exoplanets (Wiley).
\newblock \url{https://books.google.ch/books?id=mI1SEAAAQBAJ}

\bibitem[{{Ercolano} {et~al.}(2009){Ercolano}, {Clarke}, \&
  {Drake}}]{2009ApJErcolano}
{Ercolano}, B., {Clarke}, C.~J., \& {Drake}, J.~J. 2009, \apj, 699, 1639,
  \dodoi{10.1088/0004-637X/699/2/1639}

\bibitem[{{Ercolano} {et~al.}(2021){Ercolano}, {Picogna}, {Monsch}, {Drake}, \&
  {Preibisch}}]{2021MNRASErcolano}
{Ercolano}, B., {Picogna}, G., {Monsch}, K., {Drake}, J.~J., \& {Preibisch}, T.
  2021, \mnras, 508, 1675, \dodoi{10.1093/mnras/stab2590}

\bibitem[{{Facchini} {et~al.}(2016){Facchini}, {Clarke}, \&
  {Bisbas}}]{2016MNRASFacchini}
{Facchini}, S., {Clarke}, C.~J., \& {Bisbas}, T.~G. 2016, \mnras, 457, 3593,
  \dodoi{10.1093/mnras/stw240}

\bibitem[{{Fernandes} {et~al.}(2019){Fernandes}, {Mulders}, {Pascucci},
  {Mordasini}, \& {Emsenhuber}}]{2019ApJFernandes}
{Fernandes}, R.~B., {Mulders}, G.~D., {Pascucci}, I., {Mordasini}, C., \&
  {Emsenhuber}, A. 2019, \apj, 874, 81, \dodoi{10.3847/1538-4357/ab0300}

\bibitem[{{Fischer} \& {Valenti}(2005)}]{2005ApJFischer}
{Fischer}, D.~A., \& {Valenti}, J. 2005, \apj, 622, 1102,
  \dodoi{10.1086/428383}

\bibitem[{{Ford} \& {Rasio}(2008)}]{2008ApJFordRasio}
{Ford}, E.~B., \& {Rasio}, F.~A. 2008, \apj, 686, 621, \dodoi{10.1086/590926}

\bibitem[{{Forgan} {et~al.}(2018){Forgan}, {Hall}, {Meru}, \&
  {Rice}}]{2018MNRASForgan}
{Forgan}, D.~H., {Hall}, C., {Meru}, F., \& {Rice}, W.~K.~M. 2018, \mnras, 474,
  5036, \dodoi{10.1093/mnras/stx2870}

\bibitem[{{Fortier} {et~al.}(2013){Fortier}, {Alibert}, {Carron}, {Benz}, \&
  {Dittkrist}}]{2013A&AFortier}
{Fortier}, A., {Alibert}, Y., {Carron}, F., {Benz}, W., \& {Dittkrist}, K.~M.
  2013, \aap, 549, A44, \dodoi{10.1051/0004-6361/201220241}

\bibitem[{{Franz} {et~al.}(2020){Franz}, {Picogna}, {Ercolano}, \&
  {Birnstiel}}]{2020AAFranz}
{Franz}, R., {Picogna}, G., {Ercolano}, B., \& {Birnstiel}, T. 2020, \aap, 635,
  A53, \dodoi{10.1051/0004-6361/201936615}

\bibitem[{Fray \& Schmitt(2009)}]{Fray2009}
Fray, N., \& Schmitt, B. 2009, Planetary and Space Science, 57, 2053,
  \dodoi{10.1016/J.PSS.2009.09.011}

\bibitem[{{Fressin} {et~al.}(2013){Fressin}, {Torres}, {Charbonneau}, {Bryson},
  {Christiansen}, {Dressing}, {Jenkins}, {Walkowicz}, \&
  {Batalha}}]{2013ApJFressin}
{Fressin}, F., {Torres}, G., {Charbonneau}, D., {et~al.} 2013, \apj, 766, 81,
  \dodoi{10.1088/0004-637X/766/2/81}

\bibitem[{Fromang \& Nelson(2009)}]{Fromang2009}
Fromang, S., \& Nelson, R.~P. 2009, Astronomy and Astrophysics, 496, 597,
  \dodoi{10.1051/0004-6361/200811220}

\bibitem[{{Fulton} \& {Petigura}(2018)}]{2018AJFultonPetigura}
{Fulton}, B.~J., \& {Petigura}, E.~A. 2018, \aj, 156, 264,
  \dodoi{10.3847/1538-3881/aae828}

\bibitem[{{Fulton} {et~al.}(2017){Fulton}, {Petigura}, {Howard}, {Isaacson},
  {Marcy}, {Cargile}, {Hebb}, {Weiss}, {Johnson}, {Morton}, {Sinukoff},
  {Crossfield}, \& {Hirsch}}]{2017AJFulton}
{Fulton}, B.~J., {Petigura}, E.~A., {Howard}, A.~W., {et~al.} 2017, \aj, 154,
  109, \dodoi{10.3847/1538-3881/aa80eb}

\bibitem[{{Fulton} {et~al.}(2021){Fulton}, {Rosenthal}, {Hirsch}, {Isaacson},
  {Howard}, {Dedrick}, {Sherstyuk}, {Blunt}, {Petigura}, {Knutson}, {Behmard},
  {Chontos}, {Crepp}, {Crossfield}, {Dalba}, {Fischer}, {Henry}, {Kane},
  {Kosiarek}, {Marcy}, {Rubenzahl}, {Weiss}, \& {Wright}}]{2021ApJSFulton}
{Fulton}, B.~J., {Rosenthal}, L.~J., {Hirsch}, L.~A., {et~al.} 2021, \apjs,
  255, 14, \dodoi{10.3847/1538-4365/abfcc1}

\bibitem[{{Gaidos} {et~al.}(2013){Gaidos}, {Fischer}, {Mann}, \&
  {Howard}}]{Gaidos2013}
{Gaidos}, E., {Fischer}, D.~A., {Mann}, A.~W., \& {Howard}, A.~W. 2013, \apj,
  771, 18, \dodoi{10.1088/0004-637X/771/1/18}

\bibitem[{{Gammie}(1996)}]{Gammie1996}
{Gammie}, C.~F. 1996, \apj, 457, 355, \dodoi{10.1086/176735}

\bibitem[{{Gerbig} {et~al.}(2020){Gerbig}, {Murray-Clay}, {Klahr}, \&
  {Baehr}}]{Gerbig2020}
{Gerbig}, K., {Murray-Clay}, R.~A., {Klahr}, H., \& {Baehr}, H. 2020, \apj,
  895, 91, \dodoi{10.3847/1538-4357/ab8d37}

\bibitem[{{Ghezzi} {et~al.}(2018){Ghezzi}, {Montet}, \&
  {Johnson}}]{2018ApJGhezzi}
{Ghezzi}, L., {Montet}, B.~T., \& {Johnson}, J.~A. 2018, \apj, 860, 109,
  \dodoi{10.3847/1538-4357/aac37c}

\bibitem[{{Goldreich} \& {Tremaine}(1979)}]{1979ApJGoldreichTremaine}
{Goldreich}, P., \& {Tremaine}, S. 1979, \apj, 233, 857, \dodoi{10.1086/157448}

\bibitem[{{Goldreich} \& {Tremaine}(1980)}]{1980ApJGoldreichTremaine}
---. 1980, \apj, 241, 425, \dodoi{10.1086/158356}

\bibitem[{{Goldreich} \& {Ward}(1973)}]{Goldreich1973}
{Goldreich}, P., \& {Ward}, W.~R. 1973, \apj, 183, 1051, \dodoi{10.1086/152291}

\bibitem[{{Gonzalez}(1997)}]{1997MNRASGonzalez}
{Gonzalez}, G. 1997, \mnras, 285, 403, \dodoi{10.1093/mnras/285.2.403}

\bibitem[{{Gorti} \& {Hollenbach}(2009)}]{2009ApJGorti}
{Gorti}, U., \& {Hollenbach}, D. 2009, \apj, 690, 1539,
  \dodoi{10.1088/0004-637X/690/2/1539}

\bibitem[{{Grandjean} {et~al.}(2021){Grandjean}, {Lagrange}, {Meunier},
  {Rubini}, {Desidera}, {Galland}, {Borgniet}, {Zicher}, {Messina}, {Chauvin},
  {Sterzik}, \& {Pantoja}}]{Grandjean2021}
{Grandjean}, A., {Lagrange}, A.~M., {Meunier}, N., {et~al.} 2021, \aap, 650,
  A39, \dodoi{10.1051/0004-6361/202039672}

\bibitem[{Grant {et~al.}(2023)Grant, {van Dishoeck}, Tabone, Gasman, Henning,
  Kamp, G{\"u}del, Lagage, Bettoni, Perotti, Christiaens, Samland, Arabhavi,
  Argyriou, Abergel, Absil, Barrado, Boccaletti, Bouwman, {o Garatti}, Geers,
  Glauser, Guadarrama, Jang, Kanwar, Lahuis, {Morales-Calder{\'o}n}, Mueller,
  Nehm{\'e}, Olofsson, Pantin, Pawellek, Ray, {Rodgers-Lee}, Scheithauer,
  Schreiber, Schwarz, Temmink, Vandenbussche, Vlasblom, Waters, Wright, Colina,
  Greve, Justannont, \& {\"O}stlin}]{Grant2023}
Grant, S.~L., {van Dishoeck}, E.~F., Tabone, B., {et~al.} 2023, The
  Astrophysical Journal, 947, L6, \dodoi{10.3847/2041-8213/acc44b}

\bibitem[{{Greenzweig} \& {Lissauer}(1990)}]{1990IcarusGreenzweigLissauer}
{Greenzweig}, Y., \& {Lissauer}, J.~J. 1990, \icarus, 87, 40,
  \dodoi{10.1016/0019-1035(90)90021-Z}

\bibitem[{{Greenzweig} \& {Lissauer}(1992)}]{1992IcarusGreenzweigLissauer}
---. 1992, \icarus, 100, 440, \dodoi{10.1016/0019-1035(92)90110-S}

\bibitem[{{Grether} \& {Lineweaver}(2006)}]{GretherLineweaver2006}
{Grether}, D., \& {Lineweaver}, C.~H. 2006, \apj, 640, 1051,
  \dodoi{10.1086/500161}

\bibitem[{{Guillot} {et~al.}(2023){Guillot}, {Fletcher}, {Helled}, {Ikoma},
  {Line}, \& {Paramentier}}]{Guillot2023}
{Guillot}, T., {Fletcher}, L.~N., {Helled}, R., {et~al.} 2023, in Astronomical
  Society of the Pacific Conference Series, Vol. 534, Astronomical Society of
  the Pacific Conference Series, ed. S.~{Inutsuka}, Y.~{Aikawa}, T.~{Muto},
  K.~{Tomida}, \& M.~{Tamura}, 947

\bibitem[{{Guillot} \& {Gautier}(2015)}]{GuillotGautier2015}
{Guillot}, T., \& {Gautier}, D. 2015, in Treatise on Geophysics, ed.
  G.~{Schubert}, 529--557, \dodoi{10.1016/B978-0-444-53802-4.00176-7}

\bibitem[{{Guillot} \& {Hueso}(2006)}]{GuillotHueso2006}
{Guillot}, T., \& {Hueso}, R. 2006, \mnras, 367, L47,
  \dodoi{10.1111/j.1745-3933.2006.00137.x}

\bibitem[{{Gundlach} {et~al.}(2018){Gundlach}, {Schmidt}, {Kreuzig},
  {Bischoff}, {Rezaei}, {Kothe}, {Blum}, {Grzesik}, \&
  {Stoll}}]{2018MNRASGundlach}
{Gundlach}, B., {Schmidt}, K.~P., {Kreuzig}, C., {et~al.} 2018, \mnras, 479,
  1273, \dodoi{10.1093/mnras/sty1550}

\bibitem[{{Guzm{\'a}n-Mesa} {et~al.}(2022){Guzm{\'a}n-Mesa}, {Kitzmann},
  {Mordasini}, \& {Heng}}]{Guzman2022MNRAS}
{Guzm{\'a}n-Mesa}, A., {Kitzmann}, D., {Mordasini}, C., \& {Heng}, K. 2022,
  \mnras, 513, 4015, \dodoi{10.1093/mnras/stac1066}

\bibitem[{{Haffert} {et~al.}(2019{\natexlab{a}}){Haffert}, {Bohn}, {de Boer},
  {Snellen}, {Brinchmann}, {Girard}, {Keller}, \& {Bacon}}]{2019NatAsHaffert}
{Haffert}, S.~Y., {Bohn}, A.~J., {de Boer}, J., {et~al.} 2019{\natexlab{a}},
  Nature Astronomy, 3, 749, \dodoi{10.1038/s41550-019-0780-5}

\bibitem[{{Haffert} {et~al.}(2019{\natexlab{b}}){Haffert}, {Bohn}, {de Boer},
  {Snellen}, {Brinchmann}, {Girard}, {Keller}, \& {Bacon}}]{Haffert2019}
---. 2019{\natexlab{b}}, Nature Astronomy, 3, 749,
  \dodoi{10.1038/s41550-019-0780-5}

\bibitem[{{Haisch} {et~al.}(2001){Haisch}, {Lada}, \& {Lada}}]{2001ApJHaisch}
{Haisch}, Karl~E., J., {Lada}, E.~A., \& {Lada}, C.~J. 2001, \apjl, 553, L153,
  \dodoi{10.1086/320685}

\bibitem[{{Hara} {et~al.}(2019){Hara}, {Bou{\'e}}, {Laskar}, {Delisle}, \&
  {Unger}}]{2019MNRASHara}
{Hara}, N.~C., {Bou{\'e}}, G., {Laskar}, J., {Delisle}, J.~B., \& {Unger}, N.
  2019, \mnras, 489, 738, \dodoi{10.1093/mnras/stz1849}

\bibitem[{{Hasegawa} \& {Pudritz}(2013)}]{Hasegawa2013}
{Hasegawa}, Y., \& {Pudritz}, R.~E. 2013, \apj, 778, 78,
  \dodoi{10.1088/0004-637X/778/1/78}

\bibitem[{{Hashimoto} {et~al.}(2020){Hashimoto}, {Aoyama}, {Konishi}, {Uyama},
  {Takasao}, {Ikoma}, \& {Tanigawa}}]{Hashimoto2020}
{Hashimoto}, J., {Aoyama}, Y., {Konishi}, M., {et~al.} 2020, \aj, 159, 222,
  \dodoi{10.3847/1538-3881/ab811e}

\bibitem[{{Haworth} {et~al.}(2018){Haworth}, {Clarke}, {Rahman}, {Winter}, \&
  {Facchini}}]{2018MNRASHaworth}
{Haworth}, T.~J., {Clarke}, C.~J., {Rahman}, W., {Winter}, A.~J., \&
  {Facchini}, S. 2018, \mnras, 481, 452, \dodoi{10.1093/mnras/sty2323}

\bibitem[{{Hayashi}(1981)}]{1981PThPSHayashi}
{Hayashi}, C. 1981, Progress of Theoretical Physics Supplement, 70, 35,
  \dodoi{10.1143/PTPS.70.35}

\bibitem[{{Hellary} \& {Nelson}(2012)}]{2012MNRASHellaryNelson}
{Hellary}, P., \& {Nelson}, R.~P. 2012, \mnras, 419, 2737,
  \dodoi{10.1111/j.1365-2966.2011.19815.x}

\bibitem[{{Helled} {et~al.}(2020){Helled}, {Mazzola}, \& {Redmer}}]{2020Helled}
{Helled}, R., {Mazzola}, G., \& {Redmer}, R. 2020, Nature Reviews Physics, 2,
  562, \dodoi{10.1038/s42254-020-0223-3}

\bibitem[{{Helled} {et~al.}(2022){Helled}, {Stevenson}, {Lunine}, {Bolton},
  {Nettelmann}, {Atreya}, {Guillot}, {Militzer}, {Miguel}, \&
  {Hubbard}}]{Helled2022Icarus}
{Helled}, R., {Stevenson}, D.~J., {Lunine}, J.~I., {et~al.} 2022, \icarus, 378,
  114937, \dodoi{10.1016/j.icarus.2022.114937}

\bibitem[{{Hollenbach} {et~al.}(1994){Hollenbach}, {Johnstone}, {Lizano}, \&
  {Shu}}]{1994ApJHollenbach}
{Hollenbach}, D., {Johnstone}, D., {Lizano}, S., \& {Shu}, F. 1994, \apj, 428,
  654, \dodoi{10.1086/174276}

\bibitem[{{Howard} {et~al.}(2010){Howard}, {Marcy}, {Johnson}, {Fischer},
  {Wright}, {Isaacson}, {Valenti}, {Anderson}, {Lin}, \&
  {Ida}}]{2010ScienceHoward}
{Howard}, A.~W., {Marcy}, G.~W., {Johnson}, J.~A., {et~al.} 2010, Science, 330,
  653, \dodoi{10.1126/science.1194854}

\bibitem[{{Howard} \& {Guillot}(2023)}]{2023HowardGuillot}
{Howard}, S., \& {Guillot}, T. 2023, \aap, 672, L1,
  \dodoi{10.1051/0004-6361/202244851}

\bibitem[{{Huang} {et~al.}(2018){Huang}, {Andrews}, {Dullemond}, {Isella},
  {P{\'e}rez}, {Guzm{\'a}n}, {{\"O}berg}, {Zhu}, {Zhang}, {Bai}, {Benisty},
  {Birnstiel}, {Carpenter}, {Hughes}, {Ricci}, {Weaver}, \&
  {Wilner}}]{HuangAndrews2018}
{Huang}, J., {Andrews}, S.~M., {Dullemond}, C.~P., {et~al.} 2018, \apjl, 869,
  L42, \dodoi{10.3847/2041-8213/aaf740}

\bibitem[{{Hueso} \& {Guillot}(2005)}]{2005A&AHueso}
{Hueso}, R., \& {Guillot}, T. 2005, \aap, 442, 703,
  \dodoi{10.1051/0004-6361:20041905}

\bibitem[{{Hutchison} {et~al.}(2016){Hutchison}, {Laibe}, \&
  {Maddison}}]{Hutchison2016}
{Hutchison}, M.~A., {Laibe}, G., \& {Maddison}, S.~T. 2016, \mnras, 463, 2725,
  \dodoi{10.1093/mnras/stw2191}

\bibitem[{{Ida}(1990)}]{1990IcarusIda}
{Ida}, S. 1990, \icarus, 88, 129, \dodoi{10.1016/0019-1035(90)90182-9}

\bibitem[{{Ida} \& {Lin}(2004)}]{2004ApJIda1}
{Ida}, S., \& {Lin}, D.~N.~C. 2004, \apj, 604, 388, \dodoi{10.1086/381724}

\bibitem[{{Ida} \& {Lin}(2005)}]{2005ApJIdaLin}
---. 2005, \apj, 626, 1045, \dodoi{10.1086/429953}

\bibitem[{{Ida} \& {Makino}(1993)}]{1993IcarusIdaMakino}
{Ida}, S., \& {Makino}, J. 1993, \icarus, 106, 210,
  \dodoi{10.1006/icar.1993.1167}

\bibitem[{{Ikoma} {et~al.}(2000){Ikoma}, {Nakazawa}, \& {Emori}}]{2000ApJIkoma}
{Ikoma}, M., {Nakazawa}, K., \& {Emori}, H. 2000, \apj, 537, 1013,
  \dodoi{10.1086/309050}

\bibitem[{{Inaba} \& {Ikoma}(2003)}]{2003A&AInaba}
{Inaba}, S., \& {Ikoma}, M. 2003, \aap, 410, 711,
  \dodoi{10.1051/0004-6361:20031248}

\bibitem[{{Inaba} {et~al.}(2001){Inaba}, {Tanaka}, {Nakazawa}, {Wetherill}, \&
  {Kokubo}}]{2001IcarusInaba}
{Inaba}, S., {Tanaka}, H., {Nakazawa}, K., {Wetherill}, G.~W., \& {Kokubo}, E.
  2001, \icarus, 149, 235, \dodoi{10.1006/icar.2000.6533}

\bibitem[{{Jiang} \& {Ormel}(2023)}]{JiangOrmel2023}
{Jiang}, H., \& {Ormel}, C.~W. 2023, \mnras, 518, 3877,
  \dodoi{10.1093/mnras/stac3275}

\bibitem[{{Jim{\'e}nez} \& {Masset}(2017)}]{2017MNRASJimenezMasset}
{Jim{\'e}nez}, M.~A., \& {Masset}, F.~S. 2017, \mnras, 471, 4917,
  \dodoi{10.1093/mnras/stx1946}

\bibitem[{{Jin} \& {Mordasini}(2018)}]{2018ApJJin}
{Jin}, S., \& {Mordasini}, C. 2018, \apj, 853, 163,
  \dodoi{10.3847/1538-4357/aa9f1e}

\bibitem[{{Jin} {et~al.}(2014){Jin}, {Mordasini}, {Parmentier}, {van Boekel},
  {Henning}, \& {Ji}}]{2014ApJJin}
{Jin}, S., {Mordasini}, C., {Parmentier}, V., {et~al.} 2014, \apj, 795, 65,
  \dodoi{10.1088/0004-637X/795/1/65}

\bibitem[{{Johansen} {et~al.}(2006){Johansen}, {Henning}, \&
  {Klahr}}]{Johansen2006a}
{Johansen}, A., {Henning}, T., \& {Klahr}, H. 2006, \apj, 643, 1219,
  \dodoi{10.1086/502968}

\bibitem[{Johansen \& Lambrechts(2017)}]{Johansen2017}
Johansen, A., \& Lambrechts, M. 2017, Annual Review of Earth and Planetary
  Sciences, 45, 359, \dodoi{10.1146/annurev-earth-063016-020226}

\bibitem[{{Johansen} \& {Youdin}(2007)}]{Johansen2007}
{Johansen}, A., \& {Youdin}, A. 2007, \apj, 662, 627, \dodoi{10.1086/516730}

\bibitem[{{Juri{\'c}} \& {Tremaine}(2008)}]{2008ApJJuric}
{Juri{\'c}}, M., \& {Tremaine}, S. 2008, \apj, 686, 603, \dodoi{10.1086/590047}

\bibitem[{{Kanagawa} {et~al.}(2017){Kanagawa}, {Tanaka}, {Muto}, \&
  {Tanigawa}}]{2017PASJKanagawa}
{Kanagawa}, K.~D., {Tanaka}, H., {Muto}, T., \& {Tanigawa}, T. 2017, \pasj, 69,
  97, \dodoi{10.1093/pasj/psx114}

\bibitem[{{Kanagawa} {et~al.}(2018){Kanagawa}, {Tanaka}, \&
  {Szuszkiewicz}}]{2018ApJKanagawa}
{Kanagawa}, K.~D., {Tanaka}, H., \& {Szuszkiewicz}, E. 2018, \apj, 861, 140,
  \dodoi{10.3847/1538-4357/aac8d9}

\bibitem[{{Kant}(1755)}]{Kant1755}
{Kant}, I. 1755, {Allgemeine Naturgeschichte und Theorie des Himmels}

\bibitem[{{Kaufmann} \& {Alibert}(2023)}]{KaufmannAlibert2023}
{Kaufmann}, N., \& {Alibert}, Y. 2023, \aap, 676, A46,
  \dodoi{10.1051/0004-6361/202345901}

\bibitem[{{Keppler} {et~al.}(2018){Keppler}, {Benisty}, {M{\"u}ller},
  {Henning}, {van Boekel}, {Cantalloube}, {Ginski}, {van Holstein}, {Maire},
  {Pohl}, {Samland}, {Avenhaus}, {Baudino}, {Boccaletti}, {de Boer},
  {Bonnefoy}, {Chauvin}, {Desidera}, {Langlois}, {Lazzoni}, {Marleau},
  {Mordasini}, {Pawellek}, {Stolker}, {Vigan}, {Zurlo}, {Birnstiel},
  {Brandner}, {Feldt}, {Flock}, {Girard}, {Gratton}, {Hagelberg}, {Isella},
  {Janson}, {Juhasz}, {Kemmer}, {Kral}, {Lagrange}, {Launhardt}, {Matter},
  {M{\'e}nard}, {Milli}, {Molli{\`e}re}, {Olofsson}, {P{\'e}rez}, {Pinilla},
  {Pinte}, {Quanz}, {Schmidt}, {Udry}, {Wahhaj}, {Williams}, {Buenzli},
  {Cudel}, {Dominik}, {Galicher}, {Kasper}, {Lannier}, {Mesa}, {Mouillet},
  {Peretti}, {Perrot}, {Salter}, {Sissa}, {Wildi}, {Abe}, {Antichi},
  {Augereau}, {Baruffolo}, {Baudoz}, {Bazzon}, {Beuzit}, {Blanchard}, {Brems},
  {Buey}, {De Caprio}, {Carbillet}, {Carle}, {Cascone}, {Cheetham}, {Claudi},
  {Costille}, {Delboulb{\'e}}, {Dohlen}, {Fantinel}, {Feautrier}, {Fusco},
  {Giro}, {Gluck}, {Gry}, {Hubin}, {Hugot}, {Jaquet}, {Le Mignant}, {Llored},
  {Madec}, {Magnard}, {Martinez}, {Maurel}, {Meyer}, {M{\"o}ller-Nilsson},
  {Moulin}, {Mugnier}, {Orign{\'e}}, {Pavlov}, {Perret}, {Petit}, {Pragt},
  {Puget}, {Rabou}, {Ramos}, {Rigal}, {Rochat}, {Roelfsema}, {Rousset}, {Roux},
  {Salasnich}, {Sauvage}, {Sevin}, {Soenke}, {Stadler}, {Suarez}, {Turatto}, \&
  {Weber}}]{2018AAKeppler}
{Keppler}, M., {Benisty}, M., {M{\"u}ller}, A., {et~al.} 2018, \aap, 617, A44,
  \dodoi{10.1051/0004-6361/201832957}

\bibitem[{{Kessler} \& {Alibert}(2023)}]{KesslerAlibert2023A&A}
{Kessler}, A., \& {Alibert}, Y. 2023, \aap, 674, A144,
  \dodoi{10.1051/0004-6361/202245641}

\bibitem[{Kimura \& Ikoma(2020)}]{Kimura2020}
Kimura, T., \& Ikoma, M. 2020, Monthly Notices of the Royal Astronomical
  Society, 496, 3755, \dodoi{10.1093/mnras/staa1778}

\bibitem[{{Kimura} \& {Ikoma}(2022)}]{KimuraIkoma2022NatAs}
{Kimura}, T., \& {Ikoma}, M. 2022, Nature Astronomy, 6, 1296,
  \dodoi{10.1038/s41550-022-01781-1}

\bibitem[{{Kite} {et~al.}(2019){Kite}, {Fegley}, {Schaefer}, \&
  {Ford}}]{kite2019}
{Kite}, E.~S., {Fegley}, Bruce, J., {Schaefer}, L., \& {Ford}, E.~B. 2019,
  \apjl, 887, L33, \dodoi{10.3847/2041-8213/ab59d9}

\bibitem[{{Kitzmann} {et~al.}(2024){Kitzmann}, {Stock}, \&
  {Patzer}}]{Kitzmann2024MNRAS}
{Kitzmann}, D., {Stock}, J.~W., \& {Patzer}, A. B.~C. 2024, \mnras, 527, 7263,
  \dodoi{10.1093/mnras/stad3515}

\bibitem[{Klahr \& Bodenheimer(2006)}]{Klahr2006a}
Klahr, H., \& Bodenheimer, P. 2006, The Astrophysical Journal, 639, 432,
  \dodoi{10.1086/498928}

\bibitem[{{Klahr} \& {Schreiber}(2020)}]{2020ApJKlahr}
{Klahr}, H., \& {Schreiber}, A. 2020, \apj, 901, 54,
  \dodoi{10.3847/1538-4357/abac58}

\bibitem[{{Kleine} {et~al.}(2002){Kleine}, {M{\"u}nker}, {Mezger}, \&
  {Palme}}]{2002NaturKleine}
{Kleine}, T., {M{\"u}nker}, C., {Mezger}, K., \& {Palme}, H. 2002, \nat, 418,
  952, \dodoi{10.1038/nature00982}

\bibitem[{{Kleine} {et~al.}(2009){Kleine}, {Touboul}, {Bourdon}, {Nimmo},
  {Mezger}, {Palme}, {Jacobsen}, {Yin}, \& {Halliday}}]{Kleine2009}
{Kleine}, T., {Touboul}, M., {Bourdon}, B., {et~al.} 2009, Geochimica et
  Cosmochimica Acta, 73, 5150, \dodoi{10.1016/j.gca.2008.11.047}

\bibitem[{{Kley} \& {Nelson}(2012)}]{2012ARAAKleyNelson}
{Kley}, W., \& {Nelson}, R.~P. 2012, \araa, 50, 211,
  \dodoi{10.1146/annurev-astro-081811-125523}

\bibitem[{Kobayashi \& Tanaka(2021)}]{Kobayashi2021}
Kobayashi, H., \& Tanaka, H. 2021, The Astrophysical Journal, 922, 16,
  \dodoi{10.3847/1538-4357/ac289c}

\bibitem[{{Kobayashi} {et~al.}(2016){Kobayashi}, {Tanaka}, \&
  {Okuzumi}}]{2016ApJKobayashi}
{Kobayashi}, H., {Tanaka}, H., \& {Okuzumi}, S. 2016, \apj, 817, 105,
  \dodoi{10.3847/0004-637X/817/2/105}

\bibitem[{{Krapp} {et~al.}(2019){Krapp}, {Ben{\'\i}tez-Llambay}, {Gressel}, \&
  {Pessah}}]{Krapp2019}
{Krapp}, L., {Ben{\'\i}tez-Llambay}, P., {Gressel}, O., \& {Pessah}, M.~E.
  2019, \apjl, 878, L30, \dodoi{10.3847/2041-8213/ab2596}

\bibitem[{{Kratter} \& {Lodato}(2016)}]{Kratter2016}
{Kratter}, K., \& {Lodato}, G. 2016, \araa, 54, 271,
  \dodoi{10.1146/annurev-astro-081915-023307}

\bibitem[{Krijt {et~al.}(2020)Krijt, Bosman, Zhang, Schwarz, Ciesla, \&
  Bergin}]{Krijt2020}
Krijt, S., Bosman, A.~D., Zhang, K., {et~al.} 2020, The Astrophysical Journal,
  899, 134, \dodoi{10.3847/1538-4357/aba75d}

\bibitem[{{Kruijer} {et~al.}(2017){Kruijer}, {Burkhardt}, {Budde}, \&
  {Kleine}}]{2017PNASKruijer}
{Kruijer}, T.~S., {Burkhardt}, C., {Budde}, G., \& {Kleine}, T. 2017, \pnas,
  114, 6712, \dodoi{10.1073/pnas.1704461114}

\bibitem[{{Lagrange} {et~al.}(2023){Lagrange}, {Philipot}, {Rubini}, {Meunier},
  {Kiefer}, {Kervella}, {Delorme}, \& {Beust}}]{Lagrange2023}
{Lagrange}, A.~M., {Philipot}, F., {Rubini}, P., {et~al.} 2023, \aap, 677, A71,
  \dodoi{10.1051/0004-6361/202346165}

\bibitem[{{Lagrange} {et~al.}(2010){Lagrange}, {Bonnefoy}, {Chauvin}, {Apai},
  {Ehrenreich}, {Boccaletti}, {Gratadour}, {Rouan}, {Mouillet}, {Lacour}, \&
  {Kasper}}]{2010ScienceLagrange}
{Lagrange}, A.~M., {Bonnefoy}, M., {Chauvin}, G., {et~al.} 2010, \sci, 329, 57,
  \dodoi{10.1126/science.1187187}

\bibitem[{{Lambrechts} \& {Johansen}(2012)}]{2012AALambrechtsJohansen}
{Lambrechts}, M., \& {Johansen}, A. 2012, \aap, 544, A32,
  \dodoi{10.1051/0004-6361/201219127}

\bibitem[{{Lambrechts} {et~al.}(2014){Lambrechts}, {Johansen}, \&
  {Morbidelli}}]{2014AALambrechts}
{Lambrechts}, M., {Johansen}, A., \& {Morbidelli}, A. 2014, \aap, 572, A35,
  \dodoi{10.1051/0004-6361/201423814}

\bibitem[{{Lambrechts} {et~al.}(2019){Lambrechts}, {Morbidelli}, {Jacobson},
  {Johansen}, {Bitsch}, {Izidoro}, \& {Raymond}}]{2019AALambrechts}
{Lambrechts}, M., {Morbidelli}, A., {Jacobson}, S.~A., {et~al.} 2019, \aap,
  627, A83, \dodoi{10.1051/0004-6361/201834229}

\bibitem[{{Lau} {et~al.}(2022){Lau}, {Dr{\k{a}}{\.z}kowska}, {Stammler},
  {Birnstiel}, \& {Dullemond}}]{2022AALau}
{Lau}, T. C.~H., {Dr{\k{a}}{\.z}kowska}, J., {Stammler}, S.~M., {Birnstiel},
  T., \& {Dullemond}, C.~P. 2022, \aap, 668, A170,
  \dodoi{10.1051/0004-6361/202244864}

\bibitem[{Lega {et~al.}(2014)Lega, Crida, Bitsch, \& Morbidelli}]{Lega2014}
Lega, E., Crida, A., Bitsch, B., \& Morbidelli, A. 2014, Monthly Notices of the
  Royal Astronomical Society, 440, 683, \dodoi{10.1093/mnras/stu304}

\bibitem[{{Lega} {et~al.}(2022){Lega}, {Morbidelli}, {Nelson}, {Ramos},
  {Crida}, {B{\'e}thune}, \& {Batygin}}]{Lega2022}
{Lega}, E., {Morbidelli}, A., {Nelson}, R.~P., {et~al.} 2022, \aap, 658, A32,
  \dodoi{10.1051/0004-6361/202141675}

\bibitem[{{Lenz} {et~al.}(2019){Lenz}, {Klahr}, \& {Birnstiel}}]{2019ApJLenz}
{Lenz}, C.~T., {Klahr}, H., \& {Birnstiel}, T. 2019, \apj, 874, 36,
  \dodoi{10.3847/1538-4357/ab05d9}

\bibitem[{{Lesur} {et~al.}(2022){Lesur}, {Ercolano}, {Flock}, {Lin}, {Yang},
  {Barranco}, {Benitez-Llambay}, {Goodman}, {Johansen}, {Klahr}, {Laibe},
  {Lyra}, {Marcus}, {Nelson}, {Squire}, {Simon}, {Turner}, {Umurhan}, \&
  {Youdin}}]{2022PPVIILesur}
{Lesur}, G., {Ercolano}, B., {Flock}, M., {et~al.} 2022, arXiv e-prints,
  arXiv:2203.09821.
\newblock \doarXiv{2203.09821}

\bibitem[{{Levison} {et~al.}(2008){Levison}, {Morbidelli}, {Van Laerhoven},
  {Gomes}, \& {Tsiganis}}]{Levison2008Icarus}
{Levison}, H.~F., {Morbidelli}, A., {Van Laerhoven}, C., {Gomes}, R., \&
  {Tsiganis}, K. 2008, \icarus, 196, 258, \dodoi{10.1016/j.icarus.2007.11.035}

\bibitem[{Levison {et~al.}(2010)Levison, Thommes, \&
  Duncan}]{levisonthommes2010}
Levison, H.~F., Thommes, E.~W., \& Duncan, M.~J. 2010, \apj, 139, 1297

\bibitem[{Li {et~al.}(2021)Li, Bergin, Blake, Ciesla, \& Hirschmann}]{Li2021b}
Li, J., Bergin, E.~A., Blake, G.~A., Ciesla, F.~J., \& Hirschmann, M.~M. 2021,
  Science Advances, 7, eabd3632, \dodoi{10.1126/sciadv.abd3632}

\bibitem[{{Li} \& {Youdin}(2021)}]{Li2021}
{Li}, R., \& {Youdin}, A.~N. 2021, \apj, 919, 107,
  \dodoi{10.3847/1538-4357/ac0e9f}

\bibitem[{{Lichtenberg} {et~al.}(2019){Lichtenberg}, {Golabek}, {Burn},
  {Meyer}, {Alibert}, {Gerya}, \& {Mordasini}}]{Lichtenberg2019}
{Lichtenberg}, T., {Golabek}, G.~J., {Burn}, R., {et~al.} 2019, Nature
  Astronomy, 3, 307, \dodoi{10.1038/s41550-018-0688-5}

\bibitem[{Lichtenberg {et~al.}(2023)Lichtenberg, Schaefer, Nakajima, \&
  Fischer}]{Lichtenberg2023}
Lichtenberg, T., Schaefer, L.~K., Nakajima, M., \& Fischer, R.~A. 2023, in
  Protostars and {{Planets VII}}, {{Astronomical Society}} of the {{Pacific
  Conference Series}}, ed. {Inutsuka, S. and Aikawa, Y. and Muto, T. and
  Tomida, K. and Tamura, M.}, Vol. 534, {Kyoto}, 907,
  \dodoi{10.48550/arXiv.2203.10023}

\bibitem[{{Lin} \& {Papaloizou}(1986)}]{1986ApJLinPapaloizouA}
{Lin}, D.~N.~C., \& {Papaloizou}, J. 1986, \apj, 307, 395,
  \dodoi{10.1086/164426}

\bibitem[{Lin \& Youdin(2015)}]{Lin2015}
Lin, M.-K., \& Youdin, A.~N. 2015, The Astrophysical Journal, 811, 17,
  \dodoi{10.1088/0004-637X/811/1/17}

\bibitem[{{Linder} {et~al.}(2019){Linder}, {Mordasini}, {Molli{\`e}re},
  {Marleau}, {Malik}, {Quanz}, \& {Meyer}}]{2019AALinder}
{Linder}, E.~F., {Mordasini}, C., {Molli{\`e}re}, P., {et~al.} 2019, \aap, 623,
  A85, \dodoi{10.1051/0004-6361/201833873}

\bibitem[{{Lissauer}(1993)}]{Lissauer1993ARAA}
{Lissauer}, J.~J. 1993, \araa, 31, 129,
  \dodoi{10.1146/annurev.aa.31.090193.001021}

\bibitem[{{Lissauer} {et~al.}(2023){Lissauer}, {Batalha}, \&
  {Borucki}}]{Lissauer2023PPVII}
{Lissauer}, J.~J., {Batalha}, N.~M., \& {Borucki}, W.~J. 2023, in Astronomical
  Society of the Pacific Conference Series, Vol. 534, Protostars and Planets
  VII, ed. S.~{Inutsuka}, Y.~{Aikawa}, T.~{Muto}, K.~{Tomida}, \& M.~{Tamura},
  839, \dodoi{10.48550/arXiv.2311.04981}

\bibitem[{{Lissauer} {et~al.}(2009){Lissauer}, {Hubickyj}, {D'Angelo}, \&
  {Bodenheimer}}]{2009IcarusLissauer}
{Lissauer}, J.~J., {Hubickyj}, O., {D'Angelo}, G., \& {Bodenheimer}, P. 2009,
  \icarus, 199, 338, \dodoi{10.1016/j.icarus.2008.10.004}

\bibitem[{Lissauer \& Stewart(1993)}]{Lissauer1993}
Lissauer, J.~J., \& Stewart, G.~R. 1993, in Protostars and Planets {{III}}, ed.
  E.~H. Levy \& J.~I. Lunine ({University of Arizona Press}), 1061--1088.
\newblock \url{http://adsabs.harvard.edu/abs/1993prpl.conf.1061L}

\bibitem[{{Lissauer} {et~al.}(2011){Lissauer}, {Ragozzine}, {Fabrycky},
  {Steffen}, {Ford}, {Jenkins}, {Shporer}, {Holman}, {Rowe}, {Quintana},
  {Batalha}, {Borucki}, {Bryson}, {Caldwell}, {Carter}, {Ciardi}, {Dunham},
  {Fortney}, {Gautier}, {Howell}, {Koch}, {Latham}, {Marcy}, {Morehead}, \&
  {Sasselov}}]{2011ApJSLissauer}
{Lissauer}, J.~J., {Ragozzine}, D., {Fabrycky}, D.~C., {et~al.} 2011, \apjs,
  197, 8, \dodoi{10.1088/0067-0049/197/1/8}

\bibitem[{Lorek \& Johansen(2022)}]{Lorek2022}
Lorek, S., \& Johansen, A. 2022, Astronomy \& Astrophysics, 666, A108,
  \dodoi{10.1051/0004-6361/202244333}

\bibitem[{{Lovis} {et~al.}(2006){Lovis}, {Mayor}, {Pepe}, {Alibert}, {Benz},
  {Bouchy}, {Correia}, {Laskar}, {Mordasini}, {Queloz}, {Santos}, {Udry},
  {Bertaux}, \& {Sivan}}]{2006NatureLovis}
{Lovis}, C., {Mayor}, M., {Pepe}, F., {et~al.} 2006, \nat, 441, 305,
  \dodoi{10.1038/nature04828}

\bibitem[{{Lucy} \& {Sweeney}(1971)}]{Lucy1971AJ}
{Lucy}, L.~B., \& {Sweeney}, M.~A. 1971, \aj, 76, 544, \dodoi{10.1086/111159}

\bibitem[{{L{\"u}st}(1952)}]{1952ZNatALust}
{L{\"u}st}, R. 1952, Zeitschrift Naturforschung Teil A, 7, 87,
  \dodoi{10.1515/zna-1952-0118}

\bibitem[{{Lynden-Bell} \& {Pringle}(1974)}]{1974NMRASLyndenBellPringle}
{Lynden-Bell}, D., \& {Pringle}, J.~E. 1974, \mnras, 168, 603,
  \dodoi{10.1093/mnras/168.3.603}

\bibitem[{{Lyra} {et~al.}(2010){Lyra}, {Paardekooper}, \& {Mac
  Low}}]{2010ApJLyra}
{Lyra}, W., {Paardekooper}, S.-J., \& {Mac Low}, M.-M. 2010, \apjl, 715, L68,
  \dodoi{10.1088/2041-8205/715/2/L68}

\bibitem[{Machida {et~al.}(2010)Machida, Kokubo, Inutsuka, \&
  Matsumoto}]{Machida2010}
Machida, M.~N., Kokubo, E., Inutsuka, S.-i., \& Matsumoto, T. 2010, Monthly
  Notices of the Royal Astronomical Society, 405, 1227,
  \dodoi{10.1111/j.1365-2966.2010.16527.x}

\bibitem[{{Macintosh} {et~al.}(2015){Macintosh}, {Graham}, {Barman}, {De Rosa},
  {Konopacky}, {Marley}, {Marois}, {Nielsen}, {Pueyo}, {Rajan}, {Rameau},
  {Saumon}, {Wang}, {Patience}, {Ammons}, {Arriaga}, {Artigau}, {Beckwith},
  {Brewster}, {Bruzzone}, {Bulger}, {Burningham}, {Burrows}, {Chen}, {Chiang},
  {Chilcote}, {Dawson}, {Dong}, {Doyon}, {Draper}, {Duch{\^e}ne}, {Esposito},
  {Fabrycky}, {Fitzgerald}, {Follette}, {Fortney}, {Gerard}, {Goodsell},
  {Greenbaum}, {Hibon}, {Hinkley}, {Cotten}, {Hung}, {Ingraham},
  {Johnson-Groh}, {Kalas}, {Lafreniere}, {Larkin}, {Lee}, {Line}, {Long},
  {Maire}, {Marchis}, {Matthews}, {Max}, {Metchev}, {Millar-Blanchaer},
  {Mittal}, {Morley}, {Morzinski}, {Murray-Clay}, {Oppenheimer}, {Palmer},
  {Patel}, {Perrin}, {Poyneer}, {Rafikov}, {Rantakyr{\"o}}, {Rice}, {Rojo},
  {Rudy}, {Ruffio}, {Ruiz}, {Sadakuni}, {Saddlemyer}, {Salama}, {Savransky},
  {Schneider}, {Sivaramakrishnan}, {Song}, {Soummer}, {Thomas}, {Vasisht},
  {Wallace}, {Ward-Duong}, {Wiktorowicz}, {Wolff}, \&
  {Zuckerman}}]{2015ScienceMacintosh}
{Macintosh}, B., {Graham}, J.~R., {Barman}, T., {et~al.} 2015, \sci, 350, 64,
  \dodoi{10.1126/science.aac5891}

\bibitem[{{Madhusudhan} {et~al.}(2014){Madhusudhan}, {Amin}, \&
  {Kennedy}}]{Madhusudhan2014}
{Madhusudhan}, N., {Amin}, M.~A., \& {Kennedy}, G.~M. 2014, \apjl, 794, L12,
  \dodoi{10.1088/2041-8205/794/1/L12}

\bibitem[{Mah {et~al.}(2023)Mah, Bitsch, Pascucci, \& Henning}]{Mah2023}
Mah, J., Bitsch, B., Pascucci, I., \& Henning, T. 2023, Astronomy and
  Astrophysics, 677, L7, \dodoi{10.1051/0004-6361/202347169}

\bibitem[{{Manara} {et~al.}(2022){Manara}, {Ansdell}, {Rosotti}, {Hughes},
  {Armitage}, {Lodato}, \& {Williams}}]{2022PPVIIManara}
{Manara}, C.~F., {Ansdell}, M., {Rosotti}, G.~P., {et~al.} 2022, arXiv
  e-prints, arXiv:2203.09930.
\newblock \doarXiv{2203.09930}

\bibitem[{{Manger} \& {Klahr}(2018)}]{Manger2018}
{Manger}, N., \& {Klahr}, H. 2018, \mnras, 480, 2125,
  \dodoi{10.1093/mnras/sty1909}

\bibitem[{Mann {et~al.}(2016)Mann, Gaidos, Mace, Johnson, Bowler, LaCourse,
  Jacobs, Vanderburg, Kraus, Kaplan, \& Jaffe}]{Mann2016}
Mann, A.~W., Gaidos, E., Mace, G.~N., {et~al.} 2016, The Astrophysical Journal,
  818, 46, \dodoi{10.3847/0004-637X/818/1/46}

\bibitem[{{Marboeuf} {et~al.}(2014){Marboeuf}, {Thiabaud}, {Alibert}, {Cabral},
  \& {Benz}}]{2014AAMarboeufA}
{Marboeuf}, U., {Thiabaud}, A., {Alibert}, Y., {Cabral}, N., \& {Benz}, W.
  2014, \aap, 570, A35, \dodoi{10.1051/0004-6361/201322207}

\bibitem[{{Marcy} {et~al.}(2005){Marcy}, {Butler}, {Fischer}, {Vogt}, {Wright},
  {Tinney}, \& {Jones}}]{2005PThPSMarcy}
{Marcy}, G., {Butler}, R.~P., {Fischer}, D., {et~al.} 2005, Progress of
  Theoretical Physics Supplement, 158, 24, \dodoi{10.1143/PTPS.158.24}

\bibitem[{{Marleau} {et~al.}(2019){Marleau}, {Coleman}, {Leleu}, \&
  {Mordasini}}]{2019AAMarleau}
{Marleau}, G.-D., {Coleman}, G. A.~L., {Leleu}, A., \& {Mordasini}, C. 2019,
  \aap, 624, A20, \dodoi{10.1051/0004-6361/201833597}

\bibitem[{{Marleau} {et~al.}(2017){Marleau}, {Klahr}, {Kuiper}, \&
  {Mordasini}}]{2017ApJMarleau}
{Marleau}, G.-D., {Klahr}, H., {Kuiper}, R., \& {Mordasini}, C. 2017, \apj,
  836, 221, \dodoi{10.3847/1538-4357/836/2/221}

\bibitem[{{Marleau} {et~al.}(2023){Marleau}, {Kuiper}, {B{\'e}thune}, \&
  {Mordasini}}]{Marleau2023}
{Marleau}, G.-D., {Kuiper}, R., {B{\'e}thune}, W., \& {Mordasini}, C. 2023,
  \apj, 952, 89, \dodoi{10.3847/1538-4357/accf12}

\bibitem[{{Marleau} {et~al.}(2022){Marleau}, {Aoyama}, {Kuiper}, {Follette},
  {Turner}, {Cugno}, {Manara}, {Haffert}, {Kitzmann}, {Ringqvist}, {Wagner},
  {van Boekel}, {Sallum}, {Janson}, {Schmidt}, {Venuti}, {Lovis}, \&
  {Mordasini}}]{Marleau2022}
{Marleau}, G.~D., {Aoyama}, Y., {Kuiper}, R., {et~al.} 2022, \aap, 657, A38,
  \dodoi{10.1051/0004-6361/202037494}

\bibitem[{{Marois} {et~al.}(2008){Marois}, {Macintosh}, {Barman}, {Zuckerman},
  {Song}, {Patience}, {Lafreni{\`e}re}, \& {Doyon}}]{2008ScienceMarois}
{Marois}, C., {Macintosh}, B., {Barman}, T., {et~al.} 2008, Science, 322, 1348,
  \dodoi{10.1126/science.1166585}

\bibitem[{{Marois} {et~al.}(2010){Marois}, {Zuckerman}, {Konopacky},
  {Macintosh}, \& {Barman}}]{2010NatureMarois}
{Marois}, C., {Zuckerman}, B., {Konopacky}, Q.~M., {Macintosh}, B., \&
  {Barman}, T. 2010, \nat, 468, 1080, \dodoi{10.1038/nature09684}

\bibitem[{{Masset}(2001)}]{2001ApJMasset}
{Masset}, F.~S. 2001, \apj, 558, 453, \dodoi{10.1086/322446}

\bibitem[{Masset(2017)}]{Masset2017}
Masset, F.~S. 2017, Monthly Notices of the Royal Astronomical Society, 472,
  4204, \dodoi{10.1093/mnras/stx2271}

\bibitem[{Mathis {et~al.}(1977)Mathis, Rumpl, \& Nordsieck}]{Mathis1977}
Mathis, J.~S., Rumpl, W., \& Nordsieck, K.~H. 1977, The Astrophysical Journal,
  217, 425, \dodoi{10.1086/155591}

\bibitem[{{Matsumura} {et~al.}(2007){Matsumura}, {Pudritz}, \&
  {Thommes}}]{Matsumura2007}
{Matsumura}, S., {Pudritz}, R.~E., \& {Thommes}, E.~W. 2007, \apj, 660, 1609,
  \dodoi{10.1086/513175}

\bibitem[{{Mayor} \& {Queloz}(1995)}]{1995NatureMayorQueloz}
{Mayor}, M., \& {Queloz}, D. 1995, \nat, 378, 355, \dodoi{10.1038/378355a0}

\bibitem[{{Mayor} {et~al.}(2003){Mayor}, {Pepe}, {Queloz}, {Bouchy},
  {Rupprecht}, {Lo Curto}, {Avila}, {Benz}, {Bertaux}, {Bonfils}, {Dall},
  {Dekker}, {Delabre}, {Eckert}, {Fleury}, {Gilliotte}, {Gojak}, {Guzman},
  {Kohler}, {Lizon}, {Longinotti}, {Lovis}, {Megevand}, {Pasquini}, {Reyes},
  {Sivan}, {Sosnowska}, {Soto}, {Udry}, {van Kesteren}, {Weber}, \&
  {Weilenmann}}]{2003MsngrMayor}
{Mayor}, M., {Pepe}, F., {Queloz}, D., {et~al.} 2003, The Messenger, 114, 20

\bibitem[{{Mayor} {et~al.}(2011){Mayor}, {Marmier}, {Lovis}, {Udry},
  {S{\'e}gransan}, {Pepe}, {Benz}, {Bertaux}, {Bouchy}, {Dumusque}, {Lo Curto},
  {Mordasini}, {Queloz}, \& {Santos}}]{2011MayorArxiv}
{Mayor}, M., {Marmier}, M., {Lovis}, C., {et~al.} 2011, arXiv e-prints,
  arXiv:1109.2497.
\newblock \doarXiv{1109.2497}

\bibitem[{{McNally} {et~al.}(2019){McNally}, {Nelson}, {Paardekooper}, \&
  {Ben{\'\i}tez-Llambay}}]{McNally2019}
{McNally}, C.~P., {Nelson}, R.~P., {Paardekooper}, S.-J., \&
  {Ben{\'\i}tez-Llambay}, P. 2019, \mnras, 484, 728,
  \dodoi{10.1093/mnras/stz023}

\bibitem[{{Mezger} {et~al.}(2021){Mezger}, {Maltese}, \&
  {Vollstaedt}}]{Mezger2021}
{Mezger}, K., {Maltese}, A., \& {Vollstaedt}, H. 2021, \icarus, 365, 114497,
  \dodoi{10.1016/j.icarus.2021.114497}

\bibitem[{{Miguel} {et~al.}(2011){Miguel}, {Guilera}, \&
  {Brunini}}]{2011MNRASMiguel}
{Miguel}, Y., {Guilera}, O.~M., \& {Brunini}, A. 2011, \mnras, 417, 314,
  \dodoi{10.1111/j.1365-2966.2011.19264.x}

\bibitem[{{Miguel} {et~al.}(2022){Miguel}, {Bazot}, {Guillot}, {Howard},
  {Galanti}, {Kaspi}, {Hubbard}, {Militzer}, {Helled}, {Atreya}, {Connerney},
  {Durante}, {Kulowski}, {Lunine}, {Stevenson}, \& {Bolton}}]{Miguel2022}
{Miguel}, Y., {Bazot}, M., {Guillot}, T., {et~al.} 2022, \aap, 662, A18,
  \dodoi{10.1051/0004-6361/202243207}

\bibitem[{{Millholland} {et~al.}(2017){Millholland}, {Wang}, \&
  {Laughlin}}]{2017ApJMillholland}
{Millholland}, S., {Wang}, S., \& {Laughlin}, G. 2017, \apjl, 849, L33,
  \dodoi{10.3847/2041-8213/aa9714}

\bibitem[{{Miotello} {et~al.}(2022){Miotello}, {Kamp}, {Birnstiel}, {Cleeves},
  \& {Kataoka}}]{2022PPVIIMiotello}
{Miotello}, A., {Kamp}, I., {Birnstiel}, T., {Cleeves}, L.~I., \& {Kataoka}, A.
  2022, arXiv e-prints, arXiv:2203.09818.
\newblock \doarXiv{2203.09818}

\bibitem[{{Mishra} {et~al.}(2021){Mishra}, {Alibert}, {Leleu}, {Emsenhuber},
  {Mordasini}, {Burn}, {Udry}, \& {Benz}}]{2021AAMishra}
{Mishra}, L., {Alibert}, Y., {Leleu}, A., {et~al.} 2021, \aap, 656, A74,
  \dodoi{10.1051/0004-6361/202140761}

\bibitem[{{Mishra} {et~al.}(2023{\natexlab{a}}){Mishra}, {Alibert}, {Udry}, \&
  {Mordasini}}]{2023AAMishraA}
{Mishra}, L., {Alibert}, Y., {Udry}, S., \& {Mordasini}, C. 2023{\natexlab{a}},
  \aap, 670, A68, \dodoi{10.1051/0004-6361/202243751}

\bibitem[{{Mishra} {et~al.}(2023{\natexlab{b}}){Mishra}, {Alibert}, {Udry}, \&
  {Mordasini}}]{2023AAMishraB}
---. 2023{\natexlab{b}}, \aap, 670, A69, \dodoi{10.1051/0004-6361/202244705}

\bibitem[{{Mizuno}(1980)}]{1980PThPhMizuno}
{Mizuno}, H. 1980, Progress of Theoretical Physics, 64, 544,
  \dodoi{10.1143/PTP.64.544}

\bibitem[{{Moldenhauer} {et~al.}(2021){Moldenhauer}, {Kuiper}, {Kley}, \&
  {Ormel}}]{2021AAMoldenhauer}
{Moldenhauer}, T.~W., {Kuiper}, R., {Kley}, W., \& {Ormel}, C.~W. 2021, \aap,
  646, L11, \dodoi{10.1051/0004-6361/202040220}

\bibitem[{{Moldenhauer} {et~al.}(2022){Moldenhauer}, {Kuiper}, {Kley}, \&
  {Ormel}}]{2022AAMoldenhauer}
---. 2022, \aap, 661, A142, \dodoi{10.1051/0004-6361/202141955}

\bibitem[{{Molli{\`e}re} {et~al.}(2022){Molli{\`e}re}, {Molyarova}, {Bitsch},
  {Henning}, {Schneider}, {Kreidberg}, {Eistrup}, {Burn}, {Nasedkin},
  {Semenov}, {Mordasini}, {Schlecker}, {Schwarz}, {Lacour}, {Nowak}, \&
  {Schulik}}]{2022ApJMolliere}
{Molli{\`e}re}, P., {Molyarova}, T., {Bitsch}, B., {et~al.} 2022, \apj, 934,
  74, \dodoi{10.3847/1538-4357/ac6a56}

\bibitem[{{Morales} {et~al.}(2019){Morales}, {Mustill}, {Ribas}, {Davies},
  {Reiners}, {Bauer}, {Kossakowski}, {Herrero}, {Rodr{\'\i}guez},
  {L{\'o}pez-Gonz{\'a}lez}, {Rodr{\'\i}guez-L{\'o}pez}, {B{\'e}jar},
  {Gonz{\'a}lez-Cuesta}, {Luque}, {Pall{\'e}}, {Perger}, {Baroch}, {Johansen},
  {Klahr}, {Mordasini}, {Anglada-Escud{\'e}}, {Caballero},
  {Cort{\'e}s-Contreras}, {Dreizler}, {Lafarga}, {Nagel}, {Passegger},
  {Reffert}, {Rosich}, {Schweitzer}, {Tal-Or}, {Trifonov}, {Zechmeister},
  {Quirrenbach}, {Amado}, {Guenther}, {Hagen}, {Henning}, {Jeffers},
  {Kaminski}, {K{\"u}rster}, {Montes}, {Seifert}, {Abell{\'a}n}, {Abril},
  {Aceituno}, {Aceituno}, {Alonso-Floriano}, {Ammler-von Eiff}, {Antona},
  {Arroyo-Torres}, {Azzaro}, {Barrado}, {Becerril-Jarque}, {Ben{\'\i}tez},
  {Berdi{\~n}as}, {Bergond}, {Brinkm{\"o}ller}, {del Burgo}, {Burn},
  {Calvo-Ortega}, {Cano}, {C{\'a}rdenas}, {Guill{\'e}n}, {Carro}, {Casal},
  {Casanova}, {Casasayas-Barris}, {Chaturvedi}, {Cifuentes}, {Claret},
  {Colom{\'e}}, {Czesla}, {D{\'\i}ez-Alonso}, {Dorda}, {Emsenhuber},
  {Fern{\'a}ndez}, {Fern{\'a}ndez-Mart{\'\i}n}, {Ferro}, {Fuhrmeister},
  {Galad{\'\i}-Enr{\'\i}quez}, {Cava}, {Vargas}, {Garcia-Piquer}, {Gesa},
  {Gonz{\'a}lez-{\'A}lvarez}, {Hern{\'a}ndez}, {Gonz{\'a}lez-Peinado},
  {Gu{\`a}rdia}, {Guijarro}, {de Guindos}, {Hatzes}, {Hauschildt}, {Hedrosa},
  {Hermelo}, {Arabi}, {Otero}, {Hintz}, {Holgado}, {Huber}, {Huke}, {Johnson},
  {de Juan}, {Kehr}, {Kemmer}, {Kim}, {Kl{\"u}ter}, {Klutsch}, {Labarga},
  {Labiche}, {Lalitha}, {Lamp{\'o}n}, {Lara}, {Launhardt}, {L{\'a}zaro},
  {Lizon}, {Llamas}, {Lodieu}, {L{\'o}pez del Fresno}, {Salas},
  {L{\'o}pez-Santiago}, {Madinabeitia}, {Mall}, {Mancini}, {Mandel}, {Marfil},
  {Molina}, {Mart{\'\i}n}, {Mart{\'\i}n-Fern{\'a}ndez}, {Mart{\'\i}n-Ruiz},
  {Mart{\'\i}nez-Rodr{\'\i}guez}, {Marvin}, {Mirabet}, {Moya}, {Naranjo},
  {Nelson}, {Nortmann}, {Nowak}, {Ofir}, {Pascual}, {Pavlov}, {Pedraz},
  {Medialdea}, {P{\'e}rez-Calpena}, {Perryman}, {Rabaza}, {Ballesta}, {Rebolo},
  {Redondo}, {Rix}, {Rodler}, {Trinidad}, {Sabotta}, {Sadegi}, {Salz},
  {S{\'a}nchez-Blanco}, {Carrasco}, {S{\'a}nchez-L{\'o}pez}, {Sanz-Forcada},
  {Sarkis}, {Sarmiento}, {Sch{\"a}fer}, {Schlecker}, {Schmitt}, {Sch{\"o}fer},
  {Solano}, {Sota}, {Stahl}, {Stock}, {Stuber}, {St{\"u}rmer}, {Su{\'a}rez},
  {Tabernero}, {Tulloch}, {Veredas}, {Vico-Linares}, {Vilardell}, {Wagner},
  {Winkler}, {Wolthoff}, {Yan}, \& {Osorio}}]{2019ScienceMorales}
{Morales}, J.~C., {Mustill}, A.~J., {Ribas}, I., {et~al.} 2019, Science, 365,
  1441, \dodoi{10.1126/science.aax3198}

\bibitem[{{Morbidelli} {et~al.}(2009){Morbidelli}, {Bottke}, {Nesvorn{\'y}}, \&
  {Levison}}]{2009IcarusMorbidelli}
{Morbidelli}, A., {Bottke}, W.~F., {Nesvorn{\'y}}, D., \& {Levison}, H.~F.
  2009, \icarus, 204, 558, \dodoi{10.1016/j.icarus.2009.07.011}

\bibitem[{{Morbidelli} {et~al.}(2000){Morbidelli}, {Chambers}, {Lunine},
  {Petit}, {Robert}, {Valsecchi}, \& {Cyr}}]{2000MAPSMorbidelli}
{Morbidelli}, A., {Chambers}, J., {Lunine}, J.~I., {et~al.} 2000, Meteoritics
  and Planetary Science, 35, 1309, \dodoi{10.1111/j.1945-5100.2000.tb01518.x}

\bibitem[{{Morbidelli} {et~al.}(2015){Morbidelli}, {Lambrechts}, {Jacobson}, \&
  {Bitsch}}]{2015IcarusMorbidelli}
{Morbidelli}, A., {Lambrechts}, M., {Jacobson}, S., \& {Bitsch}, B. 2015,
  \icarus, 258, 418, \dodoi{10.1016/j.icarus.2015.06.003}

\bibitem[{{Mordasini}(2014)}]{2014AAMordasiniB}
{Mordasini}, C. 2014, \aap, 572, A118, \dodoi{10.1051/0004-6361/201423702}

\bibitem[{{Mordasini}(2018)}]{2018BookMordasini}
---. 2018, in Handbook of Exoplanets, ed. H.~J. Deeg \& J.~A. Belmonte
  (Springer Living Reference Work), 143, \dodoi{10.1007/978-3-319-55333-7_143}

\bibitem[{{Mordasini} {et~al.}(2006){Mordasini}, {Alibert}, \&
  {Benz}}]{2006ConfMordasini}
{Mordasini}, C., {Alibert}, Y., \& {Benz}, W. 2006, in Tenth Anniversary of 51
  Peg-b: Status of and prospects for hot Jupiter studies, ed. L.~{Arnold},
  F.~{Bouchy}, \& C.~{Moutou}, 84--86

\bibitem[{{Mordasini} {et~al.}(2012{\natexlab{a}}){Mordasini}, {Alibert},
  {Benz}, {Klahr}, \& {Henning}}]{2012A&AMordasiniA}
{Mordasini}, C., {Alibert}, Y., {Benz}, W., {Klahr}, H., \& {Henning}, T.
  2012{\natexlab{a}}, \aap, 541, A97, \dodoi{10.1051/0004-6361/201117350}

\bibitem[{{Mordasini} {et~al.}(2009){Mordasini}, {Alibert}, {Benz}, \&
  {Naef}}]{2009A&AMordasinib}
{Mordasini}, C., {Alibert}, Y., {Benz}, W., \& {Naef}, D. 2009, \aap, 501,
  1161, \dodoi{10.1051/0004-6361/200810697}

\bibitem[{{Mordasini} {et~al.}(2012{\natexlab{b}}){Mordasini}, {Alibert},
  {Georgy}, {Dittkrist}, {Klahr}, \& {Henning}}]{2012A&AMordasiniC}
{Mordasini}, C., {Alibert}, Y., {Georgy}, C., {et~al.} 2012{\natexlab{b}},
  \aap, 547, A112, \dodoi{10.1051/0004-6361/201118464}

\bibitem[{{Mordasini} {et~al.}(2012{\natexlab{c}}){Mordasini}, {Alibert},
  {Klahr}, \& {Henning}}]{2012A&AMordasiniB}
{Mordasini}, C., {Alibert}, Y., {Klahr}, H., \& {Henning}, T.
  2012{\natexlab{c}}, \aap, 547, A111, \dodoi{10.1051/0004-6361/201118457}

\bibitem[{{Mordasini} {et~al.}(2014){Mordasini}, {Klahr}, {Alibert}, {Miller},
  \& {Henning}}]{2014AAMordasiniA}
{Mordasini}, C., {Klahr}, H., {Alibert}, Y., {Miller}, N., \& {Henning}, T.
  2014, \aap, 566, A141, \dodoi{10.1051/0004-6361/201321479}

\bibitem[{{Mordasini} {et~al.}(2017){Mordasini}, {Marleau}, \&
  {Molli{\`e}re}}]{2017AAMordasini}
{Mordasini}, C., {Marleau}, G.~D., \& {Molli{\`e}re}, P. 2017, \aap, 608, A72,
  \dodoi{10.1051/0004-6361/201630077}

\bibitem[{{Mordasini} {et~al.}(2016){Mordasini}, {van Boekel}, {Molli{\`e}re},
  {Henning}, \& {Benneke}}]{2016ApJMordasini}
{Mordasini}, C., {van Boekel}, R., {Molli{\`e}re}, P., {Henning}, T., \&
  {Benneke}, B. 2016, \apj, 832, 41, \dodoi{10.3847/0004-637X/832/1/41}

\bibitem[{{Mousis} {et~al.}(2020){Mousis}, {Deleuil}, {Aguichine}, {Marcq},
  {Naar}, {Aguirre}, {Brugger}, \& {Gon{\c{c}}alves}}]{2020Mousis}
{Mousis}, O., {Deleuil}, M., {Aguichine}, A., {et~al.} 2020, \apjl, 896, L22,
  \dodoi{10.3847/2041-8213/ab9530}

\bibitem[{{Mulders} {et~al.}(2015){Mulders}, {Pascucci}, \&
  {Apai}}]{2015ApJMuldersB}
{Mulders}, G.~D., {Pascucci}, I., \& {Apai}, D. 2015, \apj, 814, 130,
  \dodoi{10.1088/0004-637X/814/2/130}

\bibitem[{{Naef} {et~al.}(2001){Naef}, {Mayor}, {Pepe}, {Queloz}, {Santos},
  {Udry}, \& {Burnet}}]{2001AANaef}
{Naef}, D., {Mayor}, M., {Pepe}, F., {et~al.} 2001, \aap, 375, 205,
  \dodoi{10.1051/0004-6361:20010841}

\bibitem[{{Nakagawa} {et~al.}(1986){Nakagawa}, {Sekiya}, \&
  {Hayashi}}]{1986IcarusNakagawa}
{Nakagawa}, Y., {Sekiya}, M., \& {Hayashi}, C. 1986, \icarus, 67, 375,
  \dodoi{10.1016/0019-1035(86)90121-1}

\bibitem[{{Narang} {et~al.}(2018){Narang}, {Manoj}, {Furlan}, {Mordasini},
  {Henning}, {Mathew}, {Banyal}, \& {Sivarani}}]{2018AJNarang}
{Narang}, M., {Manoj}, P., {Furlan}, E., {et~al.} 2018, \aj, 156, 221,
  \dodoi{10.3847/1538-3881/aae391}

\bibitem[{Nelson {et~al.}(2013)Nelson, Gressel, \& Umurhan}]{Nelson2013}
Nelson, R.~P., Gressel, O., \& Umurhan, O.~M. 2013, Monthly Notices of the
  Royal Astronomical Society, 435, 2610, \dodoi{10.1093/mnras/stt1475}

\bibitem[{Nesvorn{\'y} {et~al.}(2021)Nesvorn{\'y}, Li, Simon, Youdin,
  Richardson, Marschall, \& Grundy}]{Nesvorny2021}
Nesvorn{\'y}, D., Li, R., Simon, J.~B., {et~al.} 2021, The Planetary Science
  Journal, 2, 27, \dodoi{10.3847/PSJ/abd858}

\bibitem[{{Nettelmann} {et~al.}(2013){Nettelmann}, {Helled}, {Fortney}, \&
  {Redmer}}]{Nettelmann2013}
{Nettelmann}, N., {Helled}, R., {Fortney}, J.~J., \& {Redmer}, R. 2013,
  \planss, 77, 143, \dodoi{10.1016/j.pss.2012.06.019}

\bibitem[{{Nielsen} {et~al.}(2019){Nielsen}, {De Rosa}, {Macintosh}, {Wang},
  {Ruffio}, {Chiang}, {Marley}, {Saumon}, {Savransky}, {Ammons}, {Bailey},
  {Barman}, {Blain}, {Bulger}, {Burrows}, {Chilcote}, {Cotten}, {Czekala},
  {Doyon}, {Duch{\^e}ne}, {Esposito}, {Fabrycky}, {Fitzgerald}, {Follette},
  {Fortney}, {Gerard}, {Goodsell}, {Graham}, {Greenbaum}, {Hibon}, {Hinkley},
  {Hirsch}, {Hom}, {Hung}, {Dawson}, {Ingraham}, {Kalas}, {Konopacky},
  {Larkin}, {Lee}, {Lin}, {Maire}, {Marchis}, {Marois}, {Metchev},
  {Millar-Blanchaer}, {Morzinski}, {Oppenheimer}, {Palmer}, {Patience},
  {Perrin}, {Poyneer}, {Pueyo}, {Rafikov}, {Rajan}, {Rameau}, {Rantakyr{\"o}},
  {Ren}, {Schneider}, {Sivaramakrishnan}, {Song}, {Soummer}, {Tallis},
  {Thomas}, {Ward-Duong}, \& {Wolff}}]{2019AJNielsenA}
{Nielsen}, E.~L., {De Rosa}, R.~J., {Macintosh}, B., {et~al.} 2019, \aj, 158,
  13, \dodoi{10.3847/1538-3881/ab16e9}

\bibitem[{{\"O}berg {et~al.}(2023){\"O}berg, Facchini, \& Anderson}]{Oberg2023}
{\"O}berg, K.~I., Facchini, S., \& Anderson, D.~E. 2023, Annual Review of
  Astronomy and Astrophysics, 61, 287,
  \dodoi{10.1146/annurev-astro-022823-040820}

\bibitem[{{{\"O}berg} {et~al.}(2011){{\"O}berg}, {Murray-Clay}, \&
  {Bergin}}]{2011ApJObergB}
{{\"O}berg}, K.~I., {Murray-Clay}, R., \& {Bergin}, E.~A. 2011, \apjl, 743,
  L16, \dodoi{10.1088/2041-8205/743/1/L16}

\bibitem[{Oberg {et~al.}(2021)Oberg, Guzman, Walsh, Aikawa, Bergin, Law,
  Loomis, Alarcon, Andrews, Bae, Bergner, Boehler, Booth, Bosman, Calahan,
  Cataldi, Cleeves, Czekala, Furuya, Huang, Ilee, Kurtovic, Gal, Liu, Long,
  Menard, Nomura, Perez, Qi, Schwarz, Sierra, Teague, Tsukagoshi, Yamato,
  van~'t Hoff, Waggoner, Wilner, \& Zhang}]{Oberg2021}
Oberg, K.~I., Guzman, V.~V., Walsh, C., {et~al.} 2021, The Astrophysical
  Journal Supplement Series, 257, 1

\bibitem[{{Ogihara} {et~al.}(2018){Ogihara}, {Kokubo}, {Suzuki}, \&
  {Morbidelli}}]{2018AAOgihara}
{Ogihara}, M., {Kokubo}, E., {Suzuki}, T.~K., \& {Morbidelli}, A. 2018, \aap,
  615, A63, \dodoi{10.1051/0004-6361/201832720}

\bibitem[{{Ohno} \& {Fortney}(2023)}]{OhnoFortney2023}
{Ohno}, K., \& {Fortney}, J.~J. 2023, \apj, 946, 18,
  \dodoi{10.3847/1538-4357/acafed}

\bibitem[{Ohtsuki(1999)}]{Ohtsuki1999}
Ohtsuki, K. 1999, Icarus, 137, 152, \dodoi{10.1006/icar.1998.6041}

\bibitem[{{Ohtsuki} {et~al.}(2002){Ohtsuki}, {Stewart}, \&
  {Ida}}]{2002IcarusOhtsuki}
{Ohtsuki}, K., {Stewart}, G.~R., \& {Ida}, S. 2002, \icarus, 155, 436,
  \dodoi{10.1006/icar.2001.6741}

\bibitem[{Okuzumi {et~al.}(2012)Okuzumi, Tanaka, Kobayashi, \&
  Wada}]{Okuzumi2012}
Okuzumi, S., Tanaka, H., Kobayashi, H., \& Wada, K. 2012, The Astrophysical
  Journal, 752, 106, \dodoi{10.1088/0004-637X/752/2/106}

\bibitem[{{Onyett} {et~al.}(2023){Onyett}, {Schiller}, {Makhatadze}, {Deng},
  {Johansen}, \& {Bizzarro}}]{Onyett2023Nature}
{Onyett}, I.~J., {Schiller}, M., {Makhatadze}, G.~V., {et~al.} 2023, \nat, 619,
  539, \dodoi{10.1038/s41586-023-06135-z}

\bibitem[{{Ormel}(2014)}]{ormel2014}
{Ormel}, C.~W. 2014, \apjl, 789, L18, \dodoi{10.1088/2041-8205/789/1/L18}

\bibitem[{{Ormel}(2017)}]{2017ASSLOrmel}
{Ormel}, C.~W. 2017, in Astrophysics and Space Science Library, ed. M.~{Pessah}
  \& O.~{Gressel}, Vol. 445, 197, \dodoi{10.1007/978-3-319-60609-5_7}

\bibitem[{Ormel \& Cuzzi(2007)}]{Ormel2007}
Ormel, C.~W., \& Cuzzi, J.~N. 2007, Astronomy \& Astrophysics, 466, 413,
  \dodoi{10.1051/0004-6361:20066899}

\bibitem[{{Ormel} {et~al.}(2010){Ormel}, {Dullemond}, \&
  {Spaans}}]{2010ApJOrmel}
{Ormel}, C.~W., {Dullemond}, C.~P., \& {Spaans}, M. 2010, \apjl, 714, L103,
  \dodoi{10.1088/2041-8205/714/1/L103}

\bibitem[{{Ormel} {et~al.}(2012){Ormel}, {Ida}, \& {Tanaka}}]{ormelida2012}
{Ormel}, C.~W., {Ida}, S., \& {Tanaka}, H. 2012, \apj, 758, 80,
  \dodoi{10.1088/0004-637X/758/2/80}

\bibitem[{{Ormel} \& {Klahr}(2010)}]{2010AAOrmelKlahr}
{Ormel}, C.~W., \& {Klahr}, H.~H. 2010, \aap, 520, A43,
  \dodoi{10.1051/0004-6361/201014903}

\bibitem[{Ormel \& Kobayashi(2012)}]{Ormel2012}
Ormel, C.~W., \& Kobayashi, H. 2012, The Astrophysical Journal, 747, 115,
  \dodoi{10.1088/0004-637X/747/2/115}

\bibitem[{{Ormel} {et~al.}(2015){Ormel}, {Shi}, \& {Kuiper}}]{ormel2015}
{Ormel}, C.~W., {Shi}, J.-M., \& {Kuiper}, R. 2015, \mnras, 447, 3512,
  \dodoi{10.1093/mnras/stu2704}

\bibitem[{{Owen}(2019)}]{Owen2019AREPS}
{Owen}, J.~E. 2019, Annual Review of Earth and Planetary Sciences, 47, 67,
  \dodoi{10.1146/annurev-earth-053018-060246}

\bibitem[{{Owen} \& {Wu}(2013)}]{2013ApJOwenWu}
{Owen}, J.~E., \& {Wu}, Y. 2013, \apj, 775, 105,
  \dodoi{10.1088/0004-637X/775/2/105}

\bibitem[{Paardekooper {et~al.}(2023)Paardekooper, Dong, Duffell, Fung, Masset,
  Ogilvie, \& Tanaka}]{Paardekooper2023}
Paardekooper, S., Dong, R., Duffell, P., {et~al.} 2023, in {{ASP Conference
  Series}}, Vol. 534, Protostars and {{Planets VII}}, {{Astronomical Society}}
  of the {{Pacific Conference Series}}, ed. {Inutsuka, S. and Aikawa, Y. and
  Muto, T. and Tomida, K. and Tamura, M.}, {Kyoto}, 685.
\newblock \url{https://ui.adsabs.harvard.edu/abs/2023ASPC..534..685P}

\bibitem[{{Paardekooper}(2014)}]{Paardekooper2014}
{Paardekooper}, S.-J. 2014, \mnras, 444, 2031, \dodoi{10.1093/mnras/stu1542}

\bibitem[{{Paardekooper} {et~al.}(2010){Paardekooper}, {Baruteau}, {Crida}, \&
  {Kley}}]{2010MNRASPaardekooper}
{Paardekooper}, S.~J., {Baruteau}, C., {Crida}, A., \& {Kley}, W. 2010, \mnras,
  401, 1950, \dodoi{10.1111/j.1365-2966.2009.15782.x}

\bibitem[{{Paardekooper} {et~al.}(2011){Paardekooper}, {Baruteau}, \&
  {Kley}}]{2011MNRASPaardekooper}
{Paardekooper}, S.~J., {Baruteau}, C., \& {Kley}, W. 2011, \mnras, 410, 293,
  \dodoi{10.1111/j.1365-2966.2010.17442.x}

\bibitem[{{Pan} {et~al.}(2023){Pan}, {Liu}, {Johansen}, {Ogihara}, {Wang},
  {Ji}, {Wang}, {Feng}, \& {Riba}}]{Pan2023}
{Pan}, M., {Liu}, B., {Johansen}, A., {et~al.} 2023, arXiv e-prints,
  arXiv:2311.10317, \dodoi{10.48550/arXiv.2311.10317}

\bibitem[{Perotti {et~al.}(2023)Perotti, Christiaens, Henning, Tabone, Waters,
  Kamp, Olofsson, Grant, Gasman, Bouwman, Samland, Franceschi, {van Dishoeck},
  Schwarz, G{\"u}del, Lagage, Ray, Vandenbussche, Abergel, Absil, Arabhavi,
  Argyriou, Barrado, Boccaletti, o~Garatti, Geers, Glauser, Justannont, Lahuis,
  Mueller, Nehm{\'e}, Pantin, Scheithauer, Waelkens, Guadarrama, Jang, Kanwar,
  {Morales-Calder{\'o}n}, Pawellek, {Rodgers-Lee}, Schreiber, Colina, Greve,
  {\"O}stlin, \& Wright}]{Perotti2023}
Perotti, G., Christiaens, V., Henning, T., {et~al.} 2023, Nature,
  \dodoi{10.1038/s41586-023-06317-9}

\bibitem[{{Perri} \& {Cameron}(1974)}]{1974IcarusPerriCameron}
{Perri}, F., \& {Cameron}, A.~G.~W. 1974, \icarus, 22, 416,
  \dodoi{10.1016/0019-1035(74)90074-8}

\bibitem[{{Petigura} {et~al.}(2013){Petigura}, {Howard}, \&
  {Marcy}}]{2013PNASPetigura}
{Petigura}, E.~A., {Howard}, A.~W., \& {Marcy}, G.~W. 2013, \pnas, 110, 19273,
  \dodoi{10.1073/pnas.1319909110}

\bibitem[{{Petigura} {et~al.}(2018){Petigura}, {Marcy}, {Winn}, {Weiss},
  {Fulton}, {Howard}, {Sinukoff}, {Isaacson}, {Morton}, \&
  {Johnson}}]{2018AJPetigura}
{Petigura}, E.~A., {Marcy}, G.~W., {Winn}, J.~N., {et~al.} 2018, \aj, 155, 89,
  \dodoi{10.3847/1538-3881/aaa54c}

\bibitem[{{Pfalzner} {et~al.}(2022){Pfalzner}, {Dehghani}, \&
  {Michel}}]{2022ApJPfalzner}
{Pfalzner}, S., {Dehghani}, S., \& {Michel}, A. 2022, \apjl, 939, L10,
  \dodoi{10.3847/2041-8213/ac9839}

\bibitem[{Pfeil \& Klahr(2019)}]{Pfeil2019}
Pfeil, T., \& Klahr, H. 2019, The Astrophysical Journal, 871, 150,
  \dodoi{10.3847/1538-4357/aaf962}

\bibitem[{{Picogna} {et~al.}(2021){Picogna}, {Ercolano}, \&
  {Espaillat}}]{2021MNRASPicogna}
{Picogna}, G., {Ercolano}, B., \& {Espaillat}, C.~C. 2021, \mnras, 508, 3611,
  \dodoi{10.1093/mnras/stab2883}

\bibitem[{{Picogna} {et~al.}(2019){Picogna}, {Ercolano}, {Owen}, \&
  {Weber}}]{2019MNRASPicogna}
{Picogna}, G., {Ercolano}, B., {Owen}, J.~E., \& {Weber}, M.~L. 2019, \mnras,
  487, 691, \dodoi{10.1093/mnras/stz1166}

\bibitem[{{Pierens}(2015)}]{Pierens2015}
{Pierens}, A. 2015, \mnras, 454, 2003, \dodoi{10.1093/mnras/stv2024}

\bibitem[{{Pinte} {et~al.}(2020){Pinte}, {Price}, {M{\'e}nard}, {Duch{\^e}ne},
  {Christiaens}, {Andrews}, {Huang}, {Hill}, {van der Plas}, {Perez}, {Isella},
  {Boehler}, {Dent}, {Mentiplay}, \& {Loomis}}]{Pinte2020ApJ}
{Pinte}, C., {Price}, D.~J., {M{\'e}nard}, F., {et~al.} 2020, \apjl, 890, L9,
  \dodoi{10.3847/2041-8213/ab6dda}

\bibitem[{{Piso} \& {Youdin}(2014)}]{2014ApJPisoYoudin}
{Piso}, A.-M.~A., \& {Youdin}, A.~N. 2014, \apj, 786, 21,
  \dodoi{10.1088/0004-637X/786/1/21}

\bibitem[{{Plavchan} {et~al.}(2020){Plavchan}, {Barclay}, {Gagn{\'e}}, {Gao},
  {Cale}, {Matzko}, {Dragomir}, {Quinn}, {Feliz}, {Stassun}, {Crossfield},
  {Berardo}, {Latham}, {Tieu}, {Anglada-Escud{\'e}}, {Ricker}, {Vanderspek},
  {Seager}, {Winn}, {Jenkins}, {Rinehart}, {Krishnamurthy}, {Dynes}, {Doty},
  {Adams}, {Afanasev}, {Beichman}, {Bottom}, {Bowler}, {Brinkworth}, {Brown},
  {Cancino}, {Ciardi}, {Clampin}, {Clark}, {Collins}, {Davison},
  {Foreman-Mackey}, {Furlan}, {Gaidos}, {Geneser}, {Giddens}, {Gilbert},
  {Hall}, {Hellier}, {Henry}, {Horner}, {Howard}, {Huang}, {Huber}, {Kane},
  {Kenworthy}, {Kielkopf}, {Kipping}, {Klenke}, {Kruse}, {Latouf}, {Lowrance},
  {Mennesson}, {Mengel}, {Mills}, {Morton}, {Narita}, {Newton}, {Nishimoto},
  {Okumura}, {Palle}, {Pepper}, {Quintana}, {Roberge}, {Roccatagliata},
  {Schlieder}, {Tanner}, {Teske}, {Tinney}, {Vanderburg}, {von Braun}, {Walp},
  {Wang}, {Wang}, {Weigand}, {White}, {Wittenmyer}, {Wright}, {Youngblood},
  {Zhang}, \& {Zilberman}}]{Plavchan2020Nature}
{Plavchan}, P., {Barclay}, T., {Gagn{\'e}}, J., {et~al.} 2020, \nat, 582, 497,
  \dodoi{10.1038/s41586-020-2400-z}

\bibitem[{{Podolak} {et~al.}(2020){Podolak}, {Haghighipour}, {Bodenheimer},
  {Helled}, \& {Podolak}}]{2020ApJPodolak}
{Podolak}, M., {Haghighipour}, N., {Bodenheimer}, P., {Helled}, R., \&
  {Podolak}, E. 2020, \apj, 899, 45, \dodoi{10.3847/1538-4357/ab9ec1}

\bibitem[{{Podolak} {et~al.}(1988){Podolak}, {Pollack}, \&
  {Reynolds}}]{1988IcarusPodolak}
{Podolak}, M., {Pollack}, J.~B., \& {Reynolds}, R.~T. 1988, \icarus, 73, 163,
  \dodoi{10.1016/0019-1035(88)90090-5}

\bibitem[{Polak \& Klahr(2023)}]{Polak2023}
Polak, B., \& Klahr, H. 2023, The Astrophysical Journal, 943, 125,
  \dodoi{10.3847/1538-4357/aca58f}

\bibitem[{Pollack {et~al.}(1994)Pollack, Hollenbach, Beckwith, Simonelli,
  Roush, \& Fong}]{Pollack1994}
Pollack, J.~B., Hollenbach, D., Beckwith, S., {et~al.} 1994, The Astrophysical
  Journal, 421, 615, \dodoi{10.1086/173677}

\bibitem[{{Pollack} {et~al.}(1996){Pollack}, {Hubickyj}, {Bodenheimer},
  {Lissauer}, {Podolak}, \& {Greenzweig}}]{1996IcarusPollack}
{Pollack}, J.~B., {Hubickyj}, O., {Bodenheimer}, P., {et~al.} 1996, \icarus,
  124, 62, \dodoi{10.1006/icar.1996.0190}

\bibitem[{{Polman} {et~al.}(2023){Polman}, {Waters}, {Min}, {Miguel}, \&
  {Khorshid}}]{Polman2023A&A}
{Polman}, J., {Waters}, L.~B.~F.~M., {Min}, M., {Miguel}, Y., \& {Khorshid}, N.
  2023, \aap, 670, A161, \dodoi{10.1051/0004-6361/202244647}

\bibitem[{{Pringle}(1981)}]{1981ARAAPringle}
{Pringle}, J.~E. 1981, \araa, 19, 137,
  \dodoi{10.1146/annurev.aa.19.090181.001033}

\bibitem[{Raettig {et~al.}(2015)Raettig, Klahr, \& Lyra}]{Raettig2015}
Raettig, N., Klahr, H., \& Lyra, W. 2015, The Astrophysical Journal, 804, 35,
  \dodoi{10.1088/0004-637X/804/1/35}

\bibitem[{{Rasio} \& {Ford}(1996)}]{1996ScienceRasioFord}
{Rasio}, F.~A., \& {Ford}, E.~B. 1996, \sci, 274, 954,
  \dodoi{10.1126/science.274.5289.954}

\bibitem[{{Rauer} {et~al.}(2014){Rauer}, {Catala}, {Aerts}, {Appourchaux},
  {Benz}, {Brandeker}, {Christensen-Dalsgaard}, {Deleuil}, {Gizon}, {Goupil},
  {G{\"u}del}, {Janot-Pacheco}, {Mas-Hesse}, {Pagano}, {Piotto}, {Pollacco},
  {Santos}, {Smith}, {Su{\'a}rez}, {Szab{\'o}}, {Udry}, {Adibekyan}, {Alibert},
  {Almenara}, {Amaro-Seoane}, {Eiff}, {Asplund}, {Antonello}, {Barnes},
  {Baudin}, {Belkacem}, {Bergemann}, {Bihain}, {Birch}, {Bonfils}, {Boisse},
  {Bonomo}, {Borsa}, {Brand{\~a}o}, {Brocato}, {Brun}, {Burleigh}, {Burston},
  {Cabrera}, {Cassisi}, {Chaplin}, {Charpinet}, {Chiappini}, {Church},
  {Csizmadia}, {Cunha}, {Damasso}, {Davies}, {Deeg}, {D{\'\i}az}, {Dreizler},
  {Dreyer}, {Eggenberger}, {Ehrenreich}, {Eigm{\"u}ller}, {Erikson}, {Farmer},
  {Feltzing}, {de Oliveira Fialho}, {Figueira}, {Forveille}, {Fridlund},
  {Garc{\'\i}a}, {Giommi}, {Giuffrida}, {Godolt}, {Gomes da Silva}, {Granzer},
  {Grenfell}, {Grotsch-Noels}, {G{\"u}nther}, {Haswell}, {Hatzes},
  {H{\'e}brard}, {Hekker}, {Helled}, {Heng}, {Jenkins}, {Johansen},
  {Khodachenko}, {Kislyakova}, {Kley}, {Kolb}, {Krivova}, {Kupka}, {Lammer},
  {Lanza}, {Lebreton}, {Magrin}, {Marcos-Arenal}, {Marrese}, {Marques},
  {Martins}, {Mathis}, {Mathur}, {Messina}, {Miglio}, {Montalban}, {Montalto},
  {Monteiro}, {Moradi}, {Moravveji}, {Mordasini}, {Morel}, {Mortier},
  {Nascimbeni}, {Nelson}, {Nielsen}, {Noack}, {Norton}, {Ofir}, {Oshagh},
  {Ouazzani}, {P{\'a}pics}, {Parro}, {Petit}, {Plez}, {Poretti}, {Quirrenbach},
  {Ragazzoni}, {Raimondo}, {Rainer}, {Reese}, {Redmer}, {Reffert},
  {Rojas-Ayala}, {Roxburgh}, {Salmon}, {Santerne}, {Schneider}, {Schou},
  {Schuh}, {Schunker}, {Silva-Valio}, {Silvotti}, {Skillen}, {Snellen}, {Sohl},
  {Sousa}, {Sozzetti}, {Stello}, {Strassmeier}, {{\v{S}}vanda}, {Szab{\'o}},
  {Tkachenko}, {Valencia}, {Van Grootel}, {Vauclair}, {Ventura}, {Wagner},
  {Walton}, {Weingrill}, {Werner}, {Wheatley}, \& {Zwintz}}]{Rauer2014ExA}
{Rauer}, H., {Catala}, C., {Aerts}, C., {et~al.} 2014, Experimental Astronomy,
  38, 249, \dodoi{10.1007/s10686-014-9383-4}

\bibitem[{{Raymond} \& {Izidoro}(2017)}]{2017IcarusRaymondIzidoro}
{Raymond}, S.~N., \& {Izidoro}, A. 2017, \icarus, 297, 134,
  \dodoi{10.1016/j.icarus.2017.06.030}

\bibitem[{{Reffert} {et~al.}(2015){Reffert}, {Bergmann}, {Quirrenbach},
  {Trifonov}, \& {K{\"u}nstler}}]{Reffert2015}
{Reffert}, S., {Bergmann}, C., {Quirrenbach}, A., {Trifonov}, T., \&
  {K{\"u}nstler}, A. 2015, \aap, 574, A116, \dodoi{10.1051/0004-6361/201322360}

\bibitem[{{Ribas} {et~al.}(2023){Ribas}, {Reiners}, {Zechmeister}, {Caballero},
  {Morales}, {Sabotta}, {Baroch}, {Amado}, {Quirrenbach}, {Abril}, {Aceituno},
  {Anglada-Escud{\'e}}, {Azzaro}, {Barrado}, {B{\'e}jar}, {Ben{\'\i}tez de
  Haro}, {Bergond}, {Bluhm}, {Calvo Ortega}, {Cardona Guill{\'e}n},
  {Chaturvedi}, {Cifuentes}, {Colom{\'e}}, {Cont}, {Cort{\'e}s-Contreras},
  {Czesla}, {D{\'\i}ez-Alonso}, {Dreizler}, {Duque-Arribas}, {Espinoza},
  {Fern{\'a}ndez}, {Fuhrmeister}, {Galad{\'\i}-Enr{\'\i}quez},
  {Garc{\'\i}a-L{\'o}pez}, {Gonz{\'a}lez-{\'A}lvarez}, {Gonz{\'a}lez
  Hern{\'a}ndez}, {Guenther}, {de Guindos}, {Hatzes}, {Henning}, {Herrero},
  {Hintz}, {Huelmo}, {Jeffers}, {Johnson}, {de Juan}, {Kaminski}, {Kemmer},
  {Khaimova}, {Khalafinejad}, {Kossakowski}, {K{\"u}rster}, {Labarga},
  {Lafarga}, {Lalitha}, {Lamp{\'o}n}, {Lillo-Box}, {Lodieu}, {L{\'o}pez
  Gonz{\'a}lez}, {L{\'o}pez-Puertas}, {Luque}, {Mag{\'a}n}, {Mancini},
  {Marfil}, {Mart{\'\i}n}, {Mart{\'\i}n-Ruiz}, {Molaverdikhani}, {Montes},
  {Nagel}, {Nortmann}, {Nowak}, {Pall{\'e}}, {Passegger}, {Pavlov}, {Pedraz},
  {Perdelwitz}, {Perger}, {Ram{\'o}n-Ballesta}, {Reffert}, {Revilla},
  {Rodr{\'\i}guez}, {Rodr{\'\i}guez-L{\'o}pez}, {Sadegi}, {S{\'a}nchez
  Carrasco}, {S{\'a}nchez-L{\'o}pez}, {Sanz-Forcada}, {Sch{\"a}fer},
  {Schlecker}, {Schmitt}, {Sch{\"o}fer}, {Schweitzer}, {Seifert}, {Shan},
  {Skrzypinski}, {Solano}, {Stahl}, {Stangret}, {Stock}, {St{\"u}rmer},
  {Tabernero}, {Tal-Or}, {Trifonov}, {Vanaverbeke}, {Yan}, \& {Zapatero
  Osorio}}]{Ribas2023}
{Ribas}, I., {Reiners}, A., {Zechmeister}, M., {et~al.} 2023, \aap, 670, A139,
  \dodoi{10.1051/0004-6361/202244879}

\bibitem[{{Rosenthal} {et~al.}(2021){Rosenthal}, {Fulton}, {Hirsch},
  {Isaacson}, {Howard}, {Dedrick}, {Sherstyuk}, {Blunt}, {Petigura}, {Knutson},
  {Behmard}, {Chontos}, {Crepp}, {Crossfield}, {Dalba}, {Fischer}, {Henry},
  {Kane}, {Kosiarek}, {Marcy}, {Rubenzahl}, {Weiss}, \&
  {Wright}}]{2021ApJSRosenthal}
{Rosenthal}, L.~J., {Fulton}, B.~J., {Hirsch}, L.~A., {et~al.} 2021, \apjs,
  255, 8, \dodoi{10.3847/1538-4365/abe23c}

\bibitem[{{Sabotta} {et~al.}(2021){Sabotta}, {Schlecker}, {Chaturvedi},
  {Guenther}, {Mu{\~n}oz Rodr{\'\i}guez}, {Mu{\~n}oz S{\'a}nchez}, {Caballero},
  {Shan}, {Reffert}, {Ribas}, {Reiners}, {Hatzes}, {Amado}, {Klahr}, {Morales},
  {Quirrenbach}, {Henning}, {Dreizler}, {Pall{\'e}}, {Perger}, {Azzaro},
  {Jeffers}, {Kaminski}, {K{\"u}rster}, {Lafarga}, {Montes}, {Passegger}, \&
  {Zechmeister}}]{2021Sabotta}
{Sabotta}, S., {Schlecker}, M., {Chaturvedi}, P., {et~al.} 2021, \aap, 653,
  A114, \dodoi{10.1051/0004-6361/202140968}

\bibitem[{{Safronov}(1969)}]{1969BookSafronov}
{Safronov}, V.~S. 1969, {Evolution of the Protoplanetary Cloud and Formation of
  the Earth and the Planets} (Moscow: Nauka)

\bibitem[{{Sahlmann} {et~al.}(2011){Sahlmann}, {S{\'e}gransan}, {Queloz},
  {Udry}, {Santos}, {Marmier}, {Mayor}, {Naef}, {Pepe}, \&
  {Zucker}}]{Sahlmann2011}
{Sahlmann}, J., {S{\'e}gransan}, D., {Queloz}, D., {et~al.} 2011, \aap, 525,
  A95, \dodoi{10.1051/0004-6361/201015427}

\bibitem[{{Santos} {et~al.}(2004){Santos}, {Israelian}, \&
  {Mayor}}]{2004A&ASantos}
{Santos}, N.~C., {Israelian}, G., \& {Mayor}, M. 2004, \aap, 415, 1153,
  \dodoi{10.1051/0004-6361:20034469}

\bibitem[{{Santos} {et~al.}(2017){Santos}, {Adibekyan}, {Figueira},
  {Andreasen}, {Barros}, {Delgado-Mena}, {Demangeon}, {Faria}, {Oshagh},
  {Sousa}, {Viana}, \& {Ferreira}}]{2017A&ASantosA}
{Santos}, N.~C., {Adibekyan}, V., {Figueira}, P., {et~al.} 2017, \aap, 603,
  A30, \dodoi{10.1051/0004-6361/201730761}

\bibitem[{{Saumon} {et~al.}(1995){Saumon}, {Chabrier}, \& {van
  Horn}}]{1995ApJSSaumon}
{Saumon}, D., {Chabrier}, G., \& {van Horn}, H.~M. 1995, \apjs, 99, 713,
  \dodoi{10.1086/192204}

\bibitem[{{Sch{\"a}fer} {et~al.}(2017){Sch{\"a}fer}, {Yang}, \&
  {Johansen}}]{2017AASchaefer}
{Sch{\"a}fer}, U., {Yang}, C.-C., \& {Johansen}, A. 2017, \aap, 597, A69,
  \dodoi{10.1051/0004-6361/201629561}

\bibitem[{{Schib} {et~al.}(2022){Schib}, {Mordasini}, \&
  {Helled}}]{2022AASchib}
{Schib}, O., {Mordasini}, C., \& {Helled}, R. 2022, \aap, 664, A138,
  \dodoi{10.1051/0004-6361/202141904}

\bibitem[{{Schib} {et~al.}(2023){Schib}, {Mordasini}, \&
  {Helled}}]{2023AASchib}
---. 2023, \aap, 669, A31, \dodoi{10.1051/0004-6361/202244789}

\bibitem[{{Schib} {et~al.}(2021){Schib}, {Mordasini}, {Wenger}, {Marleau}, \&
  {Helled}}]{2021AASchib}
{Schib}, O., {Mordasini}, C., {Wenger}, N., {Marleau}, G.~D., \& {Helled}, R.
  2021, \aap, 645, A43, \dodoi{10.1051/0004-6361/202039154}

\bibitem[{{Schlecker} {et~al.}(2021){Schlecker}, {Mordasini}, {Emsenhuber},
  {Klahr}, {Henning}, {Burn}, {Alibert}, \& {Benz}}]{2021AASchleckerA}
{Schlecker}, M., {Mordasini}, C., {Emsenhuber}, A., {et~al.} 2021, \aap, 656,
  A71, \dodoi{10.1051/0004-6361/202038554}

\bibitem[{{Schlecker} {et~al.}(2022){Schlecker}, {Burn}, {Sabotta}, {Seifert},
  {Henning}, {Emsenhuber}, {Mordasini}, {Reffert}, {Shan}, \&
  {Klahr}}]{2022AASchlecker}
{Schlecker}, M., {Burn}, R., {Sabotta}, S., {et~al.} 2022, \aap, 664, A180,
  \dodoi{10.1051/0004-6361/202142543}

\bibitem[{{Schneider} \& {Bitsch}(2021)}]{Schneider2021}
{Schneider}, A.~D., \& {Bitsch}, B. 2021, \aap, 654, A71,
  \dodoi{10.1051/0004-6361/202039640}

\bibitem[{{Sch{\"o}nau} {et~al.}(2023){Sch{\"o}nau}, {Teiser}, {Demirci},
  {Joeris}, {Bila}, {Onyeagusi}, {Fritscher}, \& {Wurm}}]{Schonau2023}
{Sch{\"o}nau}, L., {Teiser}, J., {Demirci}, T., {et~al.} 2023, \aap, 672, A169,
  \dodoi{10.1051/0004-6361/202245499}

\bibitem[{{Schoonenberg} \& {Ormel}(2017)}]{Schoonenberg2017}
{Schoonenberg}, D., \& {Ormel}, C.~W. 2017, \aap, 602, A21,
  \dodoi{10.1051/0004-6361/201630013}

\bibitem[{Schr{\"a}pler {et~al.}(2022)Schr{\"a}pler, Landeck, \&
  Blum}]{Schrapler2022}
Schr{\"a}pler, R.~R., Landeck, W.~A., \& Blum, J. 2022, Monthly Notices of the
  Royal Astronomical Society, 509, 5641, \dodoi{10.1093/mnras/stab3348}

\bibitem[{Seizinger \& Kley(2013)}]{Seizinger2013}
Seizinger, A., \& Kley, W. 2013, Astronomy \& Astrophysics, 551, A65,
  \dodoi{10.1051/0004-6361/201220946}

\bibitem[{{Sekiya}(1998)}]{Sekiya1998}
{Sekiya}, M. 1998, \icarus, 133, 298, \dodoi{10.1006/icar.1998.5933}

\bibitem[{{Sellek} {et~al.}(2020){Sellek}, {Booth}, \&
  {Clarke}}]{Sellek2020Entrainment}
{Sellek}, A.~D., {Booth}, R.~A., \& {Clarke}, C.~J. 2020, \mnras, 492, 1279,
  \dodoi{10.1093/mnras/stz3528}

\bibitem[{Semenov \& Wiebe(2011)}]{Semenov2011}
Semenov, D., \& Wiebe, D. 2011, The Astrophysical Journal Supplement Series,
  196, \dodoi{10.1088/0067-0049/196/2/25}

\bibitem[{{Shakura} \& {Sunyaev}(1973)}]{1973A&AShakuraSunyaev}
{Shakura}, N.~I., \& {Sunyaev}, R.~A. 1973, \aap, 500, 33

\bibitem[{Shibata {et~al.}(2023)Shibata, Helled, \& Kobayashi}]{Shibata2023}
Shibata, S., Helled, R., \& Kobayashi, H. 2023, Monthly Notices of the Royal
  Astronomical Society, 519, 1713, \dodoi{10.1093/mnras/stac3568}

\bibitem[{{Simon} {et~al.}(2016){Simon}, {Armitage}, {Li}, \&
  {Youdin}}]{Simon2016}
{Simon}, J.~B., {Armitage}, P.~J., {Li}, R., \& {Youdin}, A.~N. 2016, \apj,
  822, 55, \dodoi{10.3847/0004-637X/822/1/55}

\bibitem[{Simon {et~al.}(2017)Simon, Armitage, Youdin, \& Li}]{Simon2017}
Simon, J.~B., Armitage, P.~J., Youdin, A.~N., \& Li, R. 2017, The Astrophysical
  Journal, 847, L12, \dodoi{10.3847/2041-8213/aa8c79}

\bibitem[{{Spohn}(2015)}]{Spohn2015trge}
{Spohn}, T. 2015, in Treatise on Geophysics, ed. G.~{Schubert}, 1--22,
  \dodoi{10.1016/B978-0-444-53802-4.00165-2}

\bibitem[{{Squire} \& {Hopkins}(2018)}]{Squire2018}
{Squire}, J., \& {Hopkins}, P.~F. 2018, \mnras, 477, 5011,
  \dodoi{10.1093/mnras/sty854}

\bibitem[{Stammler \& Birnstiel(2022)}]{Stammler2022}
Stammler, S.~M., \& Birnstiel, T. 2022, The Astrophysical Journal, 935, 35,
  \dodoi{10.3847/1538-4357/ac7d58}

\bibitem[{{Steinpilz} {et~al.}(2019){Steinpilz}, {Teiser}, \&
  {Wurm}}]{2019ApJSteinpilz}
{Steinpilz}, T., {Teiser}, J., \& {Wurm}, G. 2019, \apj, 874, 60,
  \dodoi{10.3847/1538-4357/ab07bb}

\bibitem[{{Stevenson}(1982)}]{1982PSSStevenson}
{Stevenson}, D.~J. 1982, \planss, 30, 755, \dodoi{10.1016/0032-0633(82)90108-8}

\bibitem[{{Suzuki} {et~al.}(2016{\natexlab{a}}){Suzuki}, {Bennett}, {Sumi},
  {Bond}, {Rogers}, {Abe}, {Asakura}, {Bhattacharya}, {Donachie}, {Freeman},
  {Fukui}, {Hirao}, {Itow}, {Koshimoto}, {Li}, {Ling}, {Masuda}, {Matsubara},
  {Muraki}, {Nagakane}, {Onishi}, {Oyokawa}, {Rattenbury}, {Saito}, {Sharan},
  {Shibai}, {Sullivan}, {Tristram}, {Yonehara}, \& {MOA
  Collaboration}}]{2016ApJSuzuki}
{Suzuki}, D., {Bennett}, D.~P., {Sumi}, T., {et~al.} 2016{\natexlab{a}}, \apj,
  833, 145, \dodoi{10.3847/1538-4357/833/2/145}

\bibitem[{{Suzuki} {et~al.}(2018){Suzuki}, {Bennett}, {Ida}, {Mordasini},
  {Bhattacharya}, {Bond}, {Donachie}, {Fukui}, {Hirao}, {Koshimoto},
  {Miyazaki}, {Nagakane}, {Ranc}, {Rattenbury}, {Sumi}, {Alibert}, \&
  {Lin}}]{2018ApJSuzuki}
{Suzuki}, D., {Bennett}, D.~P., {Ida}, S., {et~al.} 2018, \apjl, 869, L34,
  \dodoi{10.3847/2041-8213/aaf577}

\bibitem[{{Suzuki} {et~al.}(2016{\natexlab{b}}){Suzuki}, {Ogihara},
  {Morbidelli}, {Crida}, \& {Guillot}}]{2016AASuzuki}
{Suzuki}, T.~K., {Ogihara}, M., {Morbidelli}, A., {Crida}, A., \& {Guillot}, T.
  2016{\natexlab{b}}, \aap, 596, A74, \dodoi{10.1051/0004-6361/201628955}

\bibitem[{{Szul{\'a}gyi} \& {Mordasini}(2017)}]{2017MNRASSzulagyiMordasini}
{Szul{\'a}gyi}, J., \& {Mordasini}, C. 2017, \mnras, 465, L64,
  \dodoi{10.1093/mnrasl/slw212}

\bibitem[{Tabone {et~al.}(2022)Tabone, Rosotti, Lodato, Armitage, Cridland, \&
  {van Dishoeck}}]{Tabone2022}
Tabone, B., Rosotti, G.~P., Lodato, G., {et~al.} 2022, Monthly Notices of the
  Royal Astronomical Society, 512, L74, \dodoi{10.1093/mnrasl/slab124}

\bibitem[{Tabone {et~al.}(2023)Tabone, Bettoni, {van Dishoeck}, Arabhavi,
  Grant, Gasman, Henning, Kamp, G{\"u}del, Lagage, Ray, Vandenbussche, Abergel,
  Absil, Argyriou, Barrado, Boccaletti, Bouwman, {Caratti o Garatti}, Geers,
  Glauser, Justannont, Lahuis, Mueller, Nehm{\'e}, Olofsson, Pantin,
  Scheithauer, Waelkens, Waters, Black, Christiaens, Guadarrama,
  {Morales-Calder{\'o}n}, Jang, Kanwar, Pawellek, Perotti, Perrin,
  {Rodgers-Lee}, Samland, Schreiber, Schwarz, Colina, {\"O}stlin, \&
  Wright}]{Tabone2023}
Tabone, B., Bettoni, G., {van Dishoeck}, E.~F., {et~al.} 2023, Nature
  Astronomy, 7, 805, \dodoi{10.1038/s41550-023-01965-3}

\bibitem[{{Tanaka} {et~al.}(2002){Tanaka}, {Takeuchi}, \&
  {Ward}}]{2002ApJTanaka}
{Tanaka}, H., {Takeuchi}, T., \& {Ward}, W.~R. 2002, \apj, 565, 1257,
  \dodoi{10.1086/324713}

\bibitem[{{Teague} {et~al.}(2018){Teague}, {Bae}, {Bergin}, {Birnstiel}, \&
  {Foreman-Mackey}}]{Teague2018ApJ}
{Teague}, R., {Bae}, J., {Bergin}, E.~A., {Birnstiel}, T., \& {Foreman-Mackey},
  D. 2018, \apjl, 860, L12, \dodoi{10.3847/2041-8213/aac6d7}

\bibitem[{{Thiabaud} {et~al.}(2014){Thiabaud}, {Marboeuf}, {Alibert}, {Cabral},
  {Leya}, \& {Mezger}}]{2014AAThiabaud}
{Thiabaud}, A., {Marboeuf}, U., {Alibert}, Y., {et~al.} 2014, \aap, 562, A27,
  \dodoi{10.1051/0004-6361/201322208}

\bibitem[{{Thiabaud} {et~al.}(2015){Thiabaud}, {Marboeuf}, {Alibert}, {Leya},
  \& {Mezger}}]{2015A&AThiabaud}
{Thiabaud}, A., {Marboeuf}, U., {Alibert}, Y., {Leya}, I., \& {Mezger}, K.
  2015, \aap, 574, A138, \dodoi{10.1051/0004-6361/201424868}

\bibitem[{{Thommes} {et~al.}(2003){Thommes}, {Duncan}, \&
  {Levison}}]{2003IcarusThommes}
{Thommes}, E.~W., {Duncan}, M.~J., \& {Levison}, H.~F. 2003, \icarus, 161, 431,
  \dodoi{10.1016/S0019-1035(02)00043-X}

\bibitem[{{Tobin} {et~al.}(2020){Tobin}, {Sheehan}, {Megeath},
  {D{\'\i}az-Rodr{\'\i}guez}, {Offner}, {Murillo}, {van 't Hoff}, {van
  Dishoeck}, {Osorio}, {Anglada}, {Furlan}, {Stutz}, {Reynolds}, {Karnath},
  {Fischer}, {Persson}, {Looney}, {Li}, {Stephens}, {Chandler}, {Cox},
  {Dunham}, {Tychoniec}, {Kama}, {Kratter}, {Kounkel}, {Mazur}, {Maud},
  {Patel}, {Perez}, {Sadavoy}, {Segura-Cox}, {Sharma}, {Stephenson}, {Watson},
  \& {Wyrowski}}]{2020ApJTobinA}
{Tobin}, J.~J., {Sheehan}, P.~D., {Megeath}, S.~T., {et~al.} 2020, \apj, 890,
  130, \dodoi{10.3847/1538-4357/ab6f64}

\bibitem[{{Toomre}(1964)}]{Toomre1964}
{Toomre}, A. 1964, \apj, 139, 1217, \dodoi{10.1086/147861}

\bibitem[{{Turner} {et~al.}(2014){Turner}, {Fromang}, {Gammie}, {Klahr},
  {Lesur}, {Wardle}, \& {Bai}}]{2014PPVITurner}
{Turner}, N.~J., {Fromang}, S., {Gammie}, C., {et~al.} 2014, in Protostars and
  Planets VI, ed. H.~{Beuther}, R.~S. {Klessen}, C.~P. {Dullemond}, \&
  T.~{Henning}, 411, \dodoi{10.2458/azu\_uapress\_9780816531240-ch018}

\bibitem[{{Tychoniec} {et~al.}(2018){Tychoniec}, {Tobin}, {Karska}, {Chandler},
  {Dunham}, {Harris}, {Kratter}, {Li}, {Looney}, \&
  {Melis}}]{2018ApJSTychoniec}
{Tychoniec}, {\L}., {Tobin}, J.~J., {Karska}, A., {et~al.} 2018, \apjs, 238,
  19, \dodoi{10.3847/1538-4365/aaceae}

\bibitem[{{Udry} {et~al.}(2003){Udry}, {Mayor}, \& {Santos}}]{Udry2003}
{Udry}, S., {Mayor}, M., \& {Santos}, N.~C. 2003, \aap, 407, 369,
  \dodoi{10.1051/0004-6361:20030843}

\bibitem[{{Udry} \& {Santos}(2007)}]{2007ARAAUdrySantos}
{Udry}, S., \& {Santos}, N.~C. 2007, \araa, 45, 397,
  \dodoi{10.1146/annurev.astro.45.051806.110529}

\bibitem[{{Umurhan} {et~al.}(2020){Umurhan}, {Estrada}, \&
  {Cuzzi}}]{Umurhan2020}
{Umurhan}, O.~M., {Estrada}, P.~R., \& {Cuzzi}, J.~N. 2020, \apj, 895, 4,
  \dodoi{10.3847/1538-4357/ab899d}

\bibitem[{{van Boekel} {et~al.}(2017){van Boekel}, {Henning}, {Menu}, {de
  Boer}, {Langlois}, {M{\"u}ller}, {Avenhaus}, {Boccaletti}, {Schmid},
  {Thalmann}, {Benisty}, {Dominik}, {Ginski}, {Girard}, {Gisler}, {Lobo Gomes},
  {Menard}, {Min}, {Pavlov}, {Pohl}, {Quanz}, {Rabou}, {Roelfsema}, {Sauvage},
  {Teague}, {Wildi}, \& {Zurlo}}]{vanBoekel2017}
{van Boekel}, R., {Henning}, T., {Menu}, J., {et~al.} 2017, \apj, 837, 132,
  \dodoi{10.3847/1538-4357/aa5d68}

\bibitem[{{van Dishoeck} {et~al.}(2023){van Dishoeck}, {Grant}, {Tabone}, {van
  Gelder}, {Francis}, {Tychoniec}, {Bettoni}, {Arabhavi}, {Gasman}, {Nazari},
  {Vlasblom}, {Kavanagh}, {Christiaens}, {Klaassen}, {Beuther}, {Henning}, \&
  {Kamp}}]{vanDishoeck2023}
{van Dishoeck}, E.~F., {Grant}, S., {Tabone}, B., {et~al.} 2023, Faraday
  Discussions, 245, 52, \dodoi{10.1039/D3FD00010A}

\bibitem[{{Van Eylen} {et~al.}(2018){Van Eylen}, {Agentoft}, {Lundkvist},
  {Kjeldsen}, {Owen}, {Fulton}, {Petigura}, \& {Snellen}}]{vaneylen2018}
{Van Eylen}, V., {Agentoft}, C., {Lundkvist}, M.~S., {et~al.} 2018, \mnras,
  479, 4786, \dodoi{10.1093/mnras/sty1783}

\bibitem[{{Vazan} {et~al.}(2013){Vazan}, {Kovetz}, {Podolak}, \&
  {Helled}}]{2013MNRASVazan}
{Vazan}, A., {Kovetz}, A., {Podolak}, M., \& {Helled}, R. 2013, \mnras, 434,
  3283, \dodoi{10.1093/mnras/stt1248}

\bibitem[{Vazan {et~al.}(2022)Vazan, Sari, \& Kessel}]{Vazan2022}
Vazan, A., Sari, R., \& Kessel, R. 2022, The Astrophysical Journal, 926, 150,
  \dodoi{10.3847/1538-4357/ac458c}

\bibitem[{{Venturini} {et~al.}(2020){Venturini}, {Guilera}, {Haldemann},
  {Ronco}, \& {Mordasini}}]{Venturini2020}
{Venturini}, J., {Guilera}, O.~M., {Haldemann}, J., {Ronco}, M.~P., \&
  {Mordasini}, C. 2020, \aap, 643, L1, \dodoi{10.1051/0004-6361/202039141}

\bibitem[{{Vigan} {et~al.}(2021){Vigan}, {Fontanive}, {Meyer}, {Biller},
  {Bonavita}, {Feldt}, {Desidera}, {Marleau}, {Emsenhuber}, {Galicher}, {Rice},
  {Forgan}, {Mordasini}, {Gratton}, {Le Coroller}, {Maire}, {Cantalloube},
  {Chauvin}, {Cheetham}, {Hagelberg}, {Lagrange}, {Langlois}, {Bonnefoy},
  {Beuzit}, {Boccaletti}, {D'Orazi}, {Delorme}, {Dominik}, {Henning}, {Janson},
  {Lagadec}, {Lazzoni}, {Ligi}, {Menard}, {Mesa}, {Messina}, {Moutou},
  {M{\"u}ller}, {Perrot}, {Samland}, {Schmid}, {Schmidt}, {Sissa}, {Turatto},
  {Udry}, {Zurlo}, {Abe}, {Antichi}, {Asensio-Torres}, {Baruffolo}, {Baudoz},
  {Baudrand}, {Bazzon}, {Blanchard}, {Bohn}, {Brown Sevilla}, {Carbillet},
  {Carle}, {Cascone}, {Charton}, {Claudi}, {Costille}, {De Caprio},
  {Delboulb{\'e}}, {Dohlen}, {Engler}, {Fantinel}, {Feautrier}, {Fusco},
  {Gigan}, {Girard}, {Giro}, {Gisler}, {Gluck}, {Gry}, {Hubin}, {Hugot},
  {Jaquet}, {Kasper}, {Le Mignant}, {Llored}, {Madec}, {Magnard}, {Martinez},
  {Maurel}, {M{\"o}ller-Nilsson}, {Mouillet}, {Moulin}, {Orign{\'e}}, {Pavlov},
  {Perret}, {Petit}, {Pragt}, {Puget}, {Rabou}, {Ramos}, {Rickman}, {Rigal},
  {Rochat}, {Roelfsema}, {Rousset}, {Roux}, {Salasnich}, {Sauvage}, {Sevin},
  {Soenke}, {Stadler}, {Suarez}, {Wahhaj}, {Weber}, \& {Wildi}}]{2021AAVigan}
{Vigan}, A., {Fontanive}, C., {Meyer}, M., {et~al.} 2021, \aap, 651, A72,
  \dodoi{10.1051/0004-6361/202038107}

\bibitem[{{Voelkel} {et~al.}(2021){Voelkel}, {Deienno}, {Kretke}, \&
  {Klahr}}]{2021AAVoelkel}
{Voelkel}, O., {Deienno}, R., {Kretke}, K., \& {Klahr}, H. 2021, \aap, 645,
  A132, \dodoi{10.1051/0004-6361/202039245}

\bibitem[{{Voelkel} {et~al.}(2022){Voelkel}, {Klahr}, {Mordasini}, \&
  {Emsenhuber}}]{2022AAVoelkel}
{Voelkel}, O., {Klahr}, H., {Mordasini}, C., \& {Emsenhuber}, A. 2022, \aap,
  666, A90, \dodoi{10.1051/0004-6361/202141830}

\bibitem[{{Voelkel} {et~al.}(2020){Voelkel}, {Klahr}, {Mordasini},
  {Emsenhuber}, \& {Lenz}}]{2020AAVoelkel}
{Voelkel}, O., {Klahr}, H., {Mordasini}, C., {Emsenhuber}, A., \& {Lenz}, C.
  2020, \aap, 642, A75, \dodoi{10.1051/0004-6361/202038085}

\bibitem[{{Wafflard-Fernandez} \& Lesur(2023)}]{Wafflard-Fernandez2023}
{Wafflard-Fernandez}, G., \& Lesur, G. 2023, Astronomy \& Astrophysics, 677,
  A70, \dodoi{10.1051/0004-6361/202245305}

\bibitem[{{Wagner} {et~al.}(2019){Wagner}, {Apai}, \&
  {Kratter}}]{2019ApJWagner}
{Wagner}, K., {Apai}, D., \& {Kratter}, K.~M. 2019, \apj, 877, 46,
  \dodoi{10.3847/1538-4357/ab1904}

\bibitem[{{Wahl} {et~al.}(2017){Wahl}, {Hubbard}, {Militzer}, {Guillot},
  {Miguel}, {Movshovitz}, {Kaspi}, {Helled}, {Reese}, {Galanti}, {Levin},
  {Connerney}, \& {Bolton}}]{2017GRLWahl}
{Wahl}, S.~M., {Hubbard}, W.~B., {Militzer}, B., {et~al.} 2017, \grl, 44, 4649,
  \dodoi{10.1002/2017GL073160}

\bibitem[{Walsh {et~al.}(2015)Walsh, Nomura, \& {van Dishoeck}}]{Walsh2015}
Walsh, C., Nomura, H., \& {van Dishoeck}, E. 2015, Astronomy and Astrophysics,
  582, A88, \dodoi{10.1051/0004-6361/201526751}

\bibitem[{{Walsh} {et~al.}(2011){Walsh}, {Morbidelli}, {Raymond}, {O'Brien}, \&
  {Mandell}}]{2011NatureWalsh}
{Walsh}, K.~J., {Morbidelli}, A., {Raymond}, S.~N., {O'Brien}, D.~P., \&
  {Mandell}, A.~M. 2011, \nat, 475, 206, \dodoi{10.1038/nature10201}

\bibitem[{{Wang} {et~al.}(2017){Wang}, {Weiss}, {Bai}, {Downey}, {Wang},
  {Wang}, {Suavet}, {Fu}, \& {Zucolotto}}]{2017ScienceWang}
{Wang}, H., {Weiss}, B.~P., {Bai}, X.-N., {et~al.} 2017, \sci, 355, 623,
  \dodoi{10.1126/science.aaf5043}

\bibitem[{{Wang} {et~al.}(2021){Wang}, {Vigan}, {Lacour}, {Nowak}, {Stolker},
  {De Rosa}, {Ginzburg}, {Gao}, {Abuter}, {Amorim}, {Asensio-Torres},
  {Baub{\"o}ck}, {Benisty}, {Berger}, {Beust}, {Beuzit}, {Blunt}, {Boccaletti},
  {Bohn}, {Bonnefoy}, {Bonnet}, {Brandner}, {Cantalloube}, {Caselli},
  {Charnay}, {Chauvin}, {Choquet}, {Christiaens}, {Cl{\'e}net}, {Coud{\'e} Du
  Foresto}, {Cridland}, {de Zeeuw}, {Dembet}, {Dexter}, {Drescher}, {Duvert},
  {Eckart}, {Eisenhauer}, {Facchini}, {Gao}, {Garcia}, {Garcia Lopez},
  {Gardner}, {Gendron}, {Genzel}, {Gillessen}, {Girard}, {Haubois},
  {Hei{\ss}el}, {Henning}, {Hinkley}, {Hippler}, {Horrobin}, {Houll{\'e}},
  {Hubert}, {Jim{\'e}nez-Rosales}, {Jocou}, {Kammerer}, {Keppler}, {Kervella},
  {Meyer}, {Kreidberg}, {Lagrange}, {Lapeyr{\`e}re}, {Le Bouquin}, {L{\'e}na},
  {Lutz}, {Maire}, {M{\'e}nard}, {M{\'e}rand}, {Molli{\`e}re}, {Monnier},
  {Mouillet}, {M{\"u}ller}, {Nasedkin}, {Ott}, {Otten}, {Paladini}, {Paumard},
  {Perraut}, {Perrin}, {Pfuhl}, {Pueyo}, {Rameau}, {Rodet},
  {Rodr{\'\i}guez-Coira}, {Rousset}, {Scheithauer}, {Shangguan}, {Shimizu},
  {Stadler}, {Straub}, {Straubmeier}, {Sturm}, {Tacconi}, {van Dishoeck},
  {Vincent}, {von Fellenberg}, {Ward-Duong}, {Widmann}, {Wieprecht},
  {Wiezorrek}, {Woillez}, \& {Gravity Collaboration}}]{Wang2021}
{Wang}, J.~J., {Vigan}, A., {Lacour}, S., {et~al.} 2021, \aj, 161, 148,
  \dodoi{10.3847/1538-3881/abdb2d}

\bibitem[{{Ward}(1997)}]{1997IcarusWard}
{Ward}, W.~R. 1997, \icarus, 126, 261, \dodoi{10.1006/icar.1996.5647}

\bibitem[{Weder {et~al.}(2023)Weder, Mordasini, \& Emsenhuber}]{Weder2023}
Weder, J., Mordasini, C., \& Emsenhuber, A. 2023, Astronomy and Astrophysics,
  674, A165, \dodoi{10.1051/0004-6361/202243453}

\bibitem[{{Weidenschilling}(1977)}]{1977ApSSWeidenschilling}
{Weidenschilling}, S.~J. 1977, \apss, 51, 153, \dodoi{10.1007/BF00642464}

\bibitem[{{Weiss} {et~al.}(2022){Weiss}, {Millholland}, {Petigura}, {Adams},
  {Batygin}, {Bloch}, \& {Mordasini}}]{2022PPVIIWeiss}
{Weiss}, L.~M., {Millholland}, S.~C., {Petigura}, E.~A., {et~al.} 2022, arXiv
  e-prints, arXiv:2203.10076.
\newblock \doarXiv{2203.10076}

\bibitem[{{Weiss} {et~al.}(2018){Weiss}, {Marcy}, {Petigura}, {Fulton},
  {Howard}, {Winn}, {Isaacson}, {Morton}, {Hirsch}, {Sinukoff}, {Cumming},
  {Hebb}, \& {Cargile}}]{2018AJWeiss}
{Weiss}, L.~M., {Marcy}, G.~W., {Petigura}, E.~A., {et~al.} 2018, \aj, 155, 48,
  \dodoi{10.3847/1538-3881/aa9ff6}

\bibitem[{{Wetherill} \& {Stewart}(1993)}]{1993IcarusWetherillStewart}
{Wetherill}, G.~W., \& {Stewart}, G.~R. 1993, \icarus, 106, 190,
  \dodoi{10.1006/icar.1993.1166}

\bibitem[{Whipple(1972)}]{Whipple1972}
Whipple, F.~L. 1972, From Plasma to Planet, 211

\bibitem[{{Winn} \& {Fabrycky}(2015)}]{2015ARA&AWinn}
{Winn}, J.~N., \& {Fabrycky}, D.~C. 2015, \araa, 53, 409,
  \dodoi{10.1146/annurev-astro-082214-122246}

\bibitem[{{Winter} \& {Haworth}(2022)}]{2022EPJPWinterHaworth}
{Winter}, A.~J., \& {Haworth}, T.~J. 2022, European Physical Journal Plus, 137,
  1132, \dodoi{10.1140/epjp/s13360-022-03314-1}

\bibitem[{Woitke {et~al.}(2009)Woitke, Kamp, \& Thi}]{Woitke2009}
Woitke, P., Kamp, I., \& Thi, W.~F. 2009, Astronomy and Astrophysics, 501, 383,
  \dodoi{10.1051/0004-6361/200911821}

\bibitem[{{Wolthoff} {et~al.}(2022){Wolthoff}, {Reffert}, {Quirrenbach},
  {Jones}, {Wittenmyer}, \& {Jenkins}}]{Wolthoff2022}
{Wolthoff}, V., {Reffert}, S., {Quirrenbach}, A., {et~al.} 2022, \aap, 661,
  A63, \dodoi{10.1051/0004-6361/202142501}

\bibitem[{{Woo} {et~al.}(2023){Woo}, {Morbidelli}, {Grimm}, {Stadel}, \&
  {Brasser}}]{Woo2023Icarus}
{Woo}, J.~M.~Y., {Morbidelli}, A., {Grimm}, S.~L., {Stadel}, J., \& {Brasser},
  R. 2023, \icarus, 396, 115497, \dodoi{10.1016/j.icarus.2023.115497}

\bibitem[{{Xie} {et~al.}(2016){Xie}, {Dong}, {Zhu}, {Huber}, {Zheng}, {De Cat},
  {Fu}, {Liu}, {Luo}, {Wu}, {Zhang}, {Zhang}, {Zhou}, {Cao}, {Hou}, {Wang}, \&
  {Zhang}}]{2016PNASXie}
{Xie}, J.-W., {Dong}, S., {Zhu}, Z., {et~al.} 2016, \pnas, 113, 11431,
  \dodoi{10.1073/pnas.1604692113}

\bibitem[{{Xu} \& {Bonsor}(2021)}]{XuBonsor2021Eleme}
{Xu}, S., \& {Bonsor}, A. 2021, Elements, 17, 241,
  \dodoi{10.2138/gselements.17.4.241}

\bibitem[{{Yin} {et~al.}(2002){Yin}, {Jacobsen}, {Yamashita}, {Blichert-Toft},
  {T{\'e}louk}, \& {Albar{\`e}de}}]{Yin2002Nature}
{Yin}, Q., {Jacobsen}, S.~B., {Yamashita}, K., {et~al.} 2002, \nat, 418, 949,
  \dodoi{10.1038/nature00995}

\bibitem[{{Youdin} \& {Goodman}(2005)}]{2005ApJYoudinGoodman}
{Youdin}, A.~N., \& {Goodman}, J. 2005, \apj, 620, 459, \dodoi{10.1086/426895}

\bibitem[{{Youdin} \& {Lithwick}(2007)}]{2007IcarusYoudinLithwick}
{Youdin}, A.~N., \& {Lithwick}, Y. 2007, \icarus, 192, 588,
  \dodoi{10.1016/j.icarus.2007.07.012}

\bibitem[{{Youdin} \& {Shu}(2002)}]{Youdin2002}
{Youdin}, A.~N., \& {Shu}, F.~H. 2002, \apj, 580, 494, \dodoi{10.1086/343109}

\bibitem[{{Young} {et~al.}(2023){Young}, {Shahar}, \&
  {Schlichting}}]{Young2023Nature}
{Young}, E.~D., {Shahar}, A., \& {Schlichting}, H.~E. 2023, \nat, 616, 306,
  \dodoi{10.1038/s41586-023-05823-0}

\bibitem[{{Yu} {et~al.}(2017){Yu}, {Donati}, {H{\'e}brard}, {Moutou}, {Malo},
  {Grankin}, {Hussain}, {Collier Cameron}, {Vidotto}, {Baruteau}, {Alencar},
  {Bouvier}, {Petit}, {Takami}, {Herczeg}, {Gregory}, {Jardine}, {Morin},
  {M{\'e}nard}, \& {Matysse Collaboration}}]{YuDonati2017}
{Yu}, L., {Donati}, J.~F., {H{\'e}brard}, E.~M., {et~al.} 2017, \mnras, 467,
  1342, \dodoi{10.1093/mnras/stx009}

\bibitem[{{Zhu} \& {Dong}(2021)}]{2021ARAAZhuDong}
{Zhu}, W., \& {Dong}, S. 2021, \araa, 59,
  \dodoi{10.1146/annurev-astro-112420-020055}

\bibitem[{{Zhu} {et~al.}(2018){Zhu}, {Petrovich}, {Wu}, {Dong}, \&
  {Xie}}]{ZhuPetrovich2018}
{Zhu}, W., {Petrovich}, C., {Wu}, Y., {Dong}, S., \& {Xie}, J. 2018, \apj, 860,
  101, \dodoi{10.3847/1538-4357/aac6d5}

\end{thebibliography}

\end{document}